\documentclass[structabstract]{aa}
\usepackage{graphicx}
\usepackage{longtable,lscape}
\usepackage{rotating}

\usepackage{natbib}
\usepackage{amssymb}
\usepackage{txfonts}

\usepackage{fixltx2e}

\newcommand{\hii}{\relax \ifmmode {\mbox H\,{\scshape ii}}\else H\,{\scshape ii}\fi}
\newcommand{\mi}{\relax \ifmmode {\mu{\mbox m}}\else $\mu$m\fi}
\newcommand{\ha}{\relax \ifmmode {\mbox H}\alpha\else H$\alpha$\fi}
\newcommand{\hb}{\relax \ifmmode {\mbox H}\beta\else H$\beta$\fi}
\newcommand{\hg}{\relax \ifmmode {\mbox H}\gamma\else H$\gamma$\fi}
\newcommand{\hd}{\relax \ifmmode {\mbox H}\delta\else H$\delta$\fi}

\newcommand{\sii}{\relax \ifmmode {\mbox S\,{\scshape ii}}\else S\,{\scshape ii}\fi}
\newcommand{\siii}{\relax \ifmmode {\mbox S\,{\scshape iii}}\else S\,{\scshape iii}\fi}
\newcommand{\nii}{\relax \ifmmode {\mbox N\,{\scshape ii}}\else N\,{\scshape ii}\fi}
\newcommand{\oii}{\relax \ifmmode {\mbox O\,{\scshape ii}}\else O\,{\scshape ii}\fi}
\newcommand{\oiii}{\relax \ifmmode {\mbox O\,{\scshape iii}}\else O\,{\scshape iii}\fi}
\newcommand{\neiii}{\relax \ifmmode {\mbox Ne\,{\scshape iii}}\else Ne\,{\scshape iii}\fi}
 
\newcommand{\rdostres}{\relax \ifmmode {\,\mbox{R}}_{\rm 23}\else \,\mbox{R}$_{\rm 23}$\fi}

\begin{document}

\title{Characterization of star-forming dwarf galaxies at 0.1\,$\lesssim z \lesssim$\,0.9 in  VUDS\thanks{Based on data obtained with the European Southern Observatory Very Large Telescope, Paranal, Chile, under Large Program \mbox{185.A-0791}.}: Probing the low-mass end of the mass-metallicity relation 
   }
\author{A. Calabr\`o \inst{1} 
\and R. Amor\'in\inst{1,2,3}
\and A. Fontana\inst{1}
\and E. P\'erez-Montero\inst{11}
\and B. C. Lemaux \inst{12}
\and B. Ribeiro\inst{4}
\and S. Bardelli\inst{7}
\and \\M. Castellano\inst{1}
\and T. Contini\inst{8,9}
\and S. De Barros\inst{10}
\and B. Garilli\inst{5}
\and A. Grazian\inst{1}
\and L. Guaita\inst{1}
\and N. P. Hathi\inst{4,13}
\and A. M. Koekemoer\inst{13}
\and O. Le F\`evre\inst{4}
\and D. Maccagni\inst{5}
\and L. Pentericci\inst{1}
\and D. Schaerer\inst{10,8}
\and M. Talia\inst{6}
\and L. A. M. Tasca\inst{4} 
\and E. Zucca\inst{7} 
}

\offprints{A. Calabr\`o \email{nillo2412@gmail.com}}

\institute{INAF--Osservatorio Astronomico di Roma, via di Frascati 33, I-00040,  Monte Porzio Catone, Italy 
\and
Cavendish Laboratory, University of Cambridge, 19 JJ Thomson Avenue, Cambridge, CB3 0HE, UK.
\and
Kavli Institute for Cosmology, University of Cambridge, Madingley Road, Cambridge CB3 0HA, UK
\and
Aix Marseille Universit\'e, CNRS, LAM (Laboratoire d'Astrophysique  de Marseille) UMR 7326, 13388, Marseille, France
\and
INAF--IASF, via Bassini 15, I-20133,  Milano, Italy
\and
University of Bologna, Department of Physics and Astronomy (DIFA), V.le Berti Pichat, 6/2 - 40127, Bologna
\and
INAF--Osservatorio Astronomico di Bologna, via Ranzani,1, I-40127, Bologna, Italy
\and
IRAP, Institut de Recherche en Astrophysique et Plan\'etologie, CNRS, 14 avenue Edouard Belin, 31400 Toulouse, France
\and
Universit\'e de Toulouse, UPS-OMP, 31400 Toulouse, France
\and
Geneva Observatory, University of Geneva, ch. des Maillettes 51, CH-1290 Versoix, Switzerland
\and
Instituto de Astrof\'isica de Andaluc\'ia. CSIC. Apartado de correos 3004. 18080, Granada, Spain
\and
Department of Physics, University of California, Davis, One Shields Ave., Davis, CA 95616, USA
\and
Space Telescope Science Institute, 3700 San Martin Drive, Baltimore, MD 21218, USA
}       
\date{Received 21,09,2016; Accepted 15,01,2017}  

\abstract
{The study of statistically significant samples of star-forming dwarf galaxies (SFDGs) at different cosmic epochs is essential for the detailed understanding of galaxy assembly and chemical evolution. However, the main properties of this numerous population of galaxies at intermediate redshift are still poorly known.}
{We present the discovery and spectrophotometric characterization of a large sample of 164 faint ($i_{\rm AB} \sim$\,23-25 mag) SFDGs at redshift $0.13 \leq z \leq 0.88$ selected by the presence of bright optical emission lines in the VIMOS Ultra Deep Survey (VUDS). We investigate their integrated physical properties and ionization conditions, which are used to discuss the low-mass end of the mass-metallicity relation (MZR) and other key scaling relations.}
{We use optical VUDS spectra in the COSMOS, VVDS-02h, and ECDF-S fields, as well as deep multiwavelength photometry which includes HST-ACS \textit{F814W} imaging, to derive stellar masses, extinction-corrected star formation rates (SFR) and gas-phase metallicities of SFDGs. For the latter, we use the direct method and a $T_e$-consistent approach based on the comparison of a set of observed emission lines ratios with the predictions of detailed photoionization models.} 
{The VUDS SFDGs are compact (median $r_{\rm e} \sim$\,1.2 kpc), low-mass (M$_{*} \sim$\,10$^7$-10$^9$\,M$_{\odot}$) galaxies with a wide range of star formation rates (SFR(H$\alpha$)\,$\sim$\, $10^{-3}$-$10^{1}$\ M$_{\odot}/$yr) and morphologies. Overall, they show a broad range of subsolar metallicities (12$+\log$(O/H)$=$\,$7.26$-$8.7$\ ; 0.04\,$\la Z/Z_{\odot} \la$\,1). Nearly half of the sample are extreme emission-line galaxies (EELGs) characterized by high equivalent widths and emission line ratios indicative of higher excitation and ionization conditions. The MZR of SFDGs shows a flatter slope compared to previous studies of galaxies in the same mass range and redshift. We find the scatter of the MZR partly explained in the low mass range by varying specific SFRs and gas fractions amongst the galaxies in our sample. In agreement with recent studies, we find the subclass of EELGs to be systematically offset to lower metallicity compared to SFDGs at a given stellar mass and SFR, suggesting a younger starburst phase. Compared with simple chemical evolution models we find that most SFDGs do not follow the predictions of a "closed-box" model, but those from a gas regulating model in which gas flows are considered. While strong stellar feedback may produce large-scale outflows favoring the cessation of vigorous star formation and promoting the removal of metals, younger and more metal-poor dwarfs may have recently accreted large amounts of fresh, very metal-poor gas, that is used to fuel current star formation. 
}
{} 
 

\titlerunning {Star-forming dwarfs at intermediate-z in VUDS}        
\authorrunning{A. Calabr\`o et al.}
  \maketitle


\section{Introduction}\label{introduction}

Low mass (dwarf) galaxies are the most abundant systems of the Universe at all cosmic epochs, as shown by catalogues of nearby galaxies \citep{karachentsev04} and by the steepness of the galaxy stellar mass ($M_\ast$) function at $M_\ast < 10^{10} M_\odot$ up to high redshift  \citep{fontana06,santini12,grazian15}. The most common accepted definition of dwarf galaxies refers to low mass ($M_\ast < 10^9 M_\odot$) and low luminosity systems, with $M_{i}-5$ log$h_{100}>-18$ \citep{sanchezjanssen13}. They are considered the building blocks from which more massive galaxies form \citep{pressschechter74}. This assembly process is not constant, but it peaks at $z \sim 2$ and then declines exponentially at later times \citep[e.g.][]{madaudickinson14}. Almost $25\%$ of the stellar mass observed today has been assembled after this peak, and a significant part of it formed in young low mass galaxies in strong, short-lived starbursts \citep{guzman97,kakazu07}. Some of these star-forming low mass galaxies also show bright emission lines in their optical rest-frame spectra, due to photoionization of the nebula surrounding hot massive (O type) stars. This makes them easier to identify even beyond the Local Universe in current spectroscopic surveys. Throughout this paper we will refer to this kind of faint galaxies with bright emission lines detected in the optical ([\oii],[\oiii],\hb\ and \ha) to as star-forming dwarf galaxies (SFDG). Among them, a particular subset is represented by extreme emission line galaxies (EELG), which are selected by the high equivalent width (EW) of their optical emission lines (EW[\oiii]$>$100-200 \AA ), and have more extreme properties, e.g., higher surface densities, lower starburst ages and lower gas metallicities) than the average population of star-forming dwarfs \citep{kniazev04,cardamone09,amorin10,amorin12,atek11,vanderwel11,maseda14,amorin14,amorin15}. 
While the population of EELGs constitute by their own an ideal laboratory to study star formation and chemical enrichment under extreme physical conditions, they appear as the closest environments to those believed to be more common of typical galaxies at very high redshifts \citep[$z\ga$\,3-4, e.g.][]{smit14,stark17}.

Tracing the galaxy-averaged properties of large, representative samples of star-forming dwarf galaxies (SFDGs) since $z \sim 1.5\ $ is a necessary step for a complete understanding of the evolution of low mass galaxies and the build-up of stellar mass during the last $9$-$10$ billion years. One of the unanswered questions is about how SFDGs assemble their stellar mass. Differently from high mass ($M_\ast > 10^9 M_\odot$) galaxies which show a continuous rate of star formation (SF), the most common scenario for dwarf galaxies is the cyclic bursty mode, as pointed out either by theoretical models \citep[e.g.][]{hopkins14,sparre15} and observations \citep{guo16}. Intense SF episodes produce stellar feedback through strong winds and supernova, which heat and expel the surrounding gas in outflows, eventually resulting in a temporary quenching of SF on timescales of tens of Myr \citep[e.g.][]{olmogarcia16,pelupessy04}. Then the metal-enriched gas may be accreted back to the galaxy triggering new star forming episodes. The scaling relation between stellar mass $M_\ast$ and star formation rate (SFR=$M_\ast$ produced per year) can be used to compare the assembly dynamics of different types of galaxies. For high mass ($M_\ast > 10^9 M_\odot$) star-forming galaxies a tight correlation was found between the two quantities at all reshift from z$=$0 to z$=$5, called the "Main Sequence" (MS) of star-formation \citep{brinchmann04,noeske07,daddi07,tasca15}. This sequence moves towards higher SFRs at higher z, though its slope remains nearly constant ($\sim1\,$) \citep{guo15}. At lower masses, SFDGs and in particular those with the strongest emission lines (EELGs), are found to have increased SFRs at fixed $M_\ast$  (by $\sim$\,1 dex) compared to the extrapolation at low mass of the MS \citep{amorin15}, suggesting that they are efficiently forming stars in strong bursts with stellar mass doubling times $<$\,1 Gyr, which can not be sustained for long. 

The mechanisms regulating galaxy growth, such as gas accretion and SF feedback are still not completely understood, and the scarcity of direct observations of outflows and gas accretion (as well as a quantification of their rate) represent a limit to a deeper insight \citep{sanchezalmeida14b}. However, since the stellar mass build-up in galaxies is accompanied by the chemical enrichment of their interstellar medium (ISM), studying the gas-phase metallicity and its relation with stellar mass and SFR can help us to investigate which of these processes are playing a major role. Thus, the gas-phase metallicity (defined as the oxygen abundance, $12$+log(O/H)) is found to tightly correlate with stellar mass \citep[e.g.][]{lequeux79,tremonti04} up to high redshift \citep[$z\sim$\,3.5, e.g.][]{maiolino08,zahid12,cullen14, troncoso14, onodera16}, with a relatively steep slope flattening above $10^{10.5} M_\odot$. The normalization of this mass-metallicity relation (MZR) appears to evolve to lower metallicities with increasing redshift at fixed stellar mass \citep{savaglio10,shapley05,erb06,maiolino08,mannucci09}, while the slope, especially in its low-mass end, is still not constrained \citep[e.g.][]{christensen12,henry13,whitaker14,salim14}. However, both the slope and normalization of the MZR have been found to depend strongly on the method used to derive metallicity \citep{kewleyellison08}. The largest discrepancies (as high as $0.7$ dex) are between metallicities measured using the direct method and strong line methods. The former is also known as $T_e$ method, because is based on the measuring of the electron temperature ($T_e$) of the ionized gas, which requires measurements of weak auroral lines, such as [\oiii]$\lambda$4363 \citep[e.g.][]{hagele08,curti16}. The latter, instead, are based on empirical or model-based calibrations of bright emission line ratios as a function of metallicity. 

At lower masses ($M_\ast < 10^9 M_\odot$), the mass-metallicity relation is still not completely characterized. Among the various studies that have tried to investigate if a correlation exists at lower masses, \citet{lee06} (L06) derived a $T_e$ consistent MZR from 27 nearby ($D \leq 5$Mpc) star-forming dwarf galaxies (down to $M_\ast \sim 10^6 M_\odot$), with a low scatter ($\pm 1$ dex). Using stacked spectra from the SDSS-DR7, \citet{andrewsmartini13} (AM13) measured $T_e$ from weak auroral lines and used the direct method to derive  metallicities over $\sim$4 decades in stellar mass and study the MZR.
They found that the MZR has a sharp increase ($O/H \propto M_\ast^{1/2}$) from $log(M_\ast)=7.4$ to $log(M_\ast)=8.9\ M_\odot$, and flattens at $log(M_\ast)=8.9 M_\odot$. Above this value, the MZR derived from the direct method reaches an asymptotic metallicity of $12+log(O/H) = 8.8$. \citet{zahid12b} studied the mass-metallicity relation down to $10^7 M_\odot$, showing that the scatter increases towards lower stellar masses. More recently, \citet{ly14,ly15} measured $T_e$-metallicities respectively for 20 emission-line galaxies at $z=0.065$-$0.90$ (with detection of O4363 \AA\ line) and 28 $z\sim0.8$\ metal-poor galaxies with median stellar mass $M_\ast =8.7$.

The correlation between stellar mass and metallicity is a natural consequence of the conversion of gas into stars within galaxies, regulated by gas exchanges with the environment through inflows or outflows, but we still don't know exactly which processes influence the shape, normalization and scatter of this relation. Besides observations, semi-analytical models and cosmological hydrodynamic simulations including chemical evolution have tried to explain the observed MZR. According to \citet{kobayashi07} and \citet{spitoni10}, the galactic winds can easily drive metals out of low-mass galaxies due to their lower potential wells. In other studies, dwarf galaxies have longer timescales of conversion (regulated by galactic winds) from gas reservoirs to stars, so they are simply less evolved and less enriched systems \citep{finlatordave08}. The models proposed by \citet{tassis08} introduce a critical density threshold for the activation of star formation, without requiring outflows to reduce the star formation efficiency and the metal content. Finally, the observed MZR can be explained assuming accretion of metal-poor gas along filaments from the cosmic web (cold-flows) \citep[e.g.][]{dalcanton04,ceverino15,sanchezalmeida14}, for which also indirect evidences have been found in recent observations \citep{sanchezalmeida15}.    

Lastly, SFDGs are important because among them we find analogues of high redshift galaxies, which typically show high sSFRs, low metallicity, high ionization in terms of [\oiii]$\lambda$5007/[\oii]$\lambda$3727 and compact sizes \citep{izotov15}. A recent work by \citet{stasinska15} shows that EELGs are characterized by [\oiii]$\lambda$5007/[\oii]$\lambda$3727 $>5$, reaching values up to 60 in some of them, allowing radiation to escape and ionize the surrounding ISM \citep{nakajima14}. Such types of galaxies at high redshift are thought to contribute significantly to the reionization of the Universe, providing up to $20 \%$ of the total ionizing flux at $z \sim 6$ \citep{robertson15,dressler15}. As far as the sizes are concerned, SFDGs selected by strong optical emission lines (in particular [\oiii]$\lambda$ 5007) are typically small isolated systems with radii of the order of $\sim 1$ kpc \citep{izotov16} and they show irregular morphologies \citep{amorin15}. 

This paper presents a characterization of the main integrated physical properties of a large and representative sample of 164 star-forming dwarf galaxies at $0.13 < z < 0.88$ drawn from the VIMOS Ultra-Deep Survey \citep[VUDS,]{lefevre15}. By construction, VUDS has two important advantages compared to previous surveys (e.g. zCOSMOS): (i) its unprecedented depth due to long integrations, which allows us to probe very faint targets $I_{AB} \sim 23-25$ at $z<1$, and (ii) a large area, covering three deep fields, COSMOS, ECDFS, and VVDS-02h, for which a wealth of ancillary multiwavelength data is available. These advantages are particularly important for the main goal of this paper: to investigate the metallicity and ionization of galaxies with $M_\ast$ as low as $10^7 M_\odot$ and the relation with their stellar mass, star formation rates, and sizes. 

We derive galaxy-averaged metallicity and ionization parameters for the entire sample using a new robust $\chi^2$ minimization code called HII-CHI-mistry \citep[HCm]{perezmontero14}, based on the comparison of detailed photoionization models and observed optical line ratios.
HCm is particularly efficient because it gives results that are consistent with the direct method in the entire metallicity range spanned by our sample, even in the absence of an auroral line detection (e.g. [\oiii]$\lambda$4363). 
For most galaxies we use space-based (HST-ACS) images to study their morphological properties and quantify galaxy sizes, allowing us to compare with other samples of SFDGs and study their impact on the mass-metallicity and the other scaling relations.    

Our paper is organized as follows. In section \ref{section2} we describe the parent VUDS sample, the parent photometric catalogues and the SDSS data used in this paper for comparison. We describe in section \ref{section3} the selection criteria adopted to compile our sample of SFDGs, followed by the details on emission line measurements and AGN removal. In section \ref{section4} we describe the main physical properties (stellar masses, star formation rates, morphology and sizes, metallicity and ionization) of the sample and the methodology used to derive all of them. In section \ref{section5} we present our main results and we study empirical relations between different properties, in particular the mass-metallicity relation. We also compare the results with similar samples of star-forming dwarf galaxies at low and intermediate redshift. In section \ref{discussion} we discuss our results, with the implications on galaxy stellar mass assembly. We compare our findings to the predictions of simple chemical evolution models, and we provide a sample of Lyman-continuum galaxy candidates for reionization studies. Finally, we show the summary and conclusions in section \ref{summaryconclusions}, while in an Appendix we provide tables with all our measurements and we compare the metallicities derived with HCm with those obtained using well-know strong-line calibrations, and study the effects on the MZR. 

Throughout this paper we adopt a standard $\Lambda$-CDM cosmology with $h$ = 0.7, $\Omega_m$ = 0.3 and $\Omega_\Lambda$ = 0.7. All equivalent widths (EWs) are presented in the rest-frame. We adopt $12$+log(O/H)$=8.69$ as the solar oxygen abundance \citep{asplund09}.

\section{Observations}\label{section2}

\subsection{The parent VUDS sample and redshift measurement}
The VIMOS Ultra Deep Survey (VUDS) \citep{lefevre15} is a spectroscopic redshift survey observing $\sim$\,10000 galaxies to study the major phase of galaxy assembly up to redshift $z \sim 6$. VUDS is one of the largest programs on the ESO-VLT with 640 hours of observing time. The survey covers a total of one square degree in three separate fields to reduce the impact of cosmic variance: the COSMOS field, the extended Chandra Deep Field South (ECDFS) and the VVDS-02h field. All the details about the survey strategy, target selection, data reduction and calibrations, and redshift measurements can be found in \citet{lefevre15} and \citet{tasca16}. Below we briefly summarize the survey features that are relevant to the present study. 

The spectroscopic observations were carried out at the VLT with the VIMOS Multi-Object Spectrograph (MOS), which has a wide field of 224 arcmin$^2$ \citep{lefevre03}. The spectrograph is equipped with slits $1 \arcsec$ wide and $10 \arcsec $ long, as well as two grisms (LRBLUE and LRRED) covering a wavelength range of 3650\,$<\lambda<$\,9350\AA\ at uniform spectral resolution of $R=180$ and $R=210$, respectively. This allows us to observe the Lyman-$\alpha$ line at $\lambda$\,1215\AA\ up to redshift $z\sim$\,6.6, and also $H\beta$, [\oii]$\lambda\lambda 3727$,$3729$, and [\oiii]$\lambda\lambda$\,4959,5007 emission lines for galaxies at $z\la$\,0.88. The integration time (on-source) is $\simeq$\,14 hours per target for each grism, which allow us to detect the continuum at 8500\AA\ for $i_{AB}=$\,25, and emission lines with an observed flux limit $F=$\,1.5$\times$\,10$^{-18}$\,erg~s$^ {-1}$~cm$^{-2}$ at $S/N\sim$\,5.

Redshift measurements in VUDS were performed using the EZ code \citep{garilli10}, both in automatic and manual modes for each spectrum, supervised independently by two persons. Different flags have been assigned to each galaxy according to the probability of the measurement to be correct, with flags 3 and 4 as those with the highest ($\geq$\,95\%) probability of being correct. The overall redshift accuracy is $dz/(1+z)=$\,0.0005-0.0007 (or an absolute velocity accuracy of 150-200 km~s$^{-1}$). The redshift distribution of VUDS parent sample shows the majority of galaxies at redshift $z_{\rm spec}>$\,2, while there is a lower redshift tail peaking at $z_{\rm spec}$ $\simeq 1.5$, which represents about 20\% of the total number of targets. 

\subsection{The parent photometric catalogues}\label{photometry}

The three fields of the VUDS survey (COSMOS, ECDFS and VVDS-02h) benefit of plenty of deep multi-wavelength data, which are fundamental in combination with the spectroscopic redshifts in order to derive important physical quantities of galaxies, such as stellar masses, absolute magnitudes and SED-based star formation rates.

In the COSMOS field \citep{scoville07} GALEX near-UV ($2310$\AA) and far-UV ($1530$\AA) magnitudes are available down to $m_{AB}=25.5$ \citep{zamojski07}. Extensive imaging observations were carried out with the Subaru Suprime-Cam in \textit{BVgriz} broad-bands by \citet{taniguchi07} down to $i_{AB} \sim 26.5$\ mag, as well as with CFHT Megacam in the u-band by \citet{boulade03}. The ULTRA-VISTA survey acquired very deep near-infrared imaging in the \textit{YJHK} bands with $5\sigma$AB depths of $\sim 25$ in \textit{Y} and $\sim 24$ in \textit{JHK} bands.

The ECDFS field has been studied by a wealth of photometric surveys as well. All the photometry in this field is taken from \citet{cardamone10}, who observed with Subaru Suprime-Cam in $18$ optical medium-band filters (down to $R_{AB} \sim 25.3$) as a part of the MUSYC survey \citep{gawiser06}. They also created a uniform catalogue combining their observations with ancillary photometric data available in MUSYC. They comprise \textit{UBVRI}$z'$ bands from \citet{gawiser06}, \textit{JHK} from \citet{taylor09} and Spitzer IRAC photometry ($3.6\mu m$,$4.5\mu m$, $5.8\mu m$, $8.0\mu m$) from the SIMPLE survey \citep{damen11}. 

The VVDS-02h field was observed with \textit{u*g'r'i'z'} filters as part of CFHT Legacy Survey (CFHTLS\footnote{Data release and associated documentation available at \textit{http://terapix.iap.fr/cplt/T0007/doc/T0007-doc.html}}) by \citet{cuillandre12}, reaching $i_{AB}=25.44$. Deep infrared photometry is also available in \textit{YJHK} bands from WIRCAM at CFHT \citep{bielby12} down to $K_{AB}=24.8$, and in the $3.6\mu m$ and $4.5 \mu m$ Spitzer bands thanks to the SERVS survey \citep{mauduit12}. For the VVDS-02h and the ECDFS fields, GALEX photometry is not available.

In addition for the COSMOS and ECDFS field, we have HST-ACS images in \textit{F814W}, \textit{F125W} and \textit{F160W} bands from \citet{koekemoer07}. The typical spatial resolution of these images are $\sim 0.09\arcsec$ for the \textit{F814W} band ($0.6\,$ kpc at z$=0.6$), with a point-source detection limit of $27.2\,$ mag at $5\sigma$. We will use HST images to derive a morphological classification of our galaxies in section \ref{morphology-sizes}. In contrast, the VVDS-02h field does not have HST coverage. The $i'$ filter images available from CFHT have a lower, seeing-limited spatial resolution of $\sim 0.8 \arcsec$ ($5.4\,$ kpc at z$=0.6$), implying that the most compact and distant galaxies are not completely resolved in VVDS-02h.

\subsection{SDSS data}\label{SDSS}

Throughout this paper we compare our results with those found in the local universe. For this comparison we use star-forming galaxies coming from the SDSS survey DR7 \citep{abazajian09} and publicly available measurements set by MPA-JHU \footnote{http://wwwmpa.mpa-garching.mpg.de/SDSS/DR7/}. We select our SDSS sample by having a redshift ranging $0.02 < z < 0.32$ and S/N$>3\,$ in the following emission lines: [\oii] $\lambda\lambda 3727+3729\ $\AA\ (hereafter [\oii]$3727$), \hb, [\oiii] $\lambda 4959\,$\AA, [\oiii] $\lambda 5007\,$\AA, \ha,
[\nii] $\lambda 6584\,$\AA\ and [\sii] $\lambda 6717+6731\,$\AA. The stellar masses are derived fitting the photometric data from \citet{stoughton02} with \citet{bruzual03} population synthesis models as described in \citet{kauffmann03}. The star formation rates are calculated from \ha\ luminosities following the method of \citet{brinchmann04} and scaled to \citet{chabrier03} IMF. They fit $ugriz$ photometry and six emission line fluxes ([\oii]$3728$, \hb, [\oiii]$4959$, [\oiii]$5007$, \ha, [\nii]$6584$ and [\sii]$6717$ requiring for all S/N$>3$) with \citet{charlotlonghetti01} models. These are a combination of \citet{bruzual03} synthetic galaxy SEDs and CLOUDY \citep{ferland13} emission line models, calibrated on the observed ratios of a representative sample of spiral, irregular, starburst and HII galaxies in the local universe. Finally, the gas-phase metallicity for the SDSS-DR7 comparison sample has been obtained by \citet{amorin10} using the N2 calibration of \citet{perezmonterocontini09}, which was derived using objects with accurate measurements of the electron temperature.

\section{Sample selection based on emission lines}\label{section3}

\subsection{The SFDGs sample selection}\label{selection}
In order to define our VUDS sample of SFDGs, we first select in VUDS database \footnote{1-d spectra fits files, spectroscopic and photometric informations are retrieved from the website \textit{http://cesam.lam.fr/vuds/}} a sample of emission-line galaxies with the following selection criteria:
\begin{enumerate}
\item high confidence redshift (at least $95 \%$ probability the redshift to be correct).
\item spectroscopic redshift in the range 0.13\,$< z_{\rm spec} <$\,0.88 and magnitudes  $i_{AB} > 23$. 
\item detection ($S/N \geq 3$) of the following emission lines: [\oii] $\lambda 3727$\AA, [\oiii] $\lambda\lambda 4959, 5007$\AA, \hb\ and/or \ha\ emission lines (we refer to Section \ref{line-measurements} for the emission-line measurements).
\end{enumerate}

\begin{figure}[t!]
    \resizebox{\hsize}{!}{\includegraphics[angle=0,width=6cm,trim={0cm 0cm 1.2cm 0.7cm},clip]{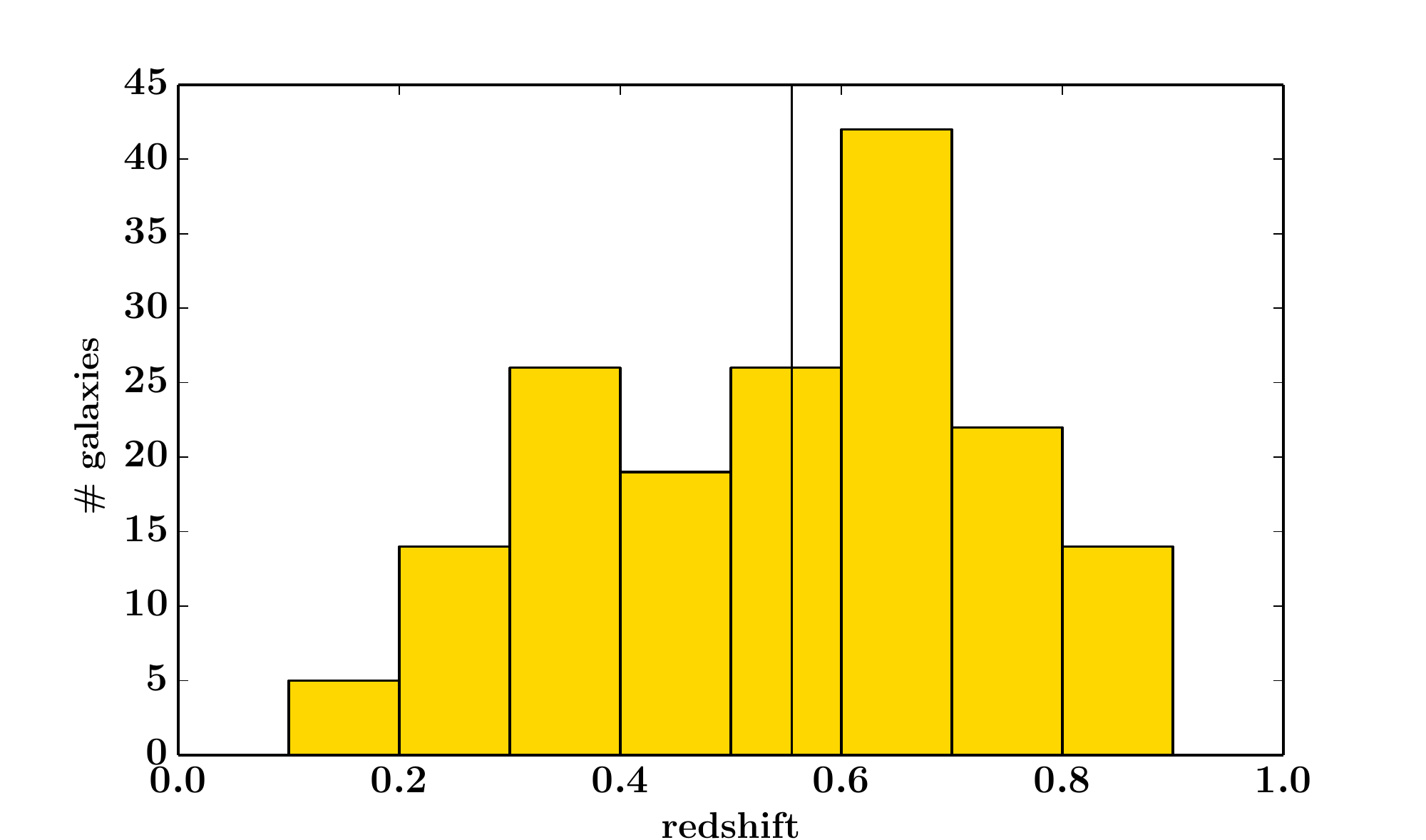}}
    \caption{\small Redshift distribution of VUDS SFDGs selected from the entire catalogue applying the criteria described in Section \ref{selection}. The galaxies are binned in intervals of 0.1 in z, and the median redshift distribution ($z_{med}=0.56$) is represented with a vertical line.}\label{redshift}
\end{figure}

The above criteria allow us to select low luminosity galaxies ($M_{\rm i}\leq$\,-13.5 mag) with at least [\oiii]$5007$\AA\ and [\oii]$3727$\AA\ included in the observed spectral range simultaneously, and derive metallicities and star formation rates (see Section \ref{section4}). In more detail, the first and second criteria yielded 302 galaxies in COSMOS, 300 in VVDS-02h and 113 in ECDFS fields. After retrieving all their spectra from the VUDS catalog we checked them visually, one by one, using IRAF. The final selection and S/N cut was done measuring the emission lines with the methodology presented in section \ref{line-measurements}, excluding the galaxy whenever one of the four emission lines mentioned above ([\oii]$3727$, [\oiii]$4959,5007$, \hb\ and/or \ha) is not detected (S/N $<3$), or appears contaminated by residual sky emission lines. Applying this procedure we obtain a final total sample of 168 SFDGs in the redshift range $0.13<z<0.88$, with median value $z_{med}=0.56$ and a vast majority of the galaxies ($54 \%$) concentrated between $0.5$ and $0.8$, as can be seen in the histogram in Fig. \ref{redshift}.

\subsection{Emission line measurements}
\label{line-measurements}

\begin{figure*}[t!]
    \centering
    \includegraphics[angle=0,width=8.4cm]{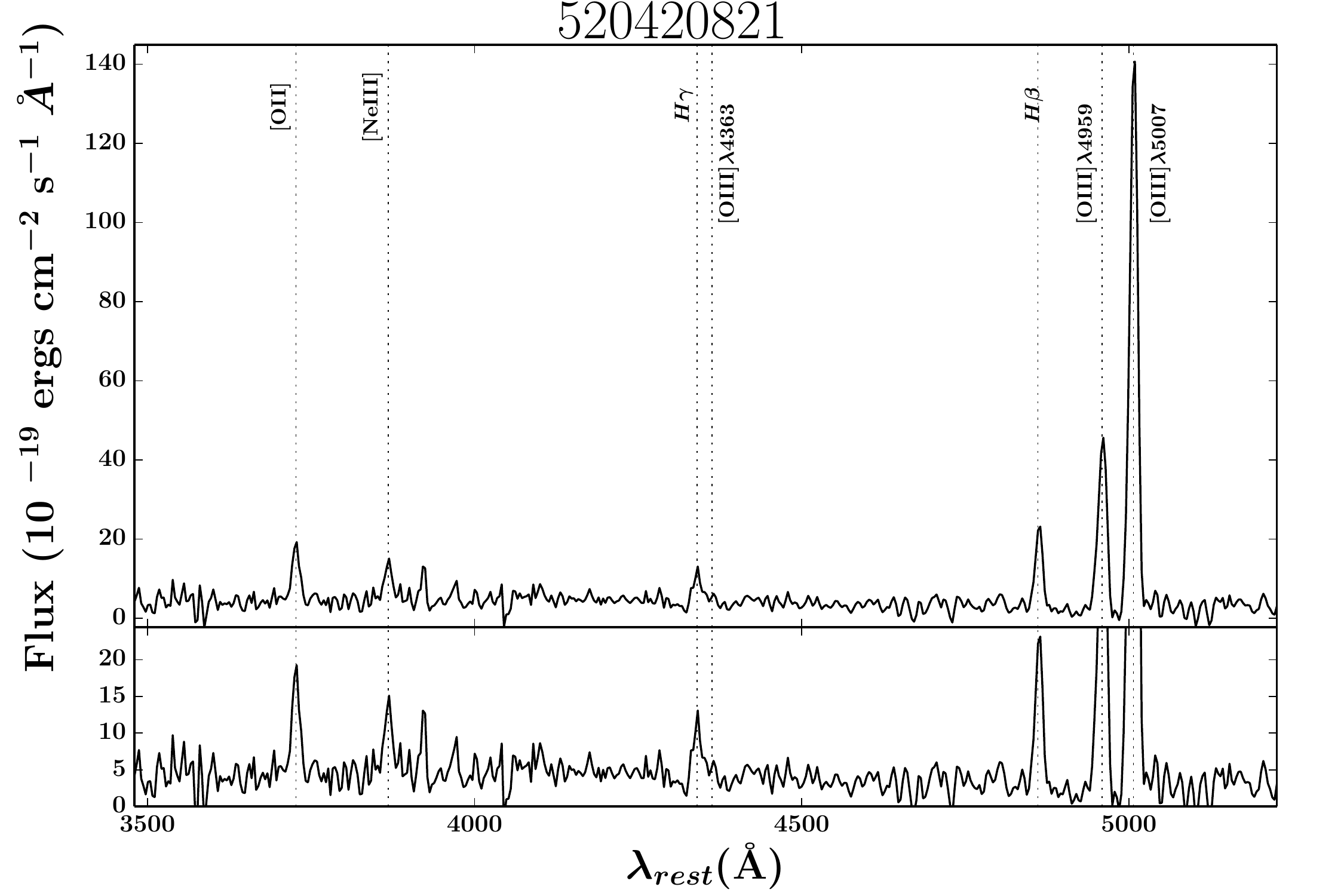}
    \includegraphics[angle=0,width=8.4cm]{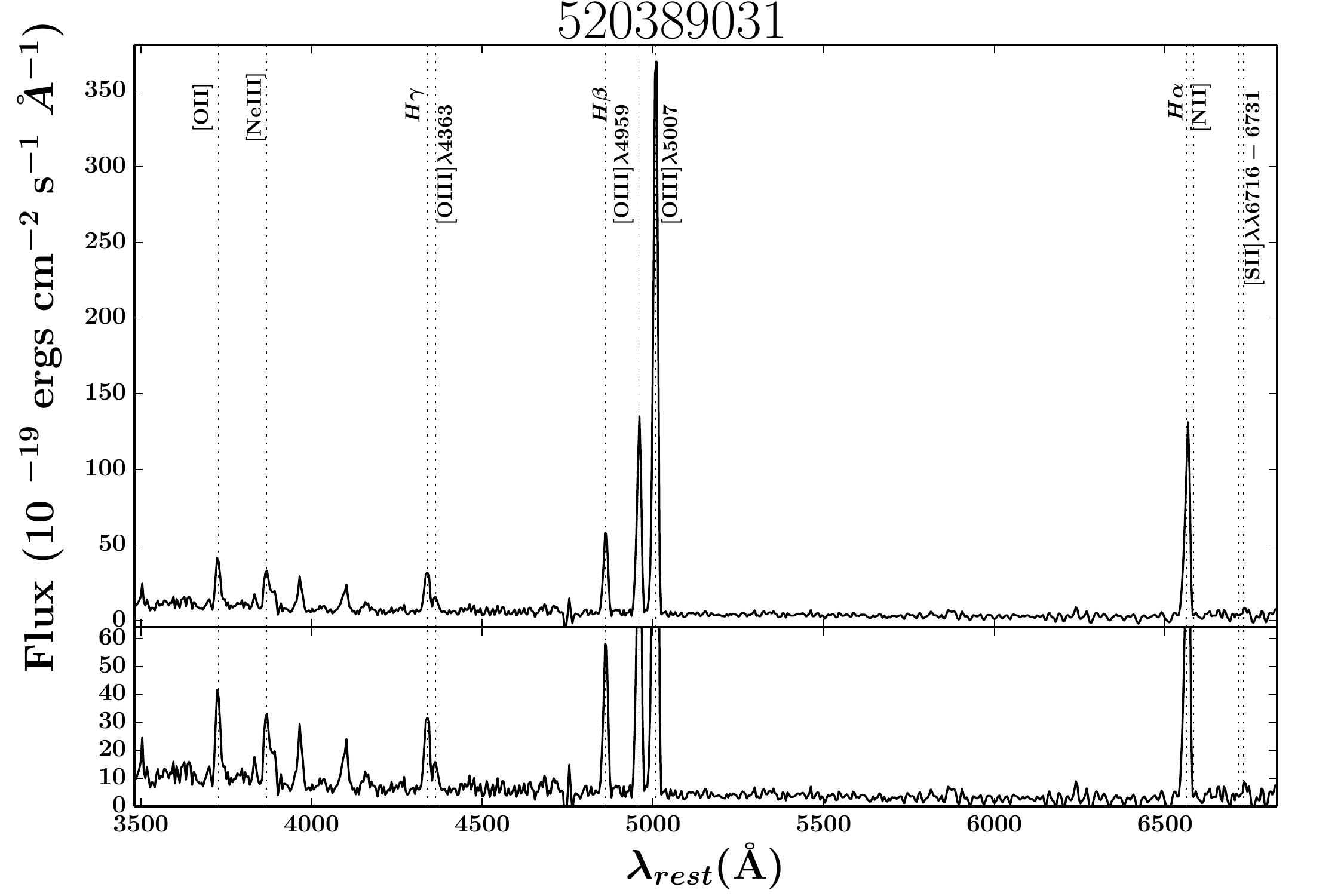}
    \caption{\small Observed spectra of two galaxies in VUDS with strong emission lines and faint continuum. The galaxy on the left and right hand side panels are at redshift $z= 0.555$ and $z= 0.173$, respectively. While for the former the  \ha\ line lies out of the wavelength range, for the latter it is still visible in the red part of the spectrum. Dotted lines and labels indicate some of the relevant emission lines detected.}\label{ELGs}
\end{figure*}

\begin{figure*}[t!]
    \centering
    \includegraphics[angle=0,width=8.4cm,trim={1cm 0.cm 1cm 1.2cm},clip]{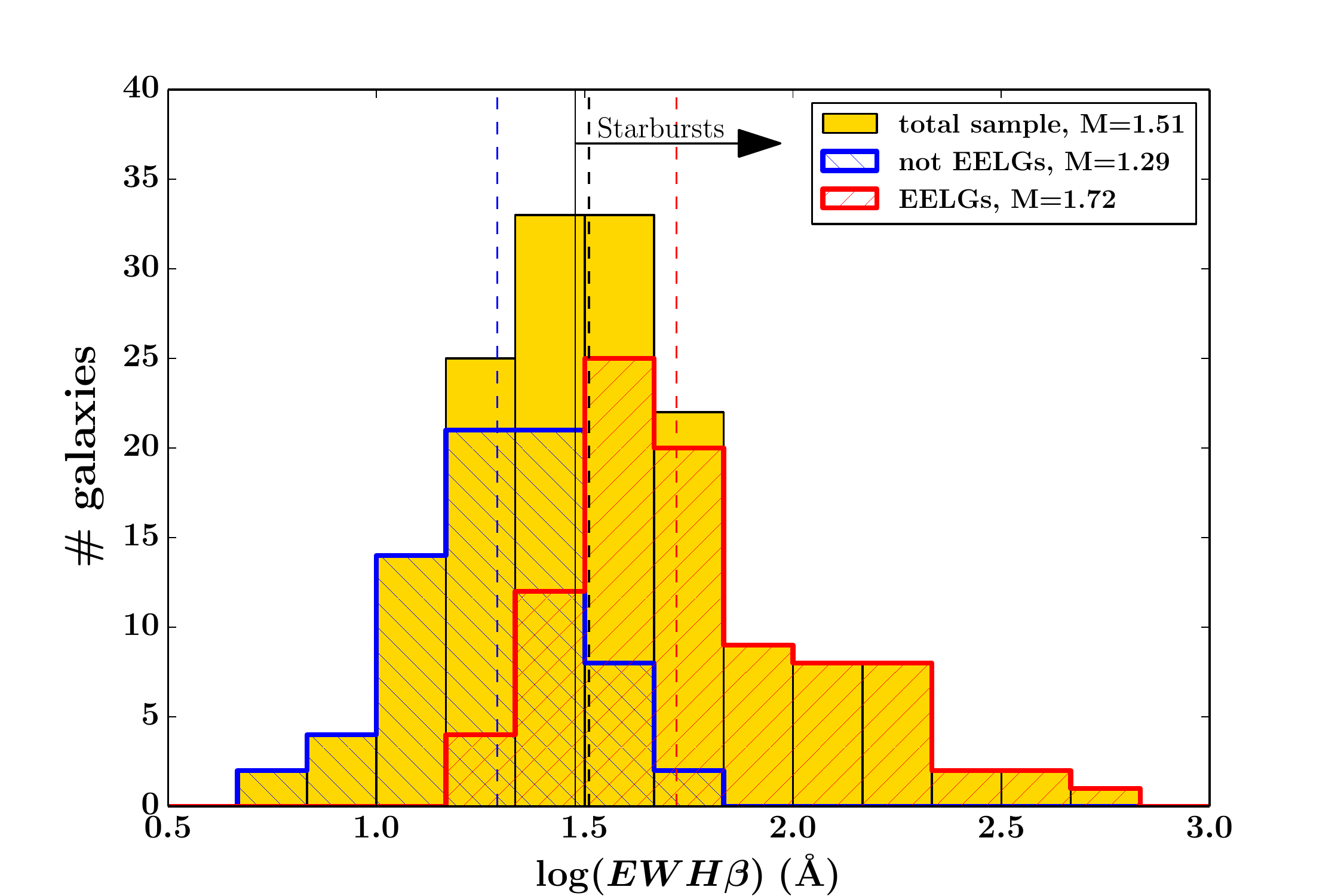}
    \includegraphics[angle=0,width=8.4cm,trim={1cm 0.cm 1.cm 1.2cm},clip]{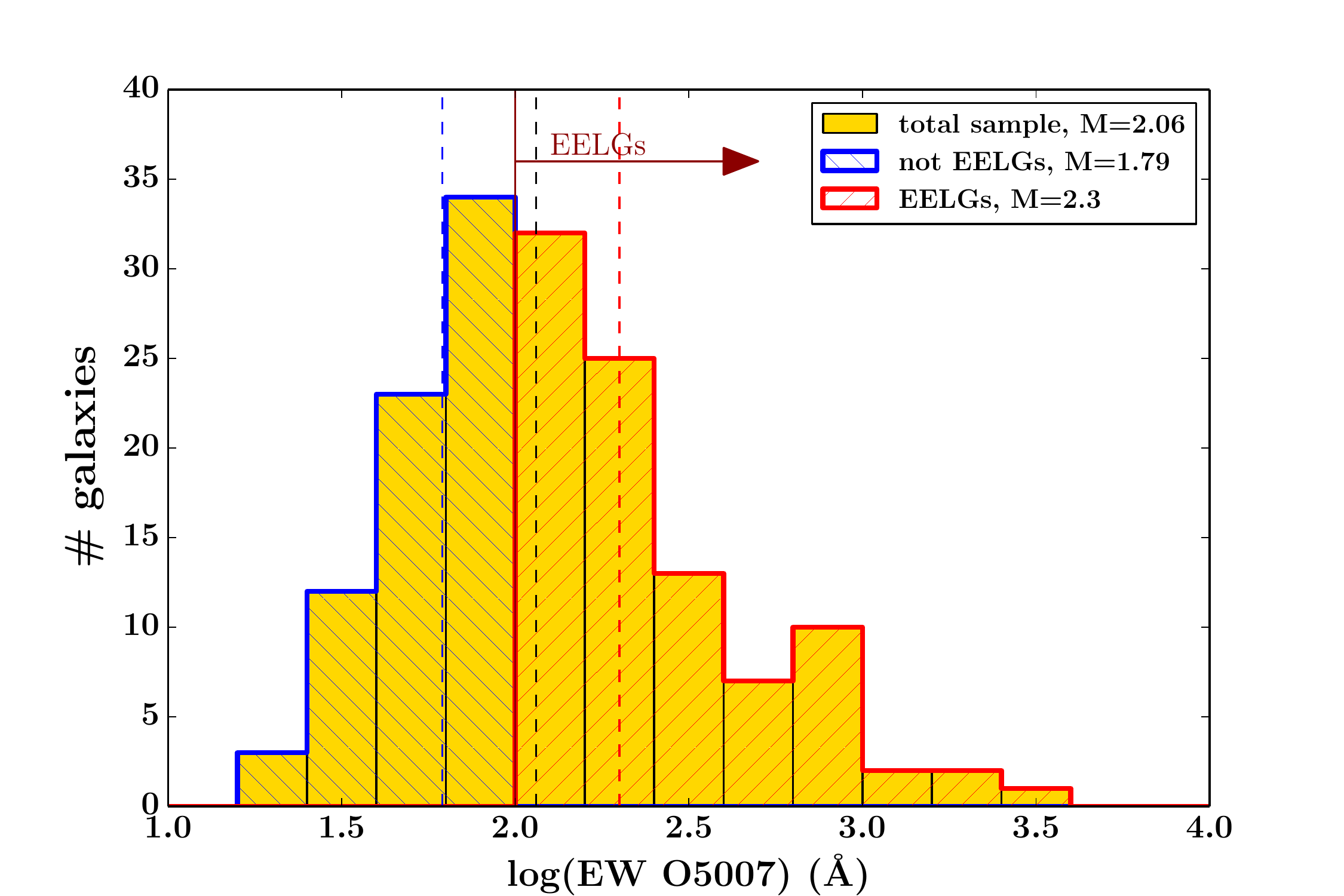}
    \caption{\small Distribution of EW(\hb) and EW(\oiii) for our sample of galaxies selected in section \ref{selection}. We lined the histogram with different colors, red for the EELG fraction (EW(\oiii)$>100$ \AA, based on Amor\'in et al. 2015) and blue for the non-EELG subset. We emphasize with vertical continuous lines the empirical limits for starburst galaxies (EW(\hb) $>30$ \AA, based on \citet{terlevich91} and EELGs, as well as with vertical dashed lines the median distribution for the whole sample (black), the EELGs (red) and non-EELGs (blue) (the values are given in the legend). The standard deviations of the entire distribution are 0.37 and 0.42 for EW(\hb) and EW(\oiii) respectively.}\label{EWHbOIII}
\end{figure*}

\begin{figure*}[t!]
    \centering
    \hspace*{-1cm}
    \includegraphics[angle=0,width=18cm,trim={2.5cm 0cm 4.0cm 1.3cm},clip]{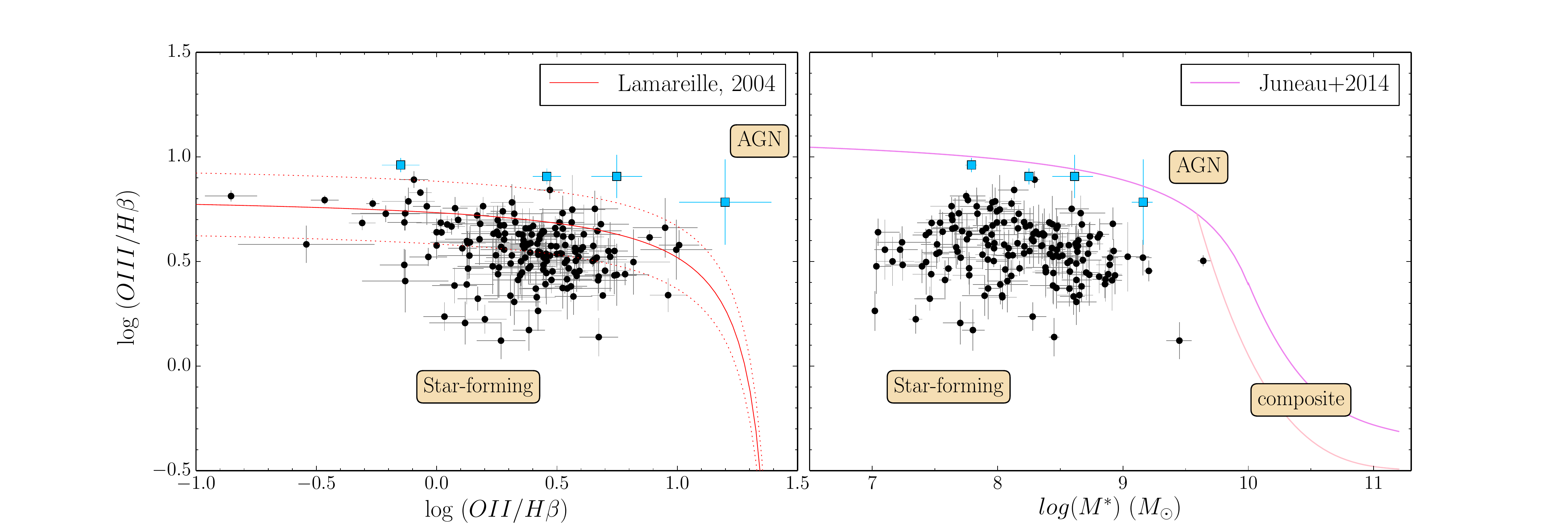}
    \caption{\small \textit{Left}: Diagnostic diagram to separate SFG, AGN and composite, comparing [\oii]$\lambda 3727$/\hb\ and [\oiii]$\lambda 5007$/\hb. The starburst region is in the bottom-left part, while AGNs lie in upper-right part. The red line is the empirical separation by \citet{lamareille04} with $0.15$ dex uncertainty. \textit{Right}: MEx diagnostic diagram for our sample of emission-line galaxies. The diagram compares the exctitation, i.e. the emission line ratio [\oiii]$\lambda 5007$/\hb, and the stellar mass $M_\ast$. The violet curve is the empirical separation limit between starburst, AGN and composite regions according to \citet{juneau14}. In both diagrams we mark with blue squares the 4 galaxies which completely lie (also including the errors) in the AGN region according to \citet{lamareille04}.}\label{BPT2}
\end{figure*}

Emission lines fluxes and equivalent widths are measured manually on a one-by-one basis using the task \textsl{splot} of \textsc{IRAF} by direct integration of the line profile after linear subtraction of the continuum, which is well detected in all cases. Additionally, we also fit the emission lines of the galaxies with a Gaussian profile. Results using the two approaches are in very good agreement for high S/N emission lines (essentially H$\alpha$, H$\beta$, [\oiii], and [\oii]), while some differences are found  for strongly asymmetric or very low S/N ($\sim$\,3-5) lines. However, their fluxes and EWs are still consistent within 1$\sigma$ uncertainties. 

We compute uncertainties in the line measurements from the dispersion of values provided by multiple measurements adopting different possible band-passes (free of lines and strong residuals from sky subtraction) for the local continuum determination, which is fitted using a second order polynomial. This approach typically gives larger uncertainties compared to those obtained from the average noise spectrum produced by the data reduction pipeline. 

For Balmer lines the presence of stellar absorption features should be considered too. Even though the faintness of the galaxies does not allow detection of significant absorptions for most of them, we have corrected upwards our measurements by $1$\AA\ in EW of \hg, \hb\ and \ha\ lines for all the galaxies \citep[e.g.][]{ly14}. In any case, our galaxies have relatively large EW of \ha\ and \hb\ lines (EW(\hb)$_{median}=33$ \AA) and the measurement errors are always higher than $1$\AA\ ($\Delta$[EW(\hb)]$_{median}=8$ \AA), so that this correction must be negligible for our work. In Fig. \ref{ELGs} we show typical spectra for two strong emission-line galaxies in the sample at low and intermediate redshift bin, respectively. 

As we have described in Section \ref{introduction}, a particular class of SFDGs have extreme properties, in particular a high EW of their optical emission lines. Galaxies with $EW(OIII)> 100$\AA\ are named extreme emission line galaxies (EELGs), according to the definition of \citet{amorin15}. From our sample, $56\%$\ qualify as EELGs (see Fig. \ref{EWHbOIII}).   

\subsection{Identification of AGNs: diagnostic diagrams}
\label{diagnostic-diagrams}

Galaxies showing emission lines in their spectra include a broad variety of astrophysical objects that can be distinguished according to their excitation mechanism, i.e. thermal (e.g. star formation) or non thermal (in e.g. AGN or shocks). AGNs are found in two main categories: broad-line (BL) and narrow-line (NL) AGNs. The former were excluded from the VUDS parent sample before we applied our selection criteria, in order to exclude from the sample any clear AGN candidate we still need to identify narrow-line AGNs, such as Seyfert 2 and LINERs. To that end, we use both a cross-correlation of our sample of galaxies with the latest catalogs of X-ray sources and a combination of two empirical diagnostic diagrams based on optical emission-line ratios.

In the ECDFS field we use the catalogs E-CDFS \citep{lehmer05} and Chandra 4MS \citep{luo08,xue12,cappelluti16}. Inside this field we can exclude AGNs of high and intermediate luminosity (i.e. those with $L_{\rm x} \geq$\,10$^{43}$\,erg~s$^{-1}$). For the COSMOS field we use the survey Chandra COSMOS \citep{elvis09,civano12} and XMM-COSMOS \citep{cappelluti09}, while for the VVDS-02h field we use the catalogs compiled by \citet{pierre04} and \citet{chiappetti13} from the XMM-LSS survey. Since we have a lower sensitivity for the COSMOS and the VVDS fields compared to the ECDFS field, we can look in those cases only for high luminosity AGNs ($L_{\rm x} \geq$\,10$^{44}$\,erg~s$^{-1}$). We find no X-ray counterpart for any of our VUDS SFDGs in each field observed by the survey. 

We inspect the spectra for the presence of very high ionization emission lines (e.g. [Ne\textsc{V}]$\lambda$\,3426\AA), very broad components in the Balmer lines and/or very red SEDs which could suggest the contribution of an AGN, and we do not find any evidence of them. 

Finally, we explore two diagnostic diagrams that are frequently used to distinguish between SF, AGN, and galaxies with a combination of different excitation mechanisms (Composites). We use adaptations of the classical BPT diagram, i.e. the diagnostics of [\oiii]$\lambda$$5007$/\hb\ vs [\nii]$\lambda$$6583$/\ha\ \citep{baldwin81}, allowing galaxy classification when \ha\ and [\nii] are no longer observable in optical spectra of intermediate redshift emission-line galaxies. The diagram in Fig. \ref{BPT2} (left) was proposed by \citet{rola97} and compares [\oii]/\hb\ and [\oiii]/\hb\ line ratios. The orange continuous line is the empirical separation (with a 1$\sigma$ uncertainty of  about 0.15 dex) derived by \citet{lamareille04} between two types of emitting sources: star-forming systems in the bottom left and AGNs in the upper right part. 

In Fig. \ref{BPT2} (right) we show the Mass-Excitation (MEx) diagram, introduced by \citet{juneau11}, which considers the galaxy stellar mass M$_{*}$ (see Section \ref{masses}) instead of the [\nii]/\ha\ ratio, which is available only for 10 galaxies of our sample. This diagram relies on the correlation between the line ratio [\nii]/\ha\ and the gas-phase metallicity in SF galaxies \citep{kewleyellison08}. The empirical relation between mass and metallicity \citep{tremonti04} provides the physical connection between the two quantities. The violet line divides the starburst and AGN regimes and was derived empirically from SDSS emission-line galaxies by \citet{juneau14}. Below this line the MEx diagram is populated by star-forming galaxies whose emission lines are powered by stellar photoionization. For this kind of objects models predict an upper limit in the excitation. In the right-upper part we find AGNs, while in the right-lower part the region between the two lines is occupied by galaxies showing both star-forming and AGN emission properties (composite). At intermediate redshift ($z \leq$\,1.5), star-forming galaxies are consistent with having normal interstellar medium (ISM) properties \citep{juneau11}, so we can in principle apply this diagnostic for all the galaxies in the sample ($z<0.88$). We also caution the reader that in AGNs the \nii\ line does not trace the metallicity \citep[e.g.][]{osterbrock89} and the connection between the BPT and MEx may not be straightforward, so we should consider the results of the MEx diagram in combination with other diagnostics which do not suffer from this drawback.

We find that all galaxies are consistent with purely star-forming systems according to the MEx diagram of \citet{juneau14}, while four galaxies in the [\oiii]/\hb\ vs [\oii]/\hb\ diagram clearly fall out of the SF region if we consider the 1$\sigma$ uncertainty ($\sim$\,0.2 dex) of the empirical relation by \citet{lamareille04}. In addition we see a small number of SFDGs with relatively higher excitation than the empirical limit of \citet{lamareille04}. It is worth noting that the excitation of objects above those limits do not necessarily require to be powered by an active nuclear source. A variety of other mechanisms can mimic the properties of AGN in these diagnostics, such as shocks, supernovae and their subsequent remnants. Indeed, SFDGs (and in particular EELGs) appear to be the preferred sites for the most powerful supernova explosions \citep{lunnan13,leloudas15,thone15}. 

From the combination of the above diagnostics we find that four galaxies clearly reside in the AGN region of the [\oiii]/\hb\ vs [\oii]/\hb\ plane of \citet{lamareille04}, so we exclude them from the following analysis. However, the exclusion or inclusion of these four possible AGN galaxies does not affect any of the results. Indeed, we find that the SFRs, metallicity and gas fractions of these four galaxies are consistent with the median values for the rest of the sample of secure star-forming systems.

After removing the AGN candidates, the subsequent analysis is carried out using $164$ SFDGs. The basic information for the galaxies, including redshift and selection magnitude, are presented in Table \ref{Table1} in Appendix \ref{appendixb}.

\section{Methodology}\label{section4}

In Table~\ref{Table2} \footnote{\label{note0}A preview table is shown in Appendix \ref{appendixb}. A complete version of this table is available {\it \emph{online.}}} we present the selected sample of 164 star-forming dwarf galaxies (SFDG) in VUDS. SFDGs are low-mass ($M_\ast < 10^9 M_\odot$), low-luminosity ($M_B > -20$ mag) galaxies that are forming stars at present, as suggested by the presence of optical emission lines, coming from the gas photoionized by newly born massive (O and B) stars. Table \ref{Table2} includes measured fluxes (non-extinction corrected) and uncertainties for the most relevant emission lines. These quantities, together with an extended ancillary multiwavelength dataset, are used to derive relevant physical properties of the galaxies, which are presented in Table~\ref{Table3}. In this section we describe in detail our methodology, and we will discuss the results in the next section. 

\subsection{Stellar masses from multiwavelength SED fitting}\label{masses}

We derive stellar masses from SED fitting performed on the available multi-wavelength photometric data using the code Le Phare \citep{ilbert06}, as described in \citet{ilbert13}. In brief, the code uses a chi-square minimization routine, fitting for each galaxy \citet{bruzual03} stellar population synthesis models to all broadband photometric data available in each VUDS field (COSMOS, ECDFS and VVDS-02h) between GALEX far-UV and Spitzer $4.5 \mu m$ band, presented in section \ref{photometry}. The models include different metallicities ($Z=0.004$, $Z=0.008$, and solar $Z=0.02$), visual reddening ranging $0<$$E(B-V)_{stellar}$$<0.7$ and exponentially declining and delayed star formation histories (SFH) with 9 different $\tau$ values from $0.1$ to $30$ Gyr.

We adopt a \citet{chabrier03} IMF and \citet{calzetti00} extinction law, while the contribution of emission lines to the stellar templates is considered in the SED fitting by using an empirical relation between the UV light and the emission line fluxes, as described in \citet{ilbert09}. The stellar masses for the entire sample of galaxies are presented in Table~\ref{Table3} in Appendix \ref{appendixb}. Typical $1\sigma$ uncertainties in $M_\ast$ are $\sim 0.2$ dex and are obtained from the median of the probability distribution function (PDF). Thus, they do not account for possible systematics (e.g. IMF variations). Besides M$_\ast$, additional physical parameters derived from SED fitting used in this paper are the extinction-uncorrected rest-frame magnitudes calculated with the method of \citet{ilbert05} and, for a sub-section of galaxies, stellar extinction $E(B-V)_{stellar}$ and star formation rates (SFR$_{SED}$).

\subsection{Extinction correction}

\begin{figure*}[t!]
    \centering
    \hspace*{-1cm}
    \includegraphics[angle=0,width=18cm,trim={2.5cm 0cm 3.5cm 1.0cm},clip]{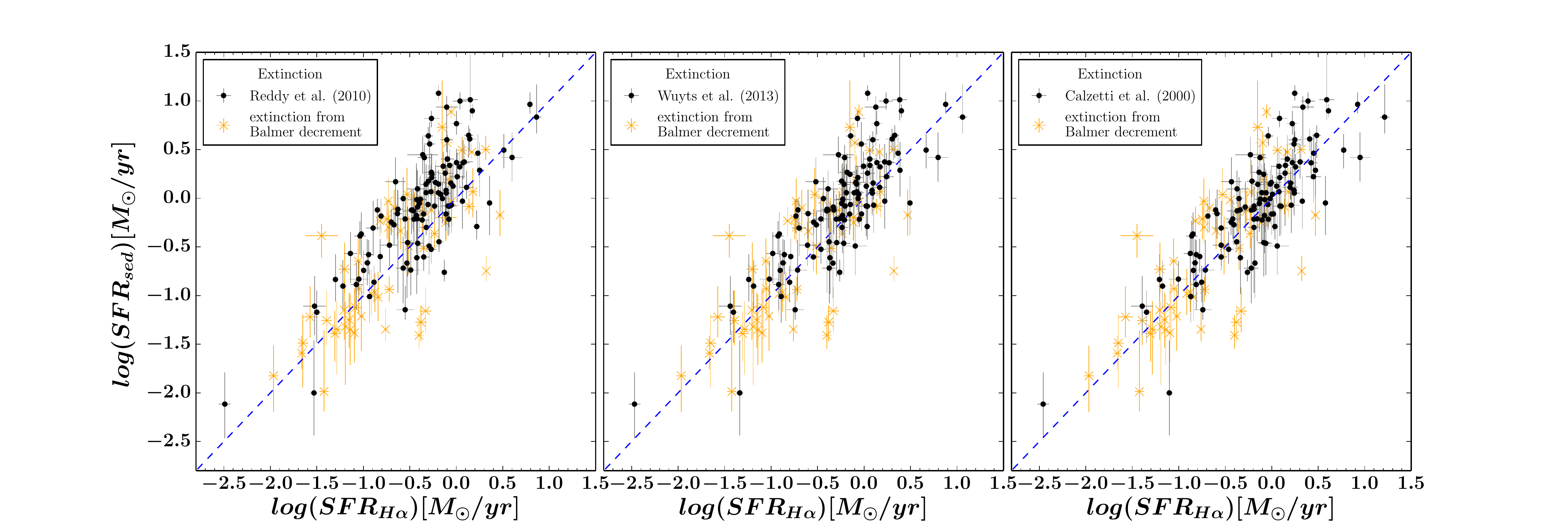}
    \caption{\small The figure illustrates, for the galaxies without Balmer decrement available, the comparison between the star formation rates derived from SED fitting ($SFR_{SED}$) and from the extinction-corrected \ha\ luminosity ($SFR_{H\alpha}$), for different calculations of the extinction coefficient c(\hb). \textit{From left to right:} $E(B-V)_{gas}=E(B-V)_{stellar}$ \citep{reddy10};  $E(B-V)_{gas}=E(B-V)_{stellar}+E(B-V)_{extra}$ \citep{wuyts13} ; $E(B-V)_{gas}=E(B-V)_{stellar}/0.44$ \citep{calzetti00}. In the last panel the orange crosses represent the galaxies with Balmer decrement available, for which the $SFR_{H\alpha}$ are consistent with the $SFR_{SED}$ and no systematic trend is observed. }\label{SFRcomparison}
\end{figure*}

In order to derive extinction-corrected luminosities we first obtain the logarithmic extinction at \hb, c(\hb), from the Balmer decrement. Using either \ha\ and \hb\ lines or \hb\ and \hg\, when \ha\ is not available, we use the following formulation, 

\begin{equation}\label{eq44}
c(\hb)_1= \log\left(\frac{\ha/\hb}{2.82}\right)/f_{H\alpha} 
\end{equation}
\begin{equation}\label{eq45}
c(\hb)_2= \log\left(\frac{\hg/\hb}{0.47}\right)/f_{H\gamma}
\end{equation}
where the observed ratios are divided by the theoretical values ($\ha/\hb = 2.82$ and $\hg/\hb = 0.47$), for case B recombination with $T_e = 2 \times 10^4 K$, $n_e = 100\ cm^{-3}$ following \citet{amorin15}, and $f_{H\alpha}$ and $f_{H\gamma}$ are the coefficients corresponding to the \citet{cardelli88} extinction curve at the wavelength of the \ha\ and \hg\ emission lines ($f_{H\alpha}=0.313$, $f_{H\gamma}=0.157$), respectively. 

For $111$ galaxies in our sample we only have one hydrogen line available or the \ha/\hb\ and \hb/\hg\ ratios are below their theoretical values. In these cases we derive the gas extinction from the stellar reddening, given by the SED fitting ($E(B-V)_{stellar}$). The reddening of the stellar and nebular components of a galaxy are generally different, and various relations between the two have been found in previous studies. \citet{calzetti00} found that the gaseous reddening is typically higher than the stellar in low redshift starburst galaxies, which are more similar to our sample, and they apply the following relation $E(B-V)_{nebular}=E(B-V)_{stellar}/0.44$. More recently, \citet{reddy10} have found that choosing $E(B-V)_{nebular}=E(B-V)_{stellar}$ is more appropriate for studying $z \sim 2$ star-forming galaxies, while \citet{wuyts13} have derived a relation for massive star-forming galaxies at $0.7<z<1.5$ : $E(B-V)_{gas}=E(B-V)_{stellar}+E(B-V)_{extra}$ , where $E(B-V)_{extra}=E(B-V)_{stellar} \times (0.9 - 0.465* E(B-V)_{stellar}) $. The latter result indicates that the nebular light is more extincted than the stellar, but the extra-correction is lower than predited by \citet{calzetti00}. 

In order to decide which relation between stellar and nebular reddening should be applied to our galaxies we analyze all the three possibilities listed above. We derive star formation rates (SFR) from extinction corrected \ha\ luminosities ($SFR_{H\alpha}$, see section \ref{SFR}) and we compare them with the SFR derived through SED fitting SFR$_{SED}$ (Fig. \ref{SFRcomparison}). 

We find that, using the relation by \citet{reddy10}, the $SFR_{H\alpha}$ are systematically lower than the $SFR_{SED}$, and the differences increase toward higher SFRs, where extinction corrections are more severe. Using the relations by \citet{wuyts13} and \citet{calzetti00}, we find a tighter correlation and a better agreement between the two SFR measurements, with the latter having the lowest offset ($\simeq -0.08$ dex) in the whole range of SFR and the lowest dispersion ($\simeq 0.32$ dex). Repeating the same procedure for the galaxies with extinction derived through Balmer decrement we see that, despite the larger scatter, there is no systematic trend between SED and \ha-based SFRs, supporting the consistency of this method. 

Overall, we decide to adopt the extinction derived through the Balmer decrement for those galaxies with at least 2 hydrogen lines available. For the remaining galaxies, we use the stellar reddening and the relation by \citet{calzetti00}; then, following the same paper, we obtain the extinction coefficient c(\hb) from the corrected reddening as: 
\begin{equation}\label{eq46}
c(H\beta)_3= E(B-V) \times 0.69
\end{equation}
The extinction from SED fitting is used for $109$ galaxies in our sample. 

\subsection{Star formation rates from Balmer lines}\label{SFR}

In order to derive the ongoing star formation rate of the galaxies (i.e. the star formation activity in the last 10--20 Myr) we adopt the standard calibration of \citet{kennicutt98},

\begin{equation}\label{eqsfr5}
\mathrm{SFR} = 7.9 \times 10^{-42} L(\ha)\ \mathrm{[erg\ s^{-1}]} 
\end{equation}  
where $L$(\ha) is the \ha\ luminosity, corrected for extinction as described in the previous section. However, for galaxies at $z \geq$\,0.42 $H\alpha$ is no longer visible in our VIMOS spectra. To overcome this limitation, we estimate the \ha\ luminosities from \hb\ fluxes by assuming the theoretical ratio $(H\alpha/H\beta)_0 = 2.82\ $, valid for case B recombination.  Following \citet{santini09}, the SFR derived this way have been scaled down by a factor of $1.7$ in order to be consistent with the \citet{chabrier03} IMF used throughout this paper.

We note that corrections for slit losses, due to the finite size of the slit ($1 \arcsec$) has already been applied to all the spectra during the calibration process, and it is as accurate as the observed-frame optical photometry. This allows to compare properly photometric (e.g. stellar masses) and spectroscopic quantities (e.g. $SFR_{H\alpha}$,metallicity) , which are investigated in the following sections. The consistency between \ha-based and SED-based SFRs, shown in Fig.\ref{SFRcomparison}, supports this procedure. 

Finally, we combine Balmer line-based SFRs with stellar masses in order to compute the specific star formation rate (sSFR=SFR/M$_{*}$).

\subsection{Morphology and sizes}\label{morphology-sizes}

\begin{figure*}[t!]
  \centering
  \includegraphics[angle=0,width=17cm,trim={0.cm 0cm 0.cm 1.2cm},clip]{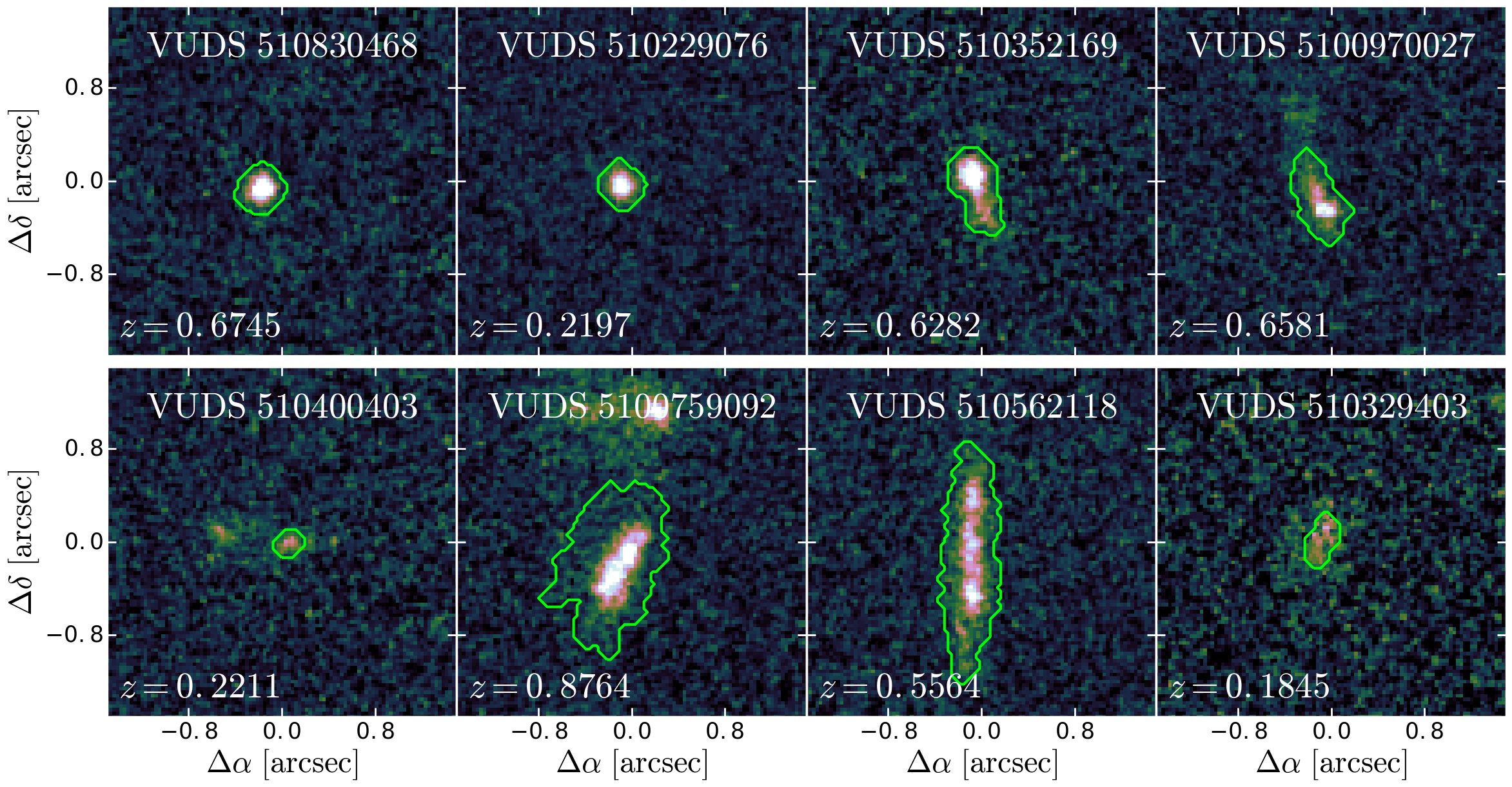} 
  \caption{\small HST i-band images of 8 galaxies in COSMOS field with different morphological properties. We have: (\textit{top line:})  two round-nucleated galaxies at different redshifts and two cometary-tadpole systems ; (\textit{bottom line:}) two interacting-merging galaxies (a faint close galaxy pair and a brighter loose couple), a clumpy-chain system and finally an example of low surface brightness dwarf. The stamps are 2 x 2 arcsec wide and the green lines represent the contours of the galaxies. 
  }\label{morphology}
   \end{figure*}

\begin{figure}[t!]
   \resizebox{\hsize}{!}{\includegraphics[angle=0,width=8.5cm,trim={0.cm 0cm 1.cm 1.2cm},clip]{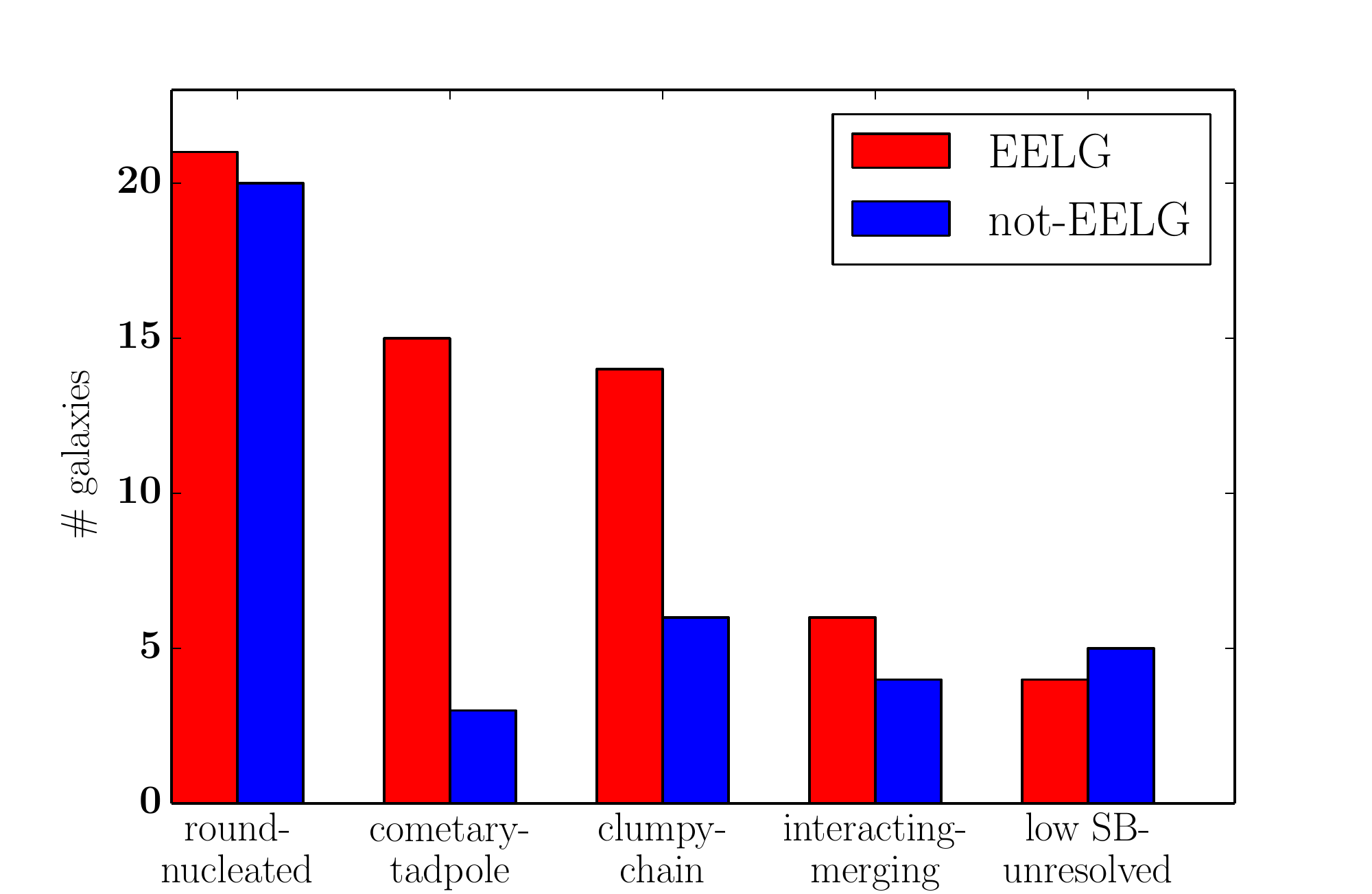}} 
  \caption{\small Bar histogram showing for EELG (red) and not-EELG (blue) subsamples the number of galaxies falling in each morphological class described in the text. We find that EELGs have a higher fraction of irregular and disturbed morphologies ($58 \%$), compared to not-EELGs ($34 \%$). Only galaxies in COSMOS and ECDFS fields are considered for this analysis. 
  }\label{hist_morphology}
   \end{figure}

We perform a visual classification of our star-forming dwarf galaxies as a first approach to study their morphological properties. The selected galaxies are nearly unresolved in ground-based CFHT images, but for most of them (those in COSMOS and ECDFS fields) space-based images are available, as presented in section \ref{photometry}. In order to have a homogeneous sample, avoiding strong biases in the classification due to the much lower resolution of CFHT images, we analyze only the galaxies in COSMOS and ECDFS (101 in total). For this subset, we do the classification based on HST-ACS \textit{F814W} band images.

Galaxies at higher reshifts typically show irregular shapes, so we cannot follow the Hubble morphological classification and we need to choose ad-hoc criteria. In this paper we divide our galaxy sample into the 4 morphological classes defined by \citet{amorin15} (A15) for low-mass EELGs (EW([\oiii])$\lambda5007>100\,$\AA) in our same redshift range, applying the criteria also to our non-EELGs. They include Round/Nucleated, Clumpy/Chain, Cometary/Tadpole and Merger/Interacting morphological types. A visual inspection of our complete sample of galaxies in COSMOS and ECDFS fields reveals that, following the classification by A15, $\simeq 40$$\%$ are round-nucleated, $\simeq 20$$\%$ are clumpy-chain, $\simeq 18$$\%$ are cometary-tadpole and $\simeq 10$$\%$ are merger-interacting systems, while for the remaining $12$$\%$ fraction we cannot determine their morphological type because they appear unresolved or have extremely low surface brightness. As an example, we show (Fig. \ref{morphology}) HST \textit{F814W} images for each galaxy morphological type, while in Fig. \ref{hist_morphology} we show the distribution of the EELG subsets. Even though there might be inevitable overlap between the last 3 classes, the latter analysis shows qualitatively that EELGs have on average more disturbed morphologies (cometary,clumpy shapes and interacting-merging systems) compared to not-EELGs, similarly to what was found in A15. 

We obtain quantitative morphological parameters for the galaxies in our sample with HST-ACS images available. The analysis has been done using GALFIT \citep[][version 3.0]{peng02,peng10}, following the methodology presented in \citet{ribeiro16}, based on fitting recursively a single 2-D S\'ersic profile \citep{sersic68} to the observed light profile of the object. Using this procedure we obtain the following quantities: the effective radius (semi-major axis) \textit{$r_e$} and \textit{q}, which is the ratio between the observed major and minor axis of the galaxy.

We do not deproject $r_e$ to avoid the introduction of additional uncertainties (given by \textit{q}), thus we may underestimate the SFR surface density for the most elongated galaxies. However, given the large fraction of round and symmetric galaxies in our sample, we consider the effective radius good enough to characterize the size of the objects for our purposes. Furthermore, we assume that half of the total star formation rate (derived from \ha\ luminosity) resides inside the effective radius. This approximation depends on the distribution of the gas relative to the stellar component and can have an opposite effect of deprojection. 

Finally, some caveats should be considered when applying GALFIT to our sample. The minimization procedure of the code weighs more the central parts of the object, where the S/N is higher and the nebular emission lines along with young stellar populations contribute the most, inducing an underestimation of the true scale length of the underlying galaxy (Cairos et al. 2007, Amor\'in et al. 2007, 2009). However, SFDGs (e.g. Blue Compact Dwarfs) are usually dominated by nebular emission lines at all radii, and the gas is typically more extended than the central stellar body \citep[][e.g. IZw18]{papaderosostlin12}, producing a compensation of the previous effect.

For the galaxies in the VVDS-02h field observed with CFHT, GALFIT does not reach a stable solution in most of the cases ($50$$\%$). Even when the fitting code converges, the effective radii suffer from systematic overestimation (difficult to quantify) due to the lower image resolution, as shown in previous VUDS works \citep{ribeiro16}. This effect can be especially important for our dwarf galaxies since they are instrinsically very small, as we will show later in section \ref{section5}. For these reasons, also for the quantitative analysis we only consider the galaxies in COSMOS and ECDFS.

\subsubsection{Gas fractions and surface densities}\label{gasfractions}

We derive analytically SFR surface densities and stellar mass surface densities, dividing SFR and $M^\ast$ by the projected area of the galaxy by assuming:
\begin{equation}\label{sigma_SFR_eq}
\Sigma_{SFR} = \frac{SFR}{2\ \pi {r_e}^2}  
\end{equation}
\begin{equation}
\Sigma_{M_*} = \frac{M_*}{2\ \pi {r_e}^2}
\end{equation}

Gas surface density and total gas mass are then computed assuming that galaxies follow the Kennicutt-Schmidt (KS) law \citep{kennicuttevans12} in the form:
\begin{equation}\label{KSeq1}
\Sigma_{SFR}=(2.5 \pm 0.7)\ \times\ 10^{-4} (\frac{\Sigma_{gas}}{1 M_\odot pc^{-2}})^{1.4\pm0.15}\ M_\odot yr^{-1} kpc^{-2} 
\end{equation}

Since this equation assumes a Salpeter IMF, we have used here SFRs consistent with this IMF. Inverting the above relation we determine the gas surface density:

\begin{equation}\label{KSeq2}
\Sigma_{gas}= [0.4 \times\ 10^{4} \times (\frac{\Sigma_{SFR}}{1 M_\odot yr^{-1} kpc^{-2}})]^{0.714}\ M_\odot pc^{-2} 
\end{equation}

Then, multiplying again by the galaxy area, we derive the total baryonic mass of the gas involved in star formation :
\begin{equation}
M_{gas}= \Sigma_{gas} \times 2\ \pi {(r_e)}^2 \times 10^6\ M_\odot
\end{equation} 

Finally, we derive the gas fractions as: 
\begin{equation}\label{fgaseq}
f_{gas}=\frac{M_{gas}}{M_{gas}+M_\ast}
\end{equation}

In previous formulae, we have used Salpeter-based SFRs to be consistent with the expression of the KS law (derived assuming a Salpeter IMF). Likewise, the stellar masses $M_\ast$ have been scaled to Salpeter IMF following \citet{bolzonella10} (log($M_\ast$)$_{Salp}$=$0.23+$log($M_\ast$)$_{Chab}$). 

Some caveats should be considered when applying the KS law to our sample. \citet{leroy05} show that dwarf galaxies and large spirals exhibit the same relationship between molecular gas and star formation rate, while \citet{filho16} find that the scatter around the KS law can be very large ($\sim 0.2$-$0.4$ dex) depending on many galaxy properties (e.g. $f_{gas}$, sSFR). In particular, extremely metal-poor (XMPs) dwarfs \citep[$Z < 1/10\ Z_\odot $][]{kunthostlin00} tend to fall below the KS relation, having unusually high HI content for their $\Sigma_{SFR}$, thus a lower star formation efficiency. On the other hand, some dwarf galaxies with enhanced SFR per unit area appear to have higher ongoing star formation efficiency \citep{amorin16}. The dispersion of the KS law, together with the less quantifiable uncertainty in the galaxy sizes, result in $f_{gas}$ errors of at least $0.2$-$0.3$ dex. Given that, the only way to have more reliable and constrained values of $f_{gas}$ and $\Sigma_{gas}$ would be to measure the gas content directly, such as using CO, dust or [CII] as $H_2$ mass tracers. 

\subsection{Derivation of chemical abundances and ionization parameter}

In order to derive the metallicity and ionization parameter for the galaxies in our sample, we use the python code HII-CHI-mistry (HCm). For a detailed description about how this program works we refer to the original paper by \citet{perezmontero14} (PM14), and we summarize here the basic principles. 

\subsubsection{$T_{\rm e}-$consistent, model-based abundances}

HCm is based on the comparison between a set of observed line ratios and the predictions of photoionization models using CLOUDY \citep{ferland13} and POPSTAR \citep{molla09}. Compared to previous methodologies, the models consider all possible ranges of physical properties for our galaxies, including variations of the ionization parameter. This quantity is defined as: 
\begin{equation}\label{logU}
log(U)= \frac{Q(H)}{4\pi r^2 n c} 
\end{equation}
where $Q(H)\ $ is the number of ionizing photons ($\lambda < 912\ $\AA) in $s^{-1}$, $r$ is the outer radius of the gas distribution in $cm$, $c$ is the speed of light in $cm/s$ and $n$ is the number density of hydrogen in $cm^{-3}$. The larger the value of $log(U)$ the more ionized the gas, even though collisional ionization can play an important role, especially when the temperature of the gas is high (hundreds of thousands of degrees).

Assuming the typical conditions in gaseous ionized nebulae, the grid probes possible values of $log(U)$ in the range $[-1.50, -4.00]$, metallicity $12+log(O/H)$ in the range $[7.1, 9.1]$ and $log(N/O)$ between $0$ and $-2$. Using a robust $\chi^2$ minimization procedure, HCm allows the derivation of three quantities, the oxygen abundance (O/H), the nitrogen over oxygen abundance (N/O) and the ionization parameter ($log(U)$), which best reproduce this set of five emission line ratios: [\oii]$\lambda 3727\,$\AA, [\oiii]$\lambda 4363\,$\AA, [\oiii]$\lambda 5007\,$\AA, [\nii]$\lambda 6584\,$\AA, and [\sii]$\lambda\lambda 6717+6731\,$\AA\ (all relative to \hb), which are provided as input parameters. 
The procedure consists of two steps: in the first step a comparison is done between the observed extinction-corrected emission-line intensities and the grid of models, providing a value of N/O. Then the nitrogen abundance is used to constrain the models and derive reliable oxygen abundance and ionization parameter with the same $\chi^2$ minimization methodology. 

HCm provides abundances which are consistent with those derived from the direct method for a large range of metallicity values and for a broad variety of galaxy types in the Local Universe. The agreement between the model-based O/H and N/O abundances and those derived using the direct method is excellent when all the lines are used. Reliable model-based estimation of O/H can also be obtained without [\oiii]$\lambda 4363\,$\AA\ detection if a limited grid of models is adopted using an empirical relation between $log(U)$ and $12+log(O/H)$. The relation between the metallicity and the ionization parameter arises from the physical properties of gaseous nebulae. It is consistent with large samples of local star-forming galaxies and HII regions (PM14), and has been tested also for the SFDGs in our sample with detected [\oiii]$\lambda 4363\,$\AA\ line, in which case no assumptions are made by the code. The role of this relation is to minimize the dispersion in the determination of O/H (PM14). Finally, when only [\oii]$\lambda 3727\,$\AA, [\oiii]$\lambda 5007\,$\AA\ and \hb\ are observed (the code is assuming a typical ratio between [\oiii]$\lambda 5007\,$\AA\ and [\oiii]$\lambda 4959\,$\AA\ of $3.0$ \citep{osterbrock89}), the $R23$ index ([\oii]$\lambda 3727\,$\AA+[\oiii]$\lambda\lambda 4959+5007\,$\AA)/\hb\ is used as a proxy to derive O/H, in combination to [\oii]$\lambda 3727\,$\AA/[\oiii]$\lambda 5007\,$\AA\ ratio to partially remove the dependence on $log(U)$. This is the most frequent case for our sample, since $93 \%$ of our SFDGs without an auroral line detection also have no detected [\nii]$6584\,$ and [\sii]$6717,6731$.

The procedure described above allowes to apply HCm for galaxies beyond the local universe, where galaxies are fainter and [\oiii]$\lambda 4363\ $\AA\ is generally not detected. HCm shows an increase of the dispersion (compared to the direct method) when a lower number of emission lines is considered, but gives results which are more consistent with the direct method than other empirical and theoretical calibrations. We find that the systematic differences with respect to HCm (and the direct method) can be as high as $\sim 0.7$\ dex at lower metallicities, depending on the adopted calibration. We refer to the Appendix \ref{appendixa} for a description of the calibrations analyzed in this paper and a more detailed comparison with our results.
{}
\subsubsection{Direct method}

\begin{figure}[t]
  \resizebox{\hsize}{!}{\includegraphics[angle=0,width=8.0cm,trim={0cm 0cm 0.0cm 0.cm},clip]{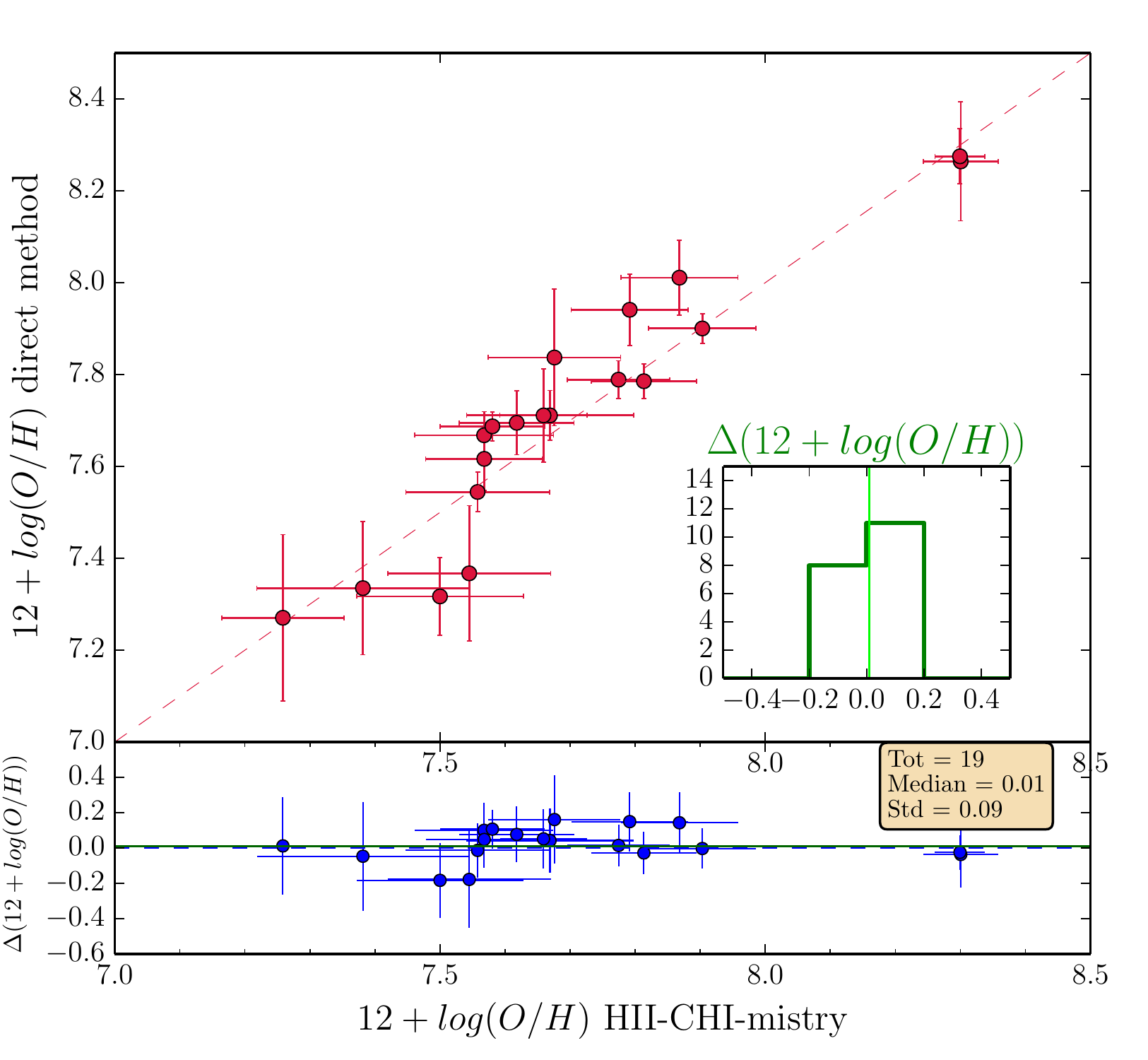}} 
  \caption{\small Comparison between the metallicity derived with the code HCm and the direct method, showing the consistency of our approach. The red dashed line is the $1:1$ relation. In the bottom side of the figure the differences between the two methods are plotted as a function of the metallicity. The blue dashed line correspond to the 0 level, while the blue continuous line is the median value. In the lower-right box we added the histogram with the distribution of the differences, with the median value line in green. 
  }\label{direct}
   \end{figure}

For a fraction of galaxies in our sample ($19$) we detect the [\oiii] $\lambda 4363$\ \AA\ auroral line, which is sensitive to the electron temperature $T_e$. This allows to use the direct method and check for these galaxies the consistency with metallicities derived through HCm. Despite the small number of galaxies, the two methods give consistent results within the errors over the entire metallicity range for the majority of them (Fig. \ref{direct}). We do not find any systematic trend in the low or high-Z regime, and the median difference between the two measurements is of $0.01$ dex. According to PM14, the consistency between HCm and the direct method, tested for an analogue sample of local star-forming galaxies, is of $\sim 0.15$\ dex. In our case, the $1\sigma$\ scatter is of the same order of magnitude ($\sim 0.09$ dex), even lower than previous value. Thus, we safely apply the code HCm to the other galaxies in the sample without auroral line detection. 

\section{Results}\label{section5}

\begin{figure}[t]
  \resizebox{\hsize}{!}{\includegraphics[angle=0,width=8.0cm,trim={0.8cm 0cm 1.9cm 1.2cm},clip]{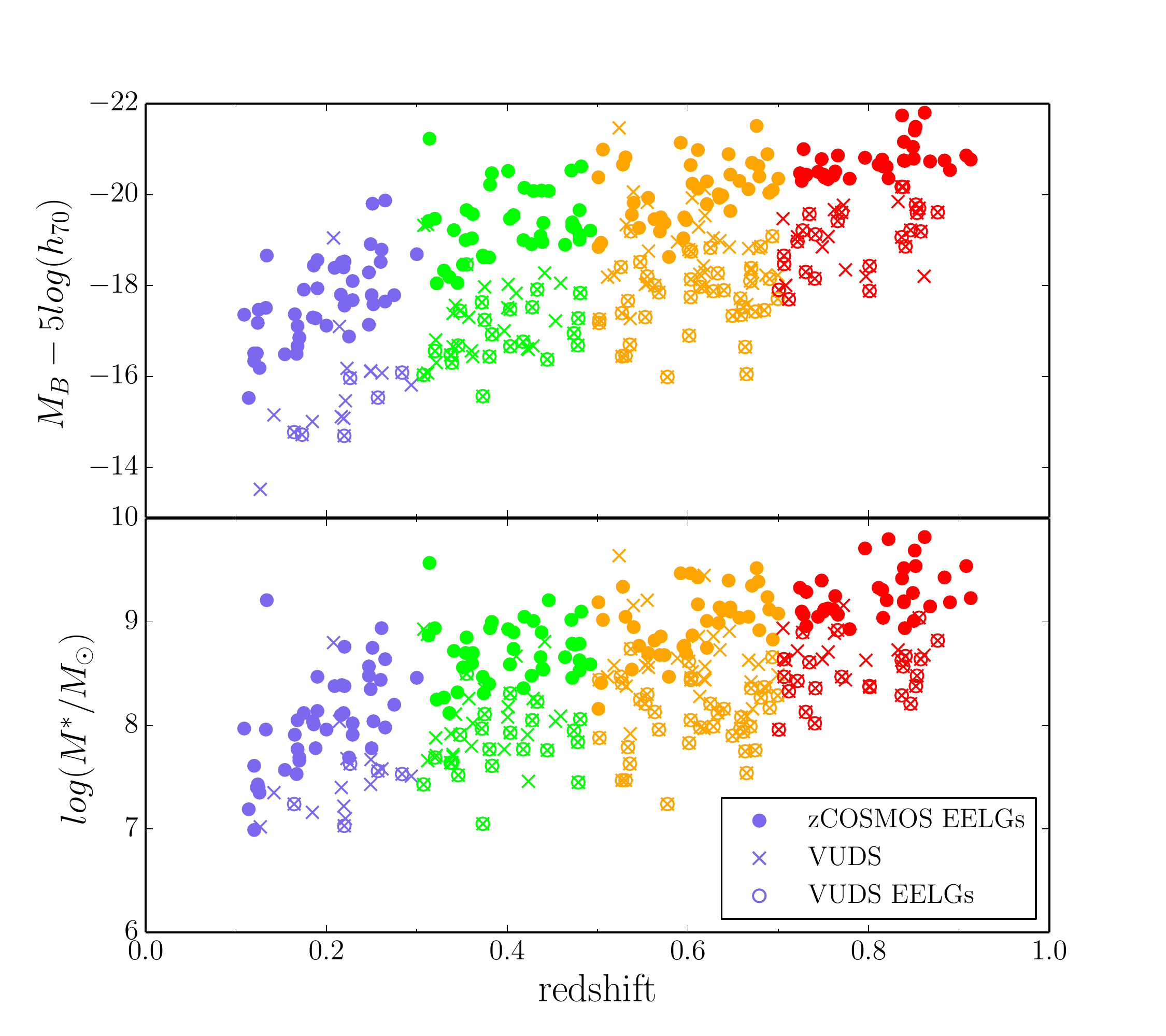}} 
  \caption{ \textit{Top:} \small $M_B-5\ logh_{70}$ vs redshift for our sample of galaxies (crosses) and for zCOSMOS EELGs by \citet{amorin15} (filled circles), color coded according to different bins of redshift ($0.1$-$0.3$,$0.3$-$0.5$,$0.5$-$0.7$ and $0.7$-$0.9$). The EELG fraction of our galaxies is evidenced by empty circles around the crosses. The B-band luminosities span a wide range, $-13 \leq M_B \leq -21.5 $, increasing with redshift. \normalsize \textit{Bottom:} \small Galaxy stellar mass vs redshift for the same galaxy samples.}
\label{BMAG-mass-z}
\end{figure}

\begin{figure*}[t]
    \centering
    \includegraphics[angle=0,width=6.cm,trim={0.8cm 0cm 1.55cm 1.2cm},clip]{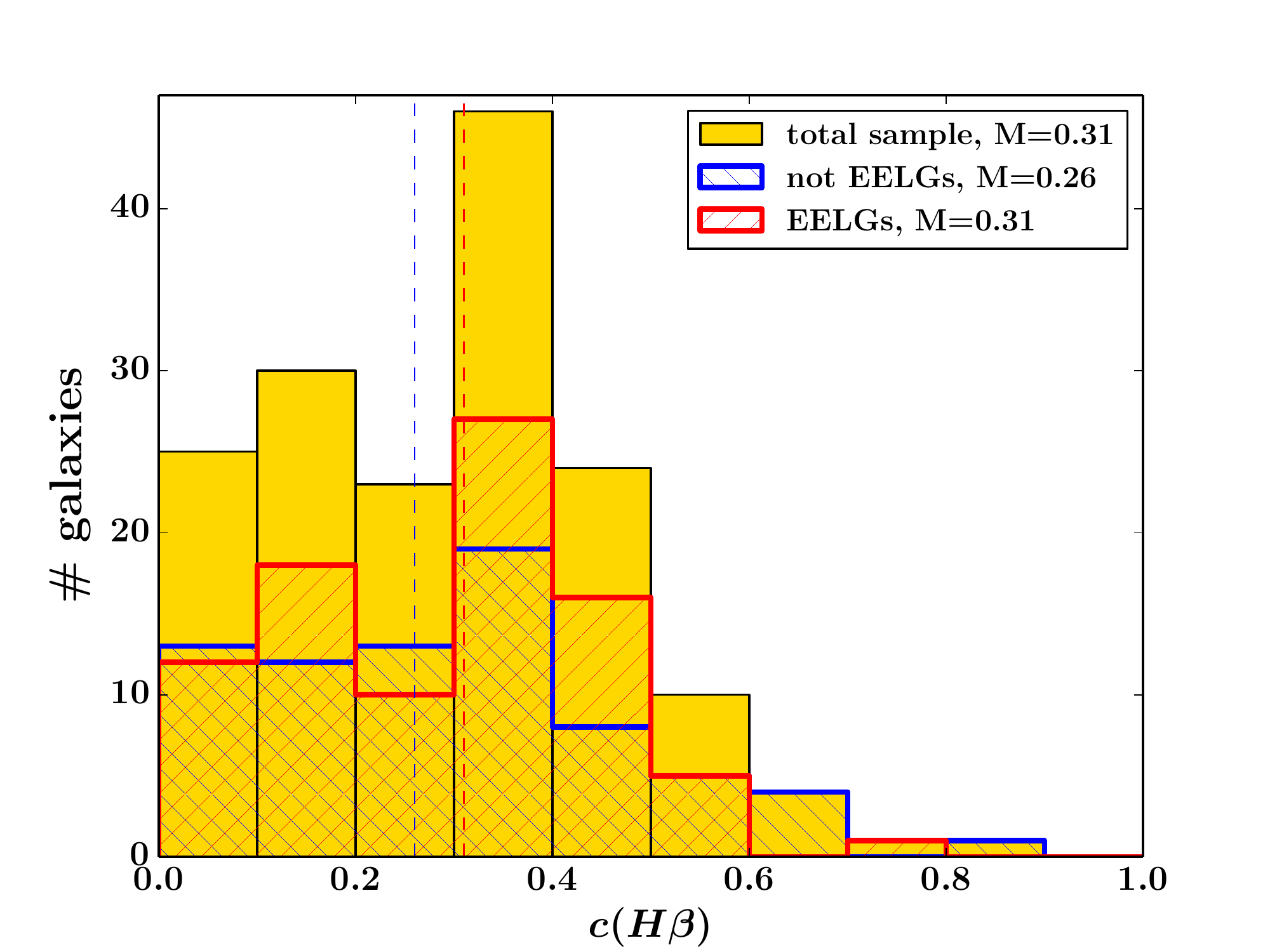}
    \includegraphics[angle=0,width=6.cm,trim={0.8cm 0cm 1.55cm 1.2cm},clip]{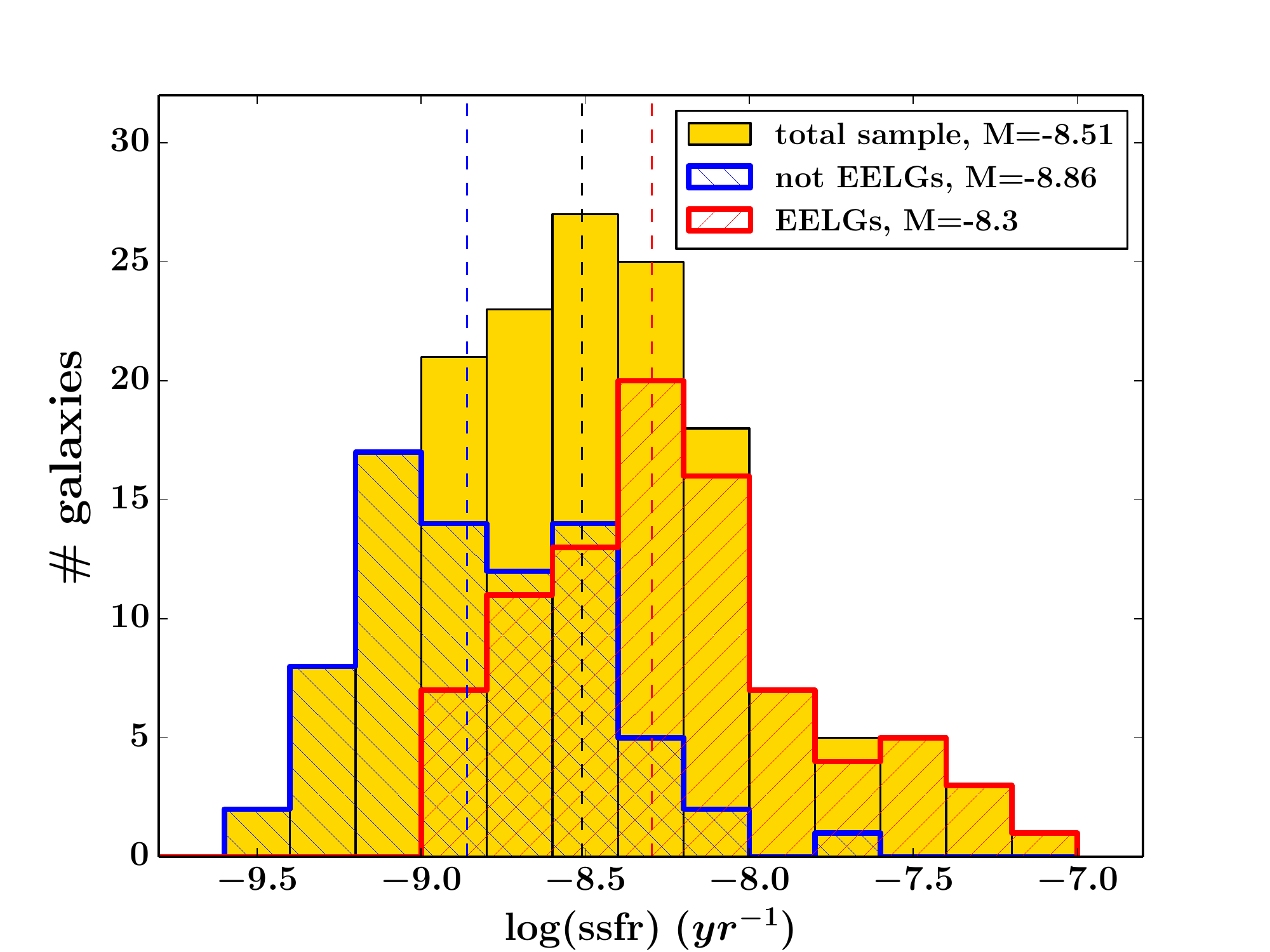}
    \includegraphics[angle=0,width=6.cm,trim={0.8cm 0cm 1.55cm 1.2cm},clip]{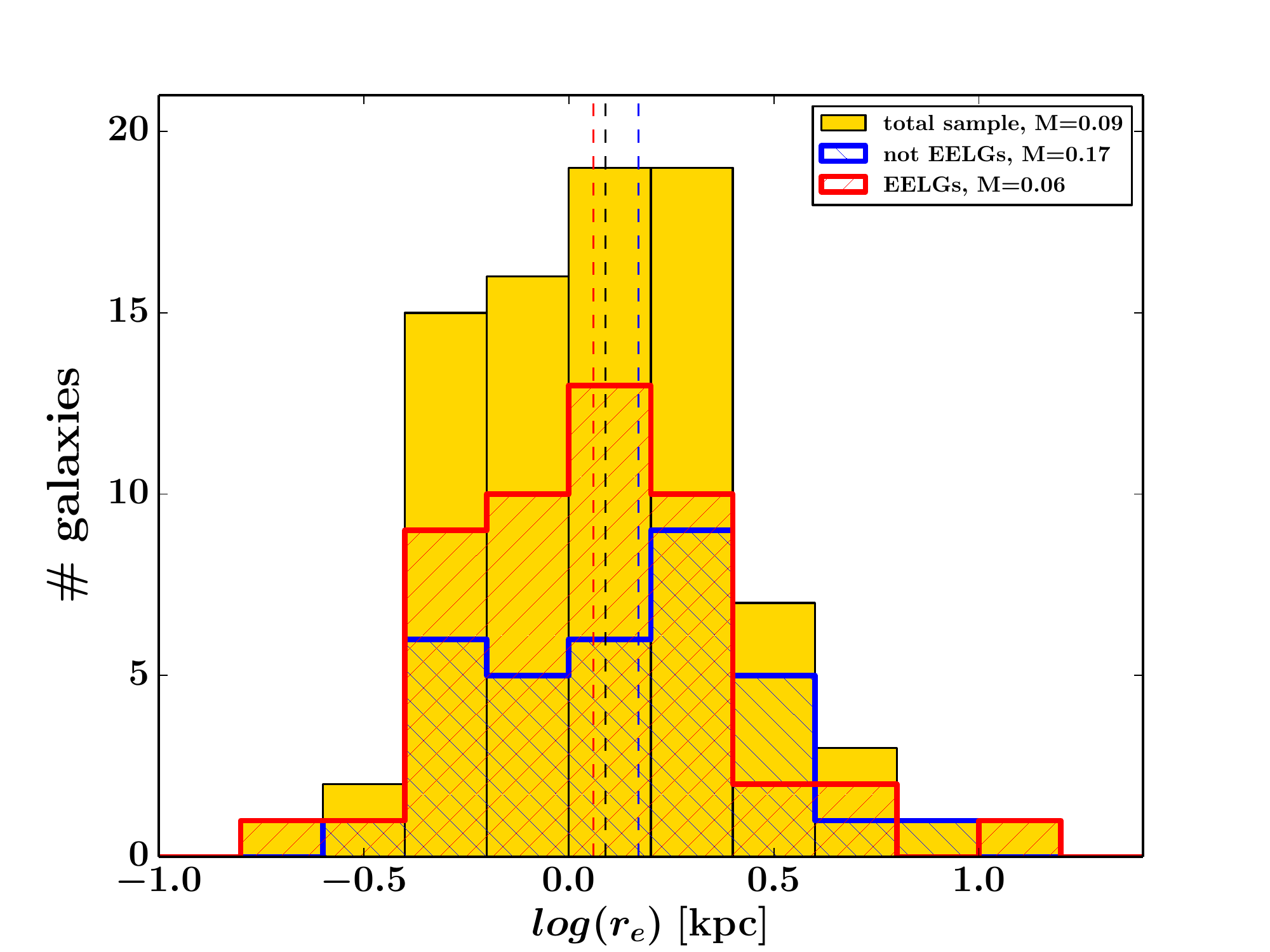}
    \includegraphics[angle=0,width=6.cm,trim={0.8cm 0cm 1.55cm 1.2cm},clip]{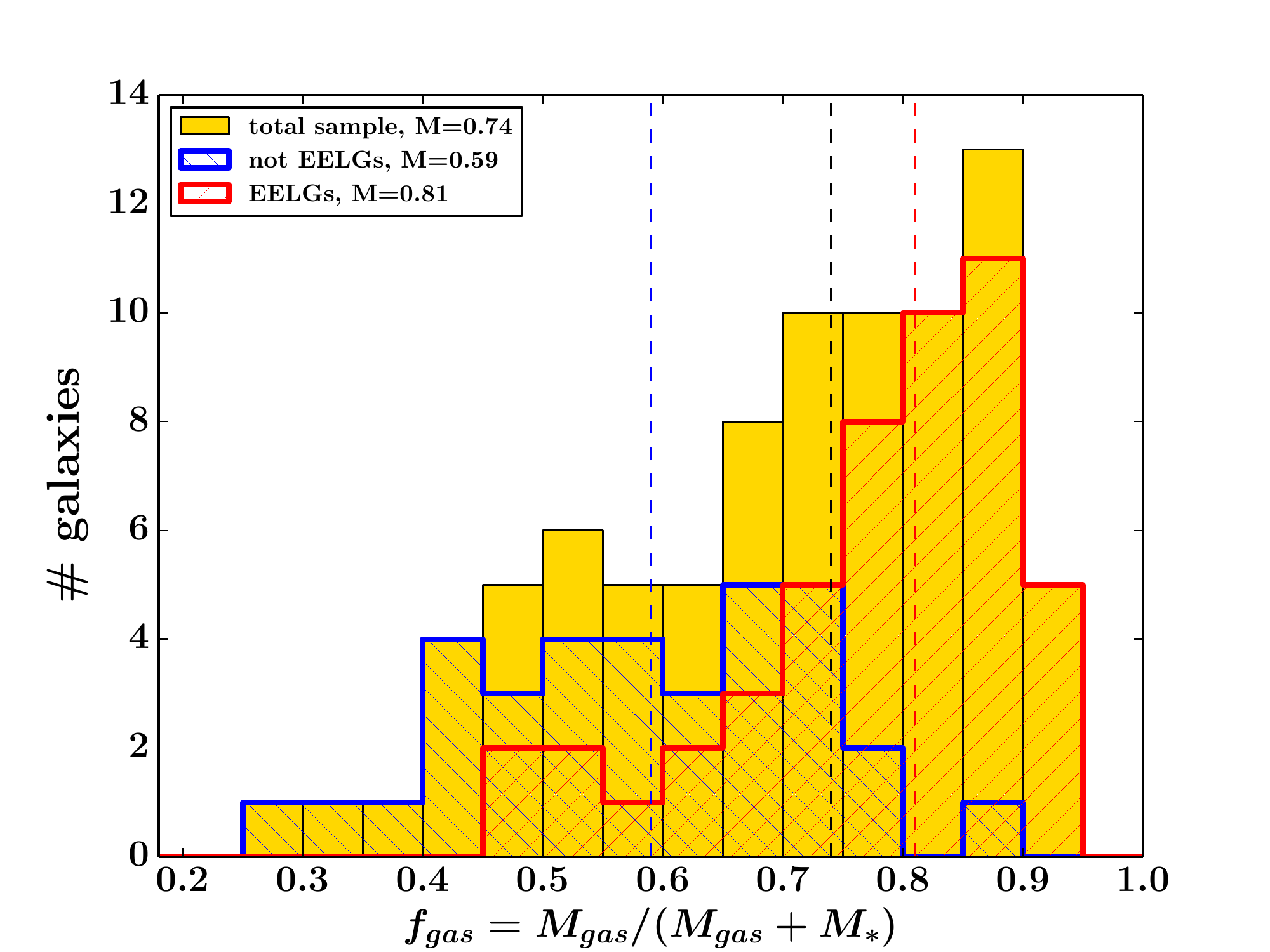}
    \includegraphics[angle=0,width=6.cm,trim={0.8cm 0cm 1.55cm 1.2cm},clip]{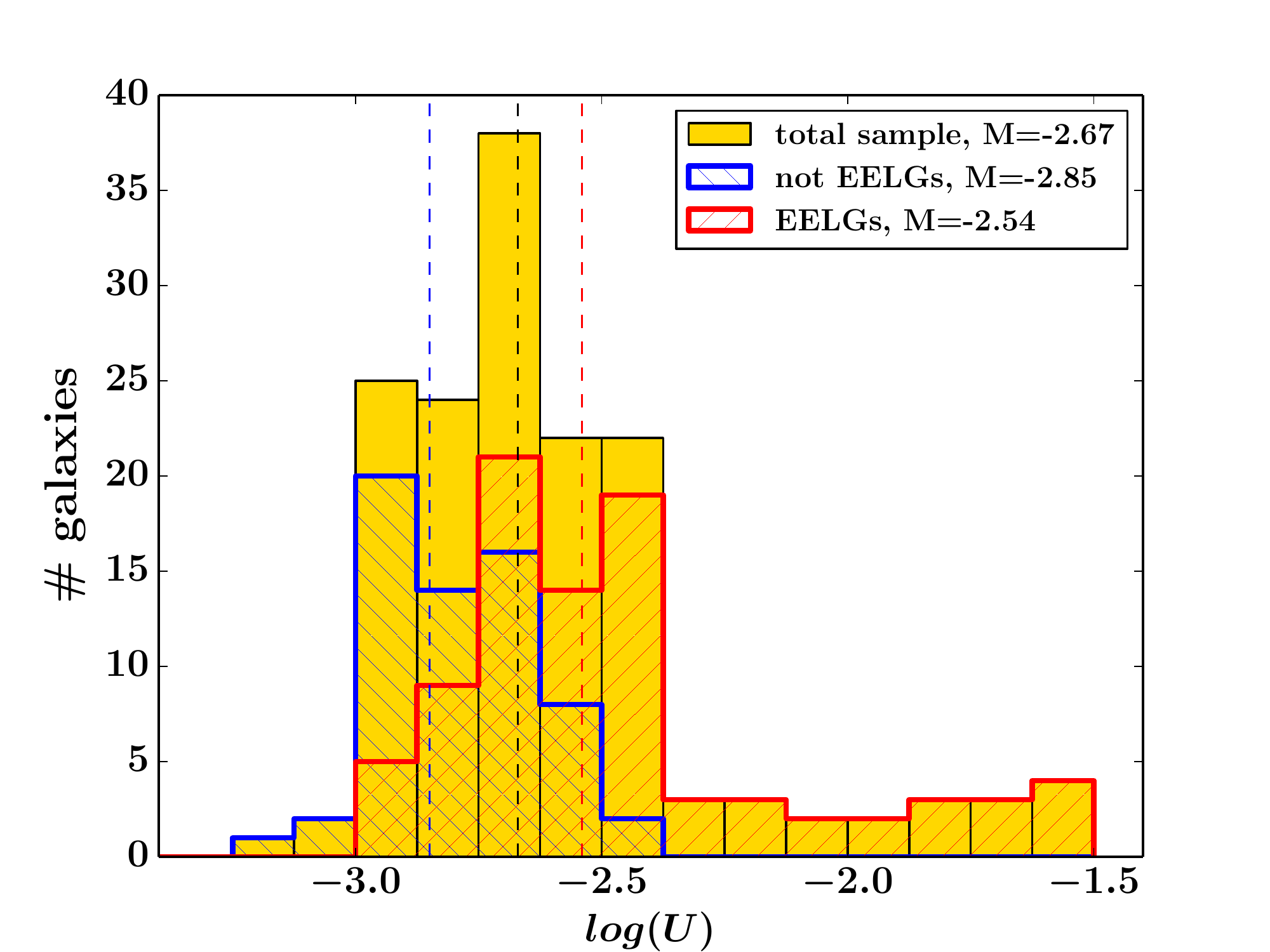}
    \includegraphics[angle=0,width=6.cm,trim={0.8cm 0cm 1.55cm 1.2cm},clip]{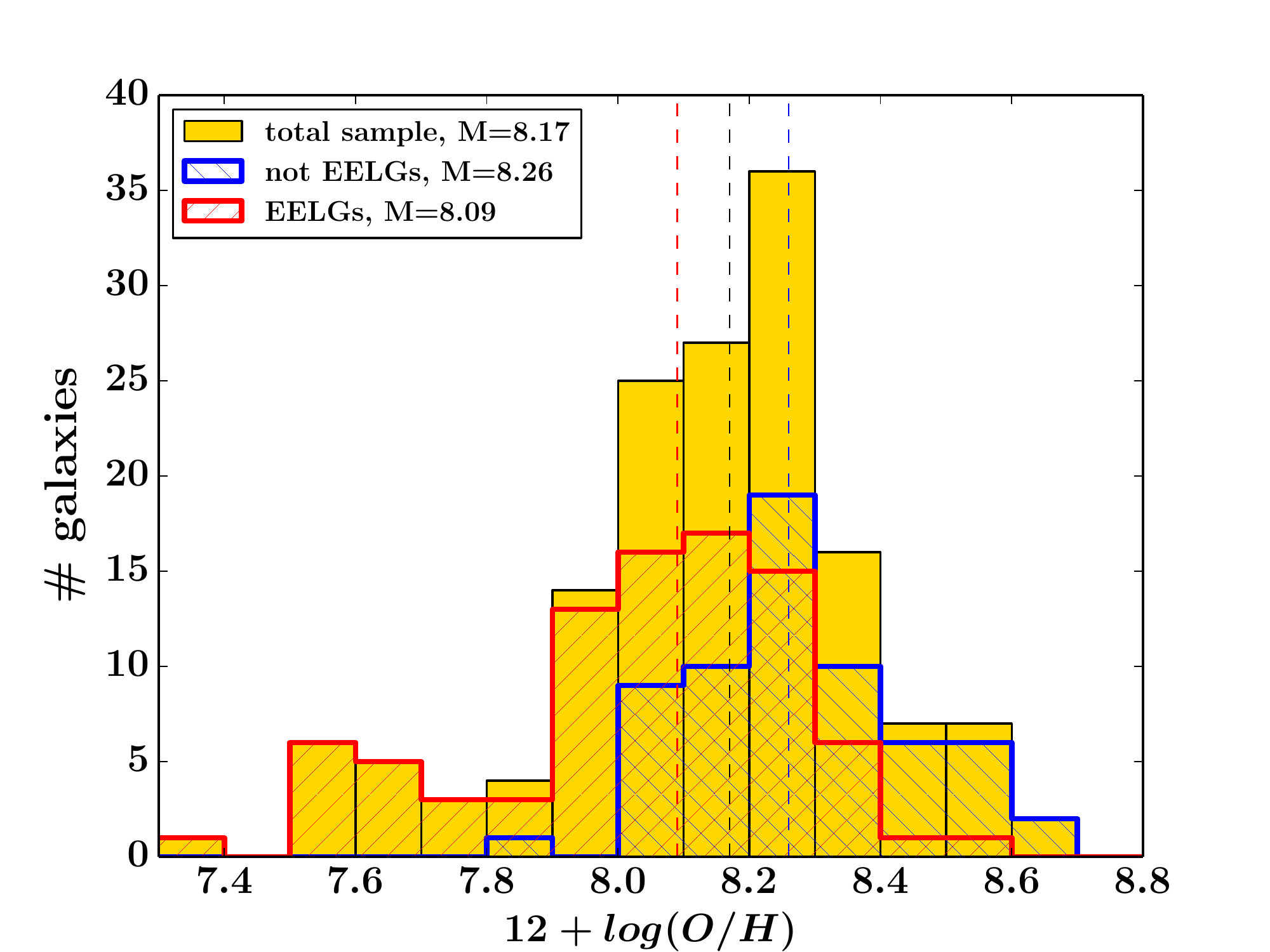}
    \caption{Histogram distribution of extinction coefficient c(\hb), sSFR and circularized effective radius $r_e$ \textit{(Top)}, gas fraction $f_{gas}$, gas-phase metallicity and ionization parameter $log(U)$ \textit{(Bottom)} for our selected galaxies. The EELG and non-EELG fractions are highlighted with red and blue lines filling the histogram. The median distribution values for each quantity are indicated in the legend for each subset of galaxies.} \label{histograms}
\end{figure*}

In this section we present the properties of the SFDG sample and we study the key scaling relations between them. The main physical quantities presented here are listed in Table \ref{Table3} in Appendix \ref{appendixb}. 

In Fig.~\ref{BMAG-mass-z} we show that the stellar masses of the galaxies, derived through SED fitting, span a wide range of values from $10^{10}$ down to $10^7$ $M_\odot$ (median value of $10^{8.2} M_\odot$), a region in the stellar mass distribution of galaxies which is still strongly underrepresented in current spectroscopic surveys at intermediate redshift. Compared to zCOSMOS EELGs (A15), which is one of the largest samples of low-mass star-forming galaxies at these redshifts \citep[see also][]{ly16}, we extend  $\sim 2$\ $mag$\ lower the intrinsic luminosity of the sources (given in rest-frame absolute magnitude $M_B$) thanks to VUDS deeper observations. As a consequence, our sample extends to lower stellar masses, though we are biased toward the higher $M_\ast$ at a given redshift.

In Fig. \ref{histograms} we present the distribution of various physical quantities: extinction correction c($\hb$), specific star formation rate (sSFR), effective radius ($r_e$), gas fraction ($f_{gas}$), ionization parameter ($log U$) and oxygen abundance ($12+logO/H$). The histogram of extinction coefficients shows that the dust extinction is generally low, with median reddening of $E(B-V)=0.45$\,mag ($\sigma=0.38$), and there is no significant difference between EELG and non-EELG distributions. The reddening values found here are consistent with previous studies on local (e.g. Kniazev 2004) and intermediate redshift samples of SFDGs \citep[e.g.][]{amorin15,ly14,ly15}.  Our galaxies have low to moderate SFRs ranging 10$^{-3}\lesssim$\,SFR\,$\lesssim$\,10$^{1}$\,M$_{\odot}$~yr$^{-1}$, with a median SFR$=$\,0.64\ M$_{\odot}$~yr$^{-1}$, and $1\sigma$ scatter of $0.6$. As a consequence, the sSFR tend to be high, spanning a broad range, 10$^{-10}$\,yr$^{-1} \lesssim$\,sSFR\,$\lesssim$\,10$^{-7}$\,yr$^{-1}$, with a tail of very high sSFR galaxies. The whole sample has a median of 10$^{-9}$\,yr$^{-1}$, well above the Milky Way integrated sSFR ($ \sim 1.5 \times 10^{-11}$ yr$^{-1}$). 

We also find that our galaxies are very compact, with effective radii (in kpc) in the range $ 0.1 \leq r_e \leq 6\ kpc $ and median value of $r_e = 1.2\ kpc$ (standard deviation of $2.3\ kpc$), comparable to BCDs \citep{papaderos96,gildepazmadore05,amorin09} and the Green Pea galaxies \citep{amorin12}. In the same figure, we show the distribution of values of the gas fraction of our SFDGs. The values range $0.25 < f_{gas} < 0.95$, and the distribution is peaked toward higher $f_{gas}$, with one half of our galaxies showing $f_{gas} > 0.74$, i.e. more than $74\%$ of the total baryonic mass resides in their gas reservoirs. The highest values are found in EELGs, which show a median $f_{gas}$ $\sim 0.2$ higher compared to non-EELGs. These results for the whole sample can be an indication that the star formation efficiency has been very low or that our galaxies are very young and still assembling most of their stellar mass. In the Local Universe, such values of gas fractions can be found in low-mass, low-metallicity and highly star-forming galaxies \citep[e.g.][]{laralopez13,filho16,amorin16}. 

\subsection{The relation between star formation rate and stellar mass}\label{SFRmass1}

\begin{figure*}[t]
  \centering
  \includegraphics[angle=0,width=16cm]{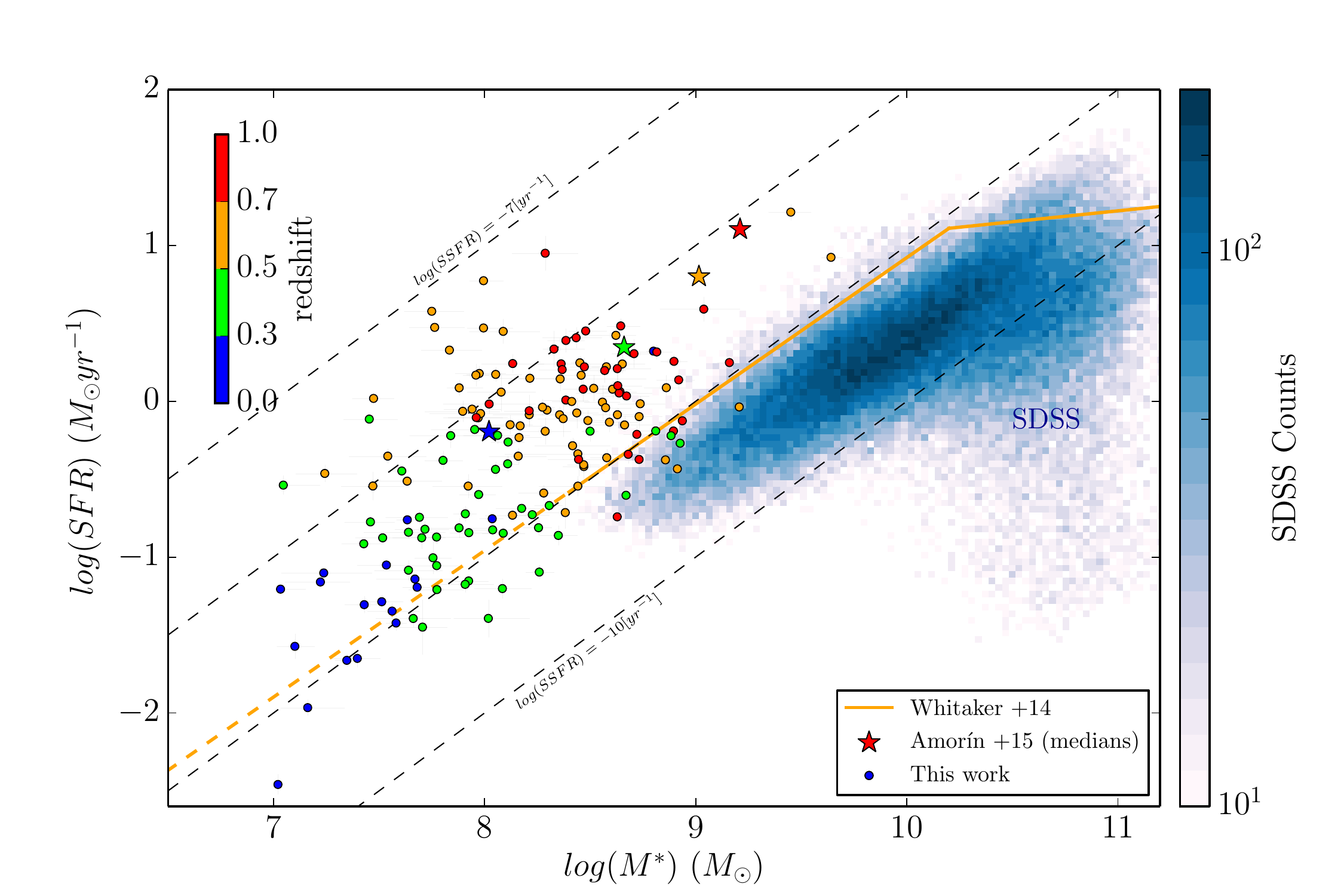} 
  \caption{\small Diagram of SFR vs $M_\ast$ for our sample. The galaxies are coded with different colors according to 4 redshift bins, and the colored stars represent the medians for zCOSMOS galaxies \citep{amorin15} in four bins of masses. The orange continuous line represents the star-forming galaxies at similar redshift $0.5 < z < 1\ $ (the so-called `Main Sequence of Star Formation') derived by Whitaker et al. (2014). We extrapolate this line to the low-mass end of the diagram (dashed line). The dashed gray lines correspond to constant SSFR values, going from $10^{-10} yr^{-1}$ to $10^{-7} yr^{-1}$. We show with a blue 2-D histogram our SDSS sample described in section \ref{SDSS}.}\label{SFRmass}
   \end{figure*}

For high-mass star-forming galaxies ($M_\ast > 10^9\ M_\odot$) a tight correlation has been found between the star formation rate and the stellar mass \citep[e.g.][]{brinchmann04,elbaz07,tasca15} up to redshift 5. This correlation, called "the star formation main sequence (MS)", is displaced towards higher values of SFR at higher z, with a nearly constant slope $\sim1\,$ \citep{guo15} and a small non-varying SFR dispersion of $\sim 0.3$ dex \citep{schreiber15}. An extension of the main sequence has been derived by \citet{whitaker14} for intermediate redshift star forming galaxies ($0.5<z<1$) down to $M_\ast \sim 10^{8} M_\odot$, but the relation remains still poorly constrained at lower masses.  

Our sample of 164 VUDS SFDGs allows to populate the low mass end of the SFR-$M_\ast$ relation (Fig.\ref{SFRmass}). Our sample has a higher average SFR per unit mass compared to the extrapolation of the MS at low masses \citep[$M_\ast < 10^{8.4} M_\odot$,][]{whitaker14} This result should be considered as a secondary effect of our selection criteria, based on a S/N $>3$ cut on hydrogen recombination lines, [\oii]$3727$ and [\oiii]$5007$ (see section \ref{selection}). At a given continuum luminosity and stellar mass we are limited to the brighter \ha\ (\hb) values (i.e. higher sSFRs), compared to a continuum S/N selected sample. We also notice that a conspicuous number of SFDGs have higher star formation rates than the median population, with starburstiness parameters \citep[defined as SFR/SFR$_{MS}$,][]{schreiber15} up to $1.8$ dex, similarly to zCOSMOS EELGs. 

\begin{figure}[t]
   \resizebox{\hsize}{!}{\includegraphics[angle=0,width=8.2cm,trim={0cm 0cm 0.2cm 1.2cm},clip]{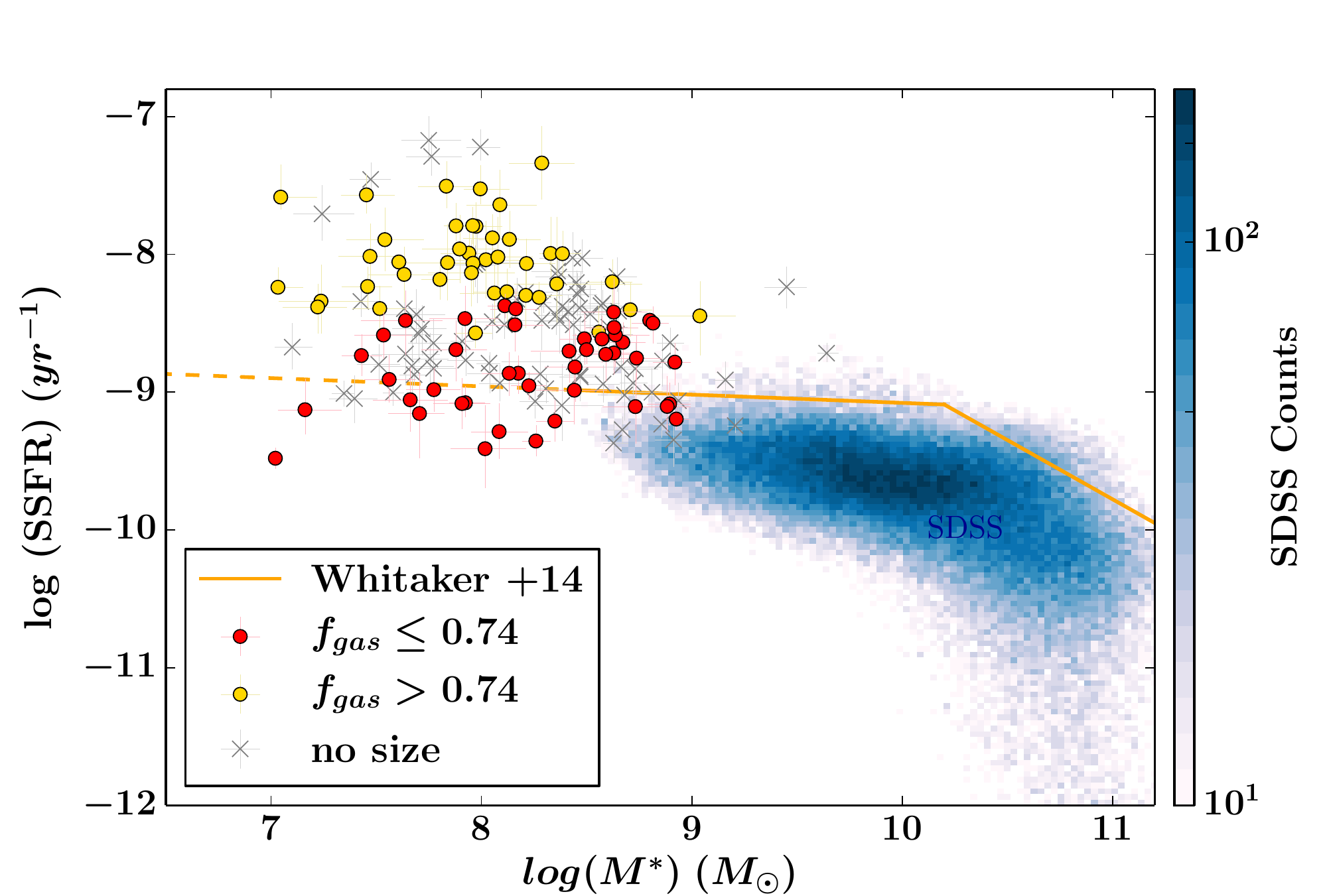} }
   \caption{\small Diagram of sSFR vs $M_\ast$ for our galaxies, coded by $f_{gas}$ (below). The segregation is evident between galaxies with higher and lower gas fraction than the median $f_{gas,med}=0.74$. The EELG fraction is evidenced with yellow color circles.}
   \label{SSFRmass2}
\end{figure}

This difference is more evident when we consider the sSFR vs $M_\ast$. The sSFR distribution of the EELG fraction is biased high ($\sim +0.6$ dex) with respect to the non-EELGs. We also find that the starburstiness parameter depends on the gas fraction. In particular, the median gas fraction of the sample ($f_{gas;med}=0.74$) is very effective to distinguish between these two classes, with more gas-rich galaxies having on average higher sSFRs ($\sim 1$\,dex) compared to the main sequence star-forming population (Fig. \ref{SSFRmass2}).

\subsection{Chemical abundances and ionization properties}

\begin{figure}[t!]
  \resizebox{\hsize}{!}{\includegraphics[angle=0,width=8.2cm]{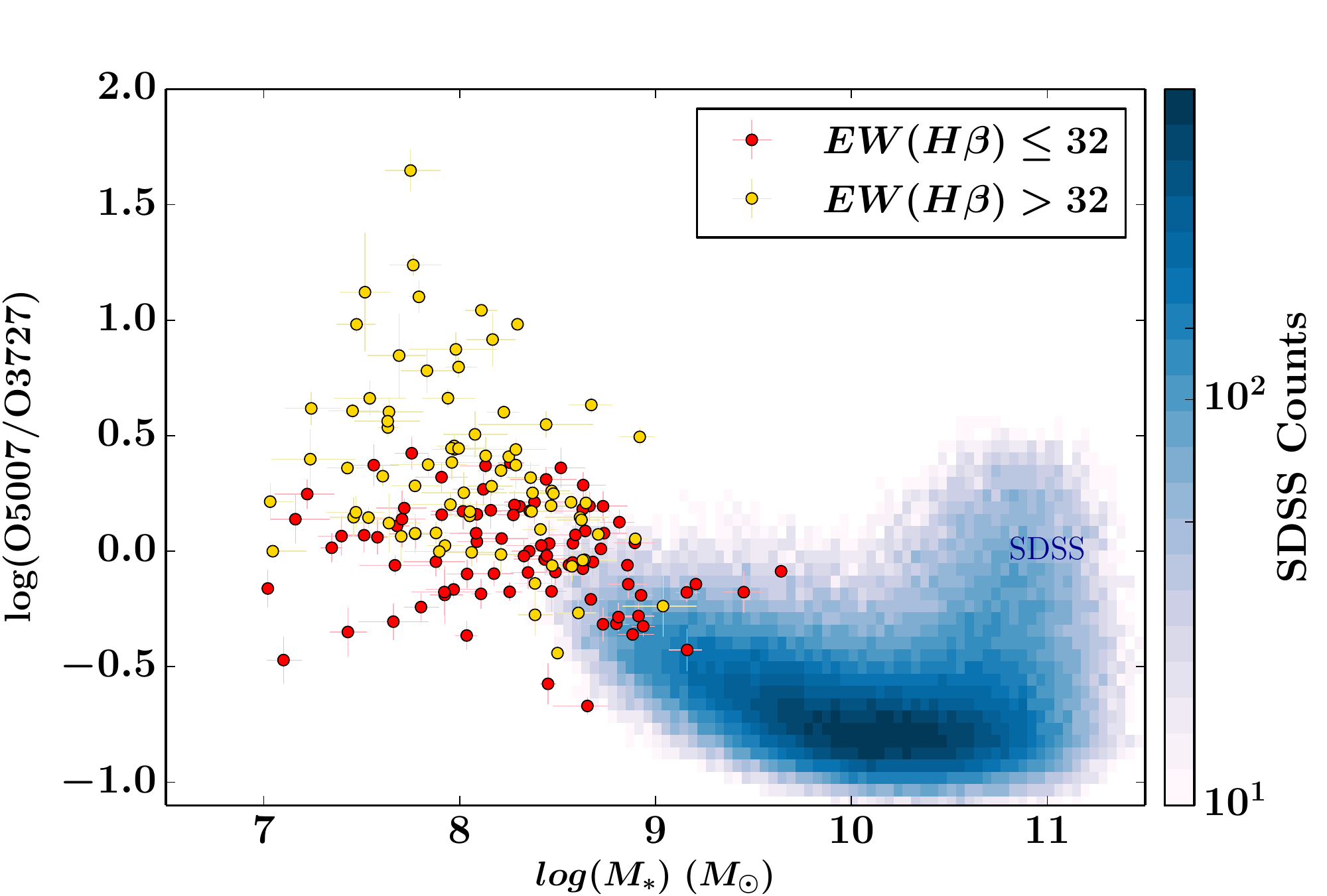}} 
  \caption{\small Comparison between extinction corrected [\oiii]/[\oii]\ and stellar mass $M_\ast\ $ for our sample of galaxies, extending to lower masses the diagram of Local star-forming galaxies (blue 2-D histogram). A color-coding is adopted according to EW(\hb) higher or lower than the sample median ($32\ \AA$).
  }\label{ioniz-mass}
   \end{figure}

In the last two panels of Fig. \ref{histograms}, we present the results of the code HCm, the ionization parameter and the metallicity. We discard from this analysis $12$ galaxies of our VUDS original sample for which some of the emission lines required by the code are not reliably detected. For the remaining $152$ galaxies we find that they span a wide range of values, respectively $-3.17 < log(U) < -1.55$ and $7.26 < 12 + log(O/H) < 8.6$.  

The ionization parameter distribution has median value $log(U)_{med} \simeq -2.6$, with an extended tail of objects toward higher ionizations. In Fig. \ref{ioniz-mass} we see that high ionization is found preferentially in lower mass objects. In this plot we compare the stellar mass to the extinction corrected emission line ratio [\oiii]$\lambda 5007$/[\oii]$\lambda 3727$, which is typically used as a proxy of the ionization parameter and is the most common ionization parameter diagnostic (Baldwin et al. 1981). Our SFDGs show a broad variety of conditions ranging $-0.5<$ [\oiii]$5007$/[\oii]$3727$ $< 1.3 $, and a mild correlation is found with $M_\ast$, though the scatter increases largely at lower masses. Compared to the bulk of local star-forming galaxies (SDSS), VUDS SFDGs show on average higher ionizations reaching log([\oiii]$5007$/[\oii]$3727$) $\sim 1$, which makes them more similar to the typical ionization conditions found at higher redshifts \citep{nakajima14}. As we will discuss later, this type of objects are important for understanding the properties of faint star-forming galaxies in reionizing the universe at $z>6$. In particular, high ionization parameters traced by [\oiii]$5007$/[\oii]$3727$ may indicate the presence of density-bounded HII regions, i.e. escaping LyC radiation \citep[e.g.][]{jaskotoey13}. 

In the metallicity distribution histogram, we see that all the galaxies in our sample have sub-solar oxygen abundances, i.e. $12+log(O/H) < 8.69$, and the median is $8.17$, consistent with zCOSMOS EELGs (($12+log(O/H))_{med} = 8.16$). 

\begin{figure}[t!]
  \resizebox{\hsize}{!}{\includegraphics[angle=0,width=8.5cm,trim={0.1cm 0cm 1.15cm 1.0cm},clip]{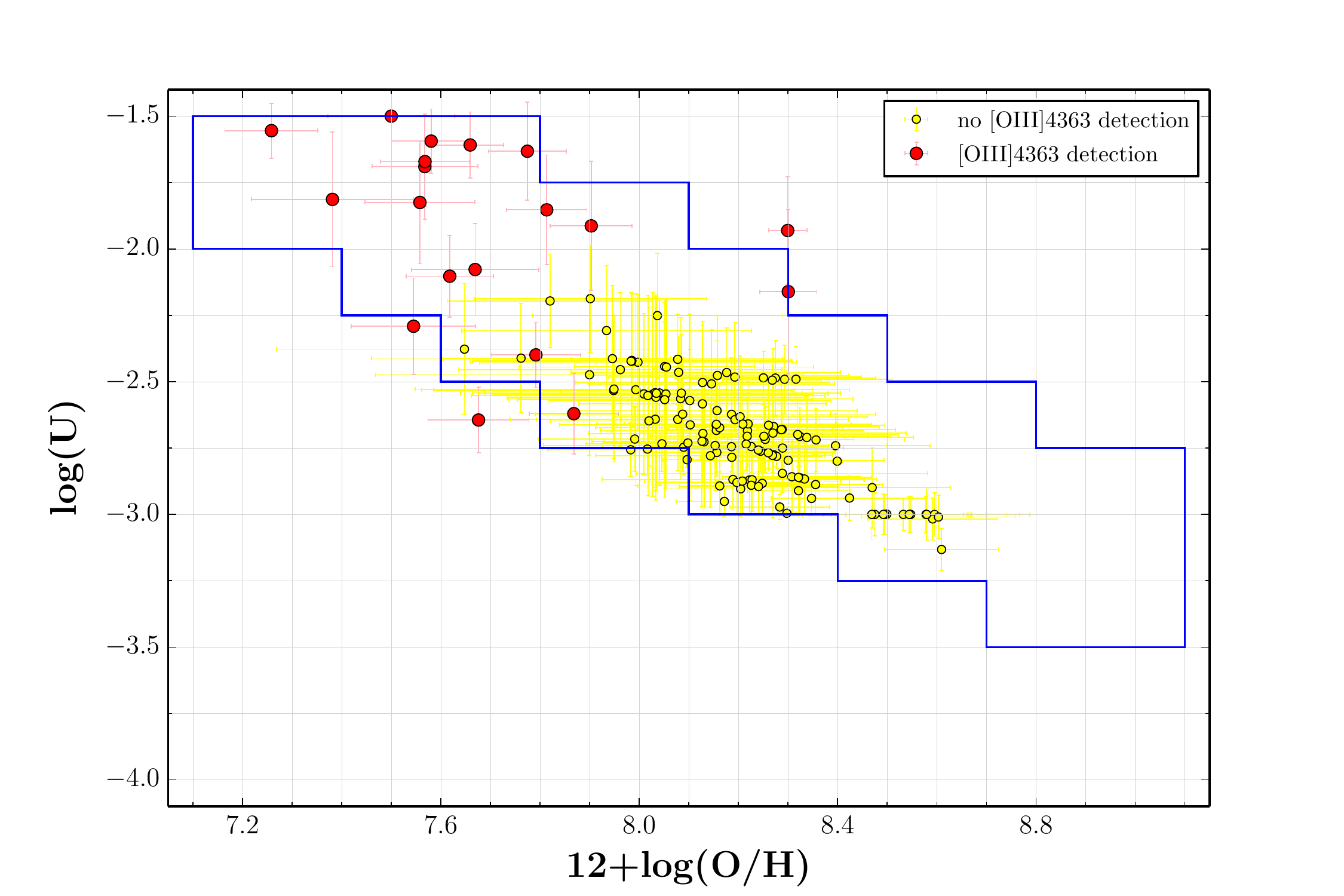}} 
  \caption{\small Comparison between the metallicity $12+log(O/H)$ and $log(U)$, both derived with the code HCm. The plot shows the anti-correlation between the two quantities, with the limited grid of models (green continuous line) considered by HCm when only [\oiii]$\lambda 3727\,$\AA\ and [\oiii]$\lambda 5007\,$\AA\ are available. Red points are the galaxies in our sample [\oiii]$4363$ detection, showing that the empirical relation is real, since HCm does not introduce any assumptions in this case. 
  }\label{OHlogU}
   \end{figure}

In Fig. \ref{OHlogU} we compare for our galaxies the metallicity and the ionization parameter, both outputs of HCm. When the auroral line [\oiii]$\lambda 4363\,$\AA\ is not detected, HCm considers a limited grid of models, which account for the anti-correlation between metallicity and ionization observed in local star-forming galaxies and giant HII regions, which origins in the physical properties of the nebulae. We show enclosed by a blue solid line the grid of models that we have used, and the relation for our [\oiii]$4363$-detected galaxies, for which no assumptions are made on the relation $12+log(O/H)$ vs $log(U)$ by HCm. Instead, we apply this constrain to the rest of the sample where auroral line is not detected, to minimize the dispersion in the determination of metallicity.

Finally, we notice that $12$ galaxies in the sample are very metal deficient and lie in the category known to as extremely metal poor (XMP) galaxies \citep[$Z < 1/10\ Z_\odot $][]{kunthostlin00}. One galaxy has a metallicity of $7.26 \pm 0.1$ (VUDS J$100045.13+022756.0$), which is below $1/20$ solar ($ < 7.4$). Its value is comparable to the most metal-poor galaxies known \citep[e.g. I Zw 18,][]{izotovthuan99}.
In this subset of XMPs, $11$ have detection of [\oiii]4363 line.

\subsection{The mass-metallicity relation}\label{MZR}

\begin{figure*}[t!]
  \centering
  \includegraphics[angle=0,width=16.cm]{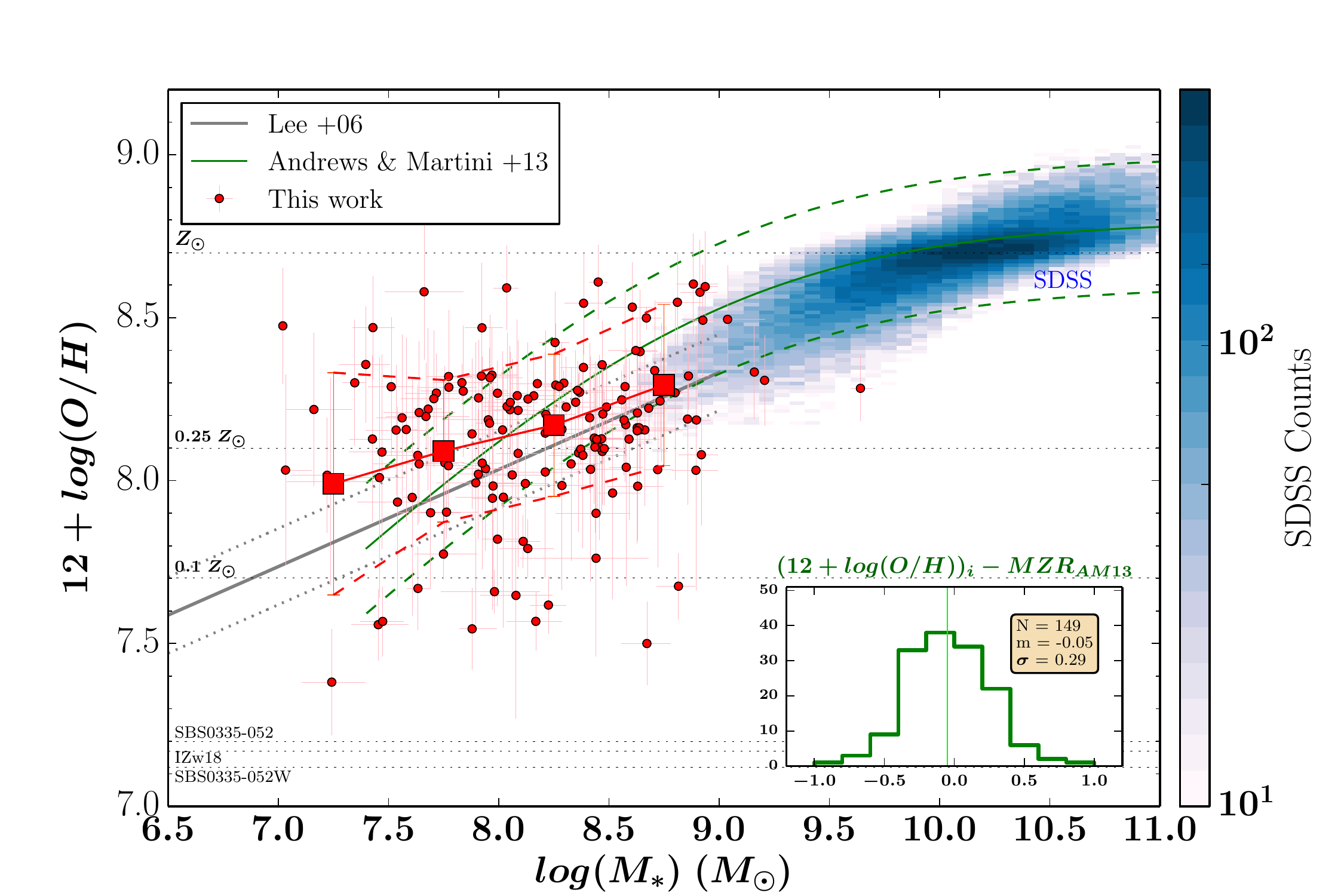} 
  \caption{\small Mass-metallicity diagram for our sample of galaxies, compared to the MZR of Andrews \& Martini 2013 and Lee et al. 2006. The green dashed lines and grey dotted lines represent the $\pm 1\sigma\ $ deviation of AM13 ($\simeq 0.2\ $ dex) and L06 ($\simeq 0.1\ $ dex), respectively. The horizontal dashed lines represent constant metallicities, referred to the Sun, and the levels of the three most metal poor galaxies known in the local universe. We show in this figure also our error-weighted mean metallicities (red squares) and $1\sigma$ standard deviations calculated in four bins of mass ($log(M_\odot)$=[7-7.5, 7.5-8, 8-8.5, 8.5-9]). The histogram in the inset shows the differences between the metallicities derived with HCm and those obtained from AM13 MZR at given stellar mass. Our sample is on average slightly below the AM13 relation (median difference of $0.05\,$dex), with a $1\sigma$ scatter around the median difference of $0.29\ $ dex. We report here the median and the $1\sigma$-std for each of the four mass bins of our MZR, in increasing order of mass: (m$=7.99$,$1\sigma$$=0.34$), (m$=8.06$,$1\sigma$$=0.22$), (m$=8.16$,$1\sigma$$=0.22$), (m$=8.29$,$1\sigma$$=0.24$). The SDSS galaxy sample is shown with a blue 2-D histogram. 
  }\label{MZR1}
   \end{figure*}

In this section we study the low-mass end of the MZR of SFGs, which can provide valuable insight into the physical processes regulating the mass assembly and chemical evolution of low-mass galaxies.  

Even though the MZR has been well-determined at higher masses ($M_\ast > 10^9 M_\odot$) \citep{tremonti04}, it is still not completely defined at lower masses, where it has been studied in the local Universe by \citet{lee06}, \citet{zahid12b} and \citet{andrewsmartini13}, but tested at intermediate redshift ($z<1$) with relatively small samples \citep[e.g.][]{henry13,ly14,ly15,ly16}.

In Fig. \ref{MZR1} we present the mass-metallicity diagram for our sample of SFDGs, which provides new observational constrains to its low mass regime. Our objects populate a large region at lower masses compared to the bulk of local star-forming galaxies, while the dispersion is high and appears to increase from solar to strongly subsolar values. Despite of the large overall scatter, we find some agreement with the local MZR derived by AM13. In particular, $50 \%$ of our galaxies follow the local $T_e$-based MZR of AM13 to within $\pm 2 \sigma\ $ ($\simeq 0.2\ $dex ), while a large number ($\sim 40 \%$) is located below the $1\sigma\ $ limit. For individual galaxies, we find a maximum difference with the metallicity that corresponds to AM13 MZR at a given mass of $\sim 0.9$\ dex.

In order to characterize the MZR for our SFDGs, we divide the sample in four bins according to their stellar mass in order to have approximately the same number of galaxies in each bin (intervals of $log(M_\ast)$ are: $7.0$-$7.5$,$7.5$-$8.0$,$8.0$-$8.5$ and $8.5$-$9.5$). For those falling in the same bin we compute the weighted mean metallicity and $1\sigma$ standard deviations (std). Finally, we connect the mean points with a continuous line and the $1\sigma$ limits with dashed lines. The slope of our MZR is lower compared to the local relation of AM13, while it is more in agreement with the MZR found by \citet{lee06} using 27 nearby SFDGs with good measurements of $T_e$ (with a small offset toward higher metallicities).
We also find that the overall dispersion of our MZR (computed as the average $1\sigma$-std of the metallicities in the four bins) is of $\sim 0.26$, though it increases by $0.1$\ dex towards very low masses, reaching $0.34$ in the last mass bin. The exact values of the median metallicity and std for each bin are reported in the description of Fig. \ref{MZR1}.

Finally, in order to check for a possible dependence of our results on the S/N used to select our sample (section \ref{selection}), we have tested three different S/N lower-limits (3, 4 and 5) on the following emission lines critical for the determination of metallicity: [\oii]$\lambda 3727$\AA\,[\oiii]$\lambda 5007$\AA\ and \hb. Choosing a higher S/N threshold, applied simultaneously to the three above emission lines, reduces the total number of galaxies in the sample (from 164 to $111$ for S/N $=5$), but does not change our results. The average metallicities in the four bins scatter by less than $0.04$ dex ($\sim 5 \%$), which is lower than the mean uncertainty of the metallicity estimations. Likewise, the overall slope and $1\sigma$ deviation remain consistent with our former result. 

\subsection{The dependence of the metallicity on the stellar mass and SFR}

\begin{figure}[t]
  \resizebox{\hsize}{!}{\includegraphics[angle=0,width=9.4cm]{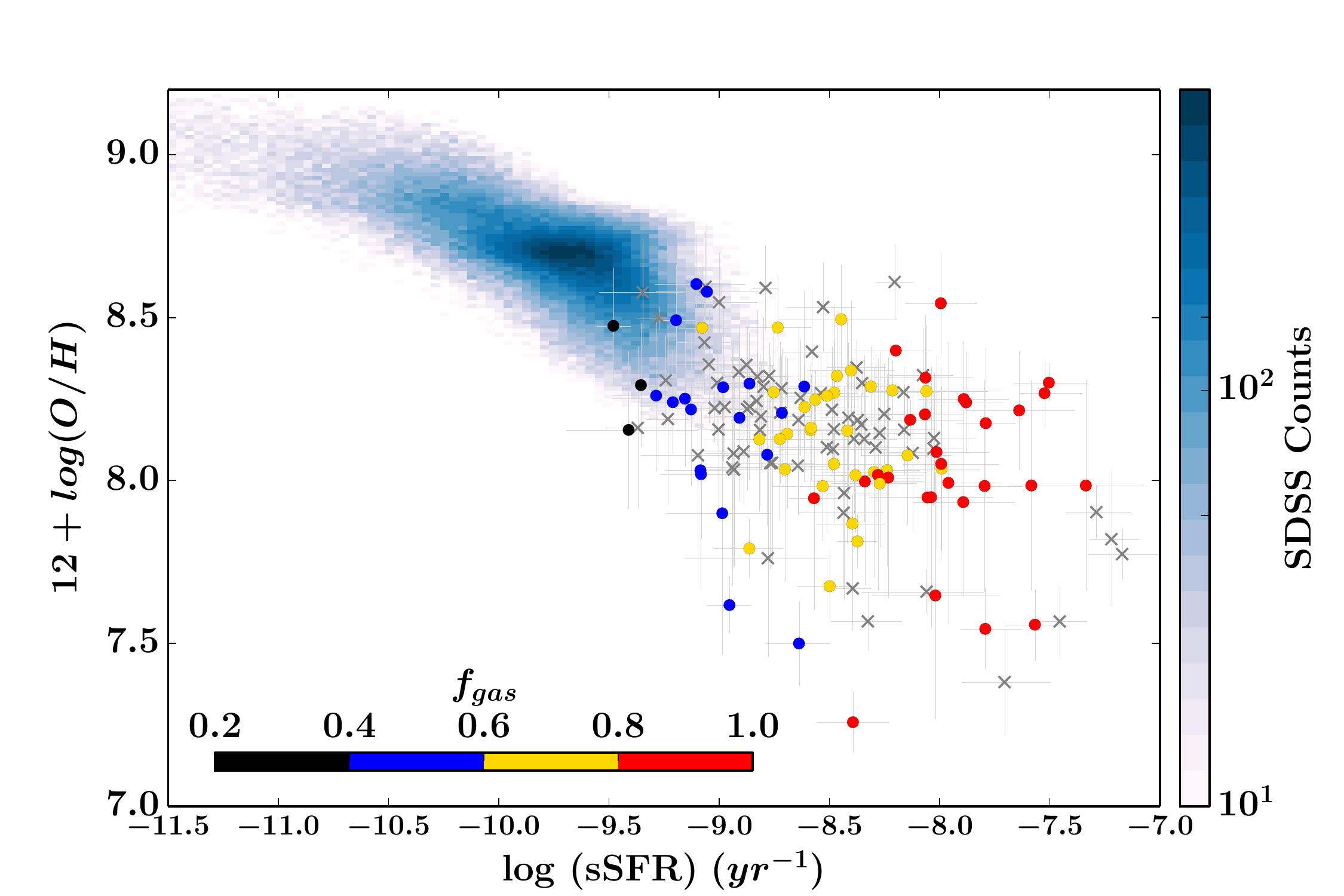}}
  \caption{\small Diagram of metallicity vs sSFR for VUDS SFDGs. Different colors represent galaxies with different gas fractions, divided in four bins in $f_{gas}$: 0.2-0.4, 0.4-0.6, 0.6-0.8 and 0.8-1. Higher SSFRs indicate higher gas fractions too. Our sample extends the trend of local star-forming galaxies (the blue histogram) towards lower metallicities, showing that the SSFR is anti-correlated with the oxygen abundance, though the dispersion increases at higher sSFR. The overall scatter of the relation is $\sim 0.5$\ dex in sSFR. 
  }\label{Z_SSFR}
\end{figure}

We investigate now the relation between the metallicity and the specific SFR proposed by \citep{laralopez13}, which appear to be modulated by the gas fraction. 

Another scaling relation has been found for local star forming galaxies relating the metallicity and the sSFR of the galaxies, with a dependence also on the gas fraction \citep{laralopez13}. For high mass galaxies, this relation has been interpreted as an indication of evolution: galaxies evolve toward lower sSFRs and lower gas fractions as they form stars, increasing at the same time their metallicity because of the gradual chemical enrichment of their ISM by supernova explosions and stellar winds. Following this idea, in Fig. \ref{Z_SSFR} we compare metallicity and sSFR for our sample of SFDGs, dividing the sample in four bins of $f_{gas}$. We see that the two quantities are still correlated in the low metallicity regime, with $12+log(O/H)$ decreasing toward higher sSFR, even though the scatter is large. It is also evident that $f_{gas}$ increases on average with sSFR (Fig. \ref{Z_SSFR}). We remind the reader that, given our definition of the gas fraction in Eq. \ref{fgaseq}, $f_{gas}$ and the sSFR are not independent quantities (i.e. $f_{gas}$ can be written as $1/(1+k \times$ SFR/M$)$, with k a constant factor), so this result is in part expected. 

\begin{figure}[t!]
  \resizebox{\hsize}{!}{\includegraphics[angle=0,width=9.4cm]{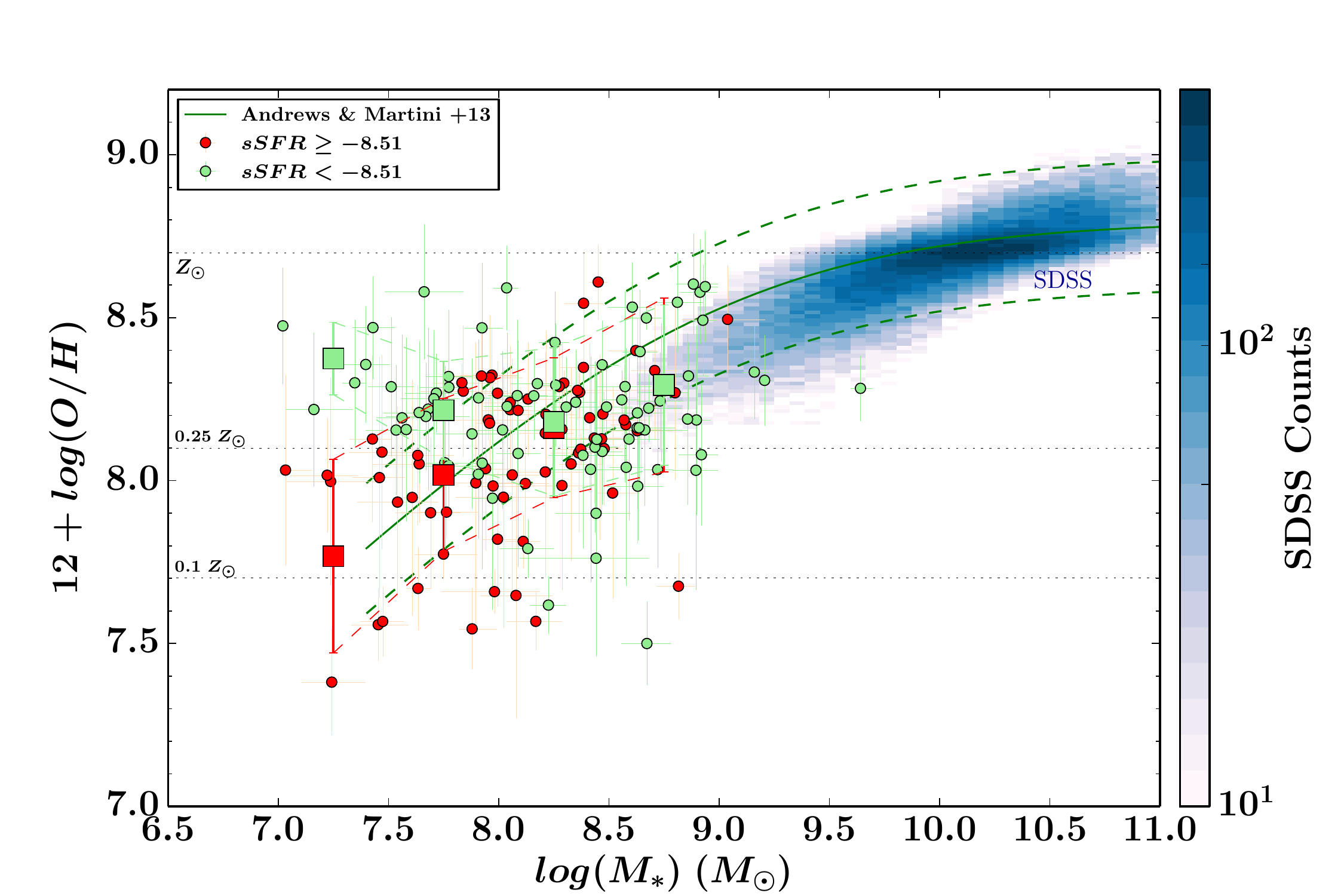}} 
  \caption{\small Mass-metallicity diagram for our galaxies, coded by the sSFR. We see that more metal-poor galaxies have on average higher sSFRs compared to those with higher oxygen abundances, and they follow different trends. The green dashed lines are $\pm 1\sigma\ $ deviation of AM13 MZR as in Fig. \ref{MZR1}.
  }\label{FMR1}
   \end{figure}

The scaling relation betwen metallicity and sSFR suggests that part of the scatter of our galaxies in the MZR can be explained by assuming at fixed mass a dependence of $12+log(O/H)$ on the sSFR. In Fig. \ref{FMR1} we divide our sample in two subsets using the median distribution value $sSFR_{med}=10^{-8.51} yr^{-1}$. We see that the dependence of our MZR on the sSFR gives two different results at $M_\ast$ higher and lower than $10^8 M_\odot$. The MZRs of the two subsets are well separated in the lower mass part, with a mean metallicity offset of $0.5$ dex, which is similar to the result found at $M_\ast > 10^{8.5} M_\odot$ by \citet{ellison08} in the local Universe. On the contrary, they become very close at higher masses, suggesting that the MZR is independent on the sSFR, which is more in agreement with the results of \citet{hughes13} and \citet{sanchez13}.

A similar result is seen for the gas fraction, as expected from the dependence between sSFR and $f_{gas}$ in Fig.\ref{Z_SSFR}. Using the median population value ($f_{gas,med}=0.74$) we have found that at $M_\ast < 10^8$ galaxies with higher $f_{gas}$ are more metal-poor, but this distinction vanishes at higher stellar masses. Even though the dependence of the MZR on the gas content is substantially unknown at low masses, a dependence of the MZR on the observed $f_{gas}$ (similar to our finding at $M_\ast < 10^8 M_\odot$) has been observed by \citet{bothwell13} for SDSS star-forming galaxies in the high mass regime ($M_\ast > 10^9 M_\odot$). 

We remind the reader that, as explained before in section \ref{SFRmass1}, we may miss a significant fraction of galaxies with very low levels of SFRs at the given stellar mass range. Our results indicate as well that at the lowest stellar masses ($M_\ast < 10^8 M_\odot$), where our dispersion is larger, galaxies with lower star-formation activity and gas fraction tend to have higher metallicity. Galaxies with similar properties and even farther (i.e. less star-forming and more metal-rich) have also been identified in the local universe \citep{peeples08} and interpreted as a connection with more evolved classes of dwarfs \citep{zahid12}. Overall, the different trends observed at lower masses may be due to the increasing role of an underlying population of metal-rich low-mass galaxies, which broaden the scatter of the relation toward lower stellar masses, as found in some previous works \citep{zahid12}.

\subsection{Comparison with the FMR}

\begin{figure}[t!]
   \resizebox{\hsize}{!}{\includegraphics[angle=0,width=9.1cm,trim={0.cm 0.0cm 0.6cm 1.15cm},clip]{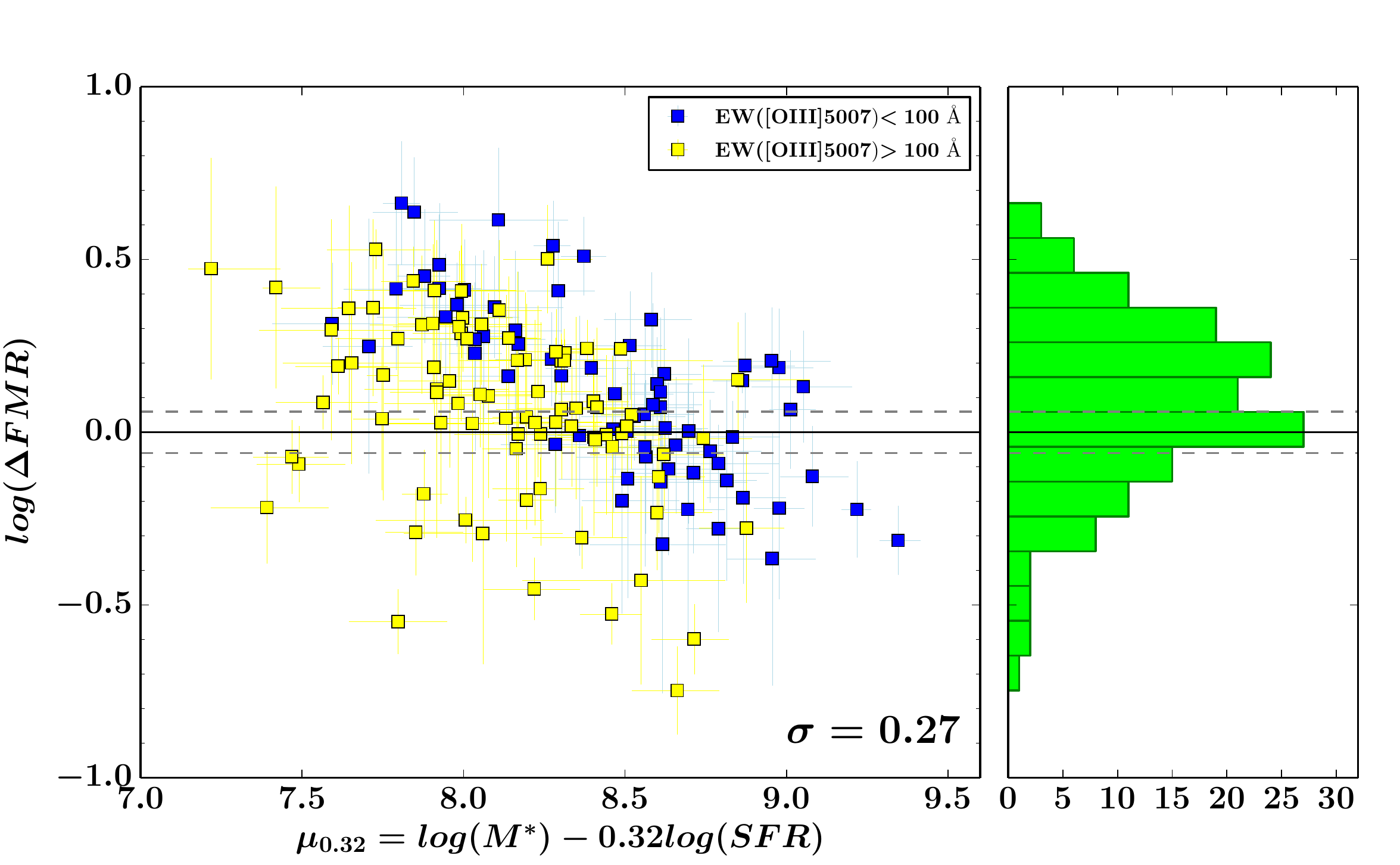}}   
   \resizebox{\hsize}{!}{\includegraphics[angle=0,width=9.1cm,trim={0.cm 0.0cm 0.6cm 1.15cm},clip]{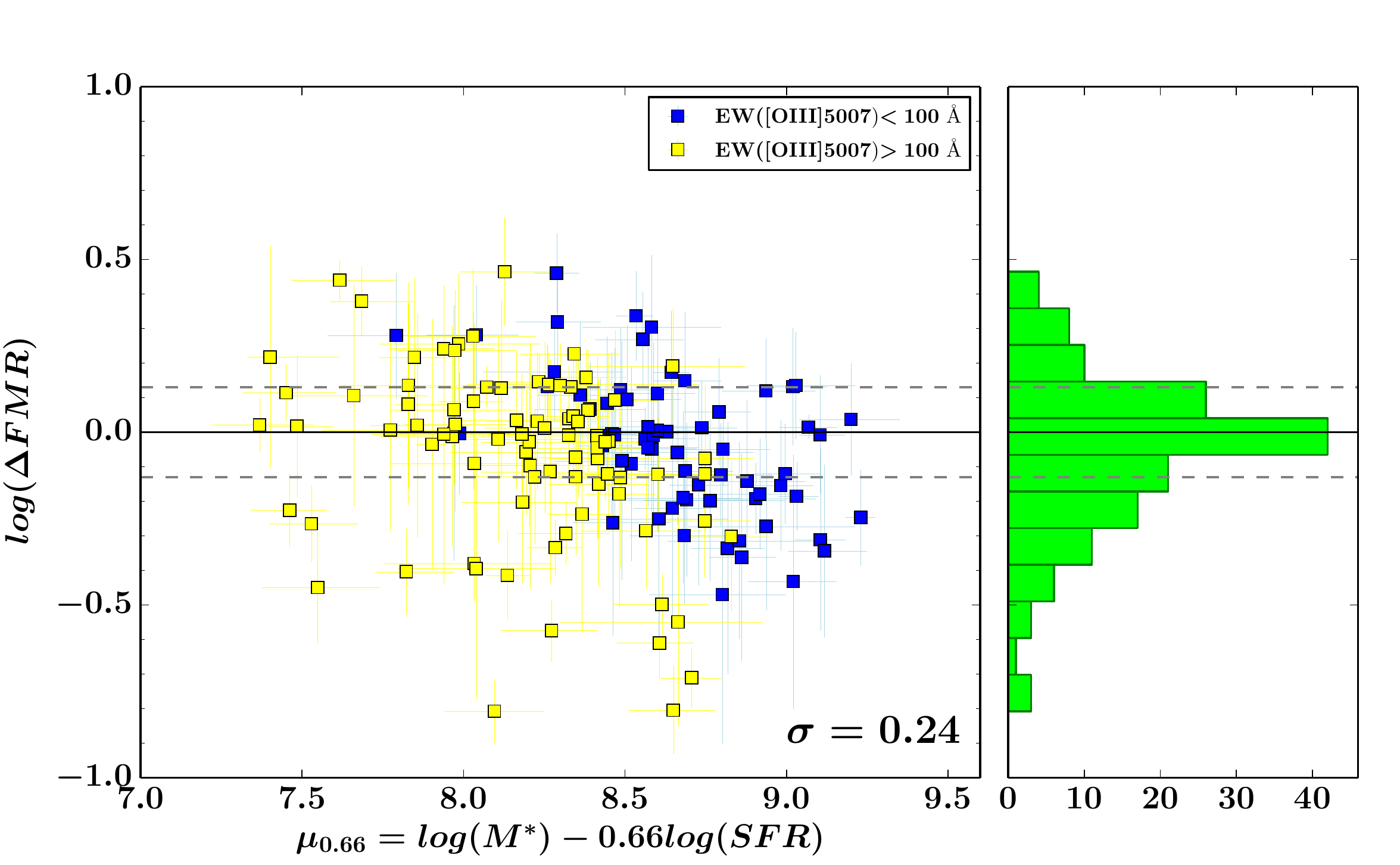}}
   \caption{\small Diagrams showing the differences between the HCm metallicities and the FMR of \citet{mannucci11} (upper panel) and \citet{andrewsmartini13} (bottom panel). The dashed lines represent $1\sigma$ deviations for these relations. The EELGs in our sample (EW([\oiii]$5007$)$>100$ \AA) show slightly larger scatter in the FMR than the whole population, in agreement with \citet{amorin14}. 
  }\label{FMRcompare}
\end{figure}

In order to study the dependence of the MZR on the SFR, previous works by \citet{mannucci10} and \citet{laralopez10} have proposed the Fundamental Metallicity Relation (FMR), which is a 3-dimensional relation in the space parameter defined by $M_\ast$, $12+log(O/H)$ and SFR. It has been found that a particular 2-D projection of the FMR (i.e. plotting the metallicity against the new parameter $\mu_\alpha = log(M^\ast) - \alpha log(SFR)$, with $\alpha \neq 0$) minimizes for a particular value of $\alpha$ the scatter of the points compared to the MZR (which corresponds to $\alpha =0$). An extension of the FMR to low masses ($\lesssim 10^9 M_\odot$) has been presented by \citet{mannucci11}.

Given that our sample is not complete in mass (due to the VUDS limiting magnitude, see section \ref{section5}) and SFR (as we have shown before in section \ref{SFRmass1}, due to the selection criteria in section \ref{selection}), our goal is not to derive a new relation representative of low-mass galaxies ($M_\ast < 10^9 M_\odot$). However, we compare our results with the FMRs found in previous studies using different values of the parameter $\alpha$. In the first panel of Fig. \ref{FMRcompare} we show the comparison with the FMR of \citet{andrewsmartini13} (AM13), who find $\alpha=0.66$ using the direct method on $M_\ast$-SFR stacks for local star-forming galaxies and down to $\mu_{0.66} \simeq 7.5$. In the second panel of Fig. \ref{FMRcompare} we compare with the extrapolation to low-masses of the FMR by \citet{mannucci11} (M11), calibrated in the low-Z regime ($\lesssim 8.4$) using objects with metallicities obtained through the direct method \citep{maiolino08}. The SFRs in both diagrams have been homogenized to the \citet{chabrier03} IMF. 

We find that, even though the position of our SFDGs are generally consistent with both the FMRs (as seen from the peak of the metallicity difference $\Delta$(FMR) distribution), the scatter is larger than reported in the two previous studies. The $1\sigma$ standard devitations of our galaxies from AM13 and M11 FMRs are respectively of $0.24$ and $0.27$ dex, and are higher than the median error on metallicity measurements from HCm. An increased dispersion with respect to the FMR has been shown for highly star-forming galaxies by \citet{amorin14}, and could be due to differences in the methods for metallicity derivation or interpreted as an effect of the large spread of star-formation histories and current SFR in the low mass range \citep[e.g.][]{zhao10}.

\subsection{MZR of galaxies with O$4363$ detection } 

\begin{figure}[t!]
  \resizebox{\hsize}{!}{\includegraphics[angle=0,width=8.1cm,trim={0.cm 0.0cm 3.2cm 1.15cm},clip]{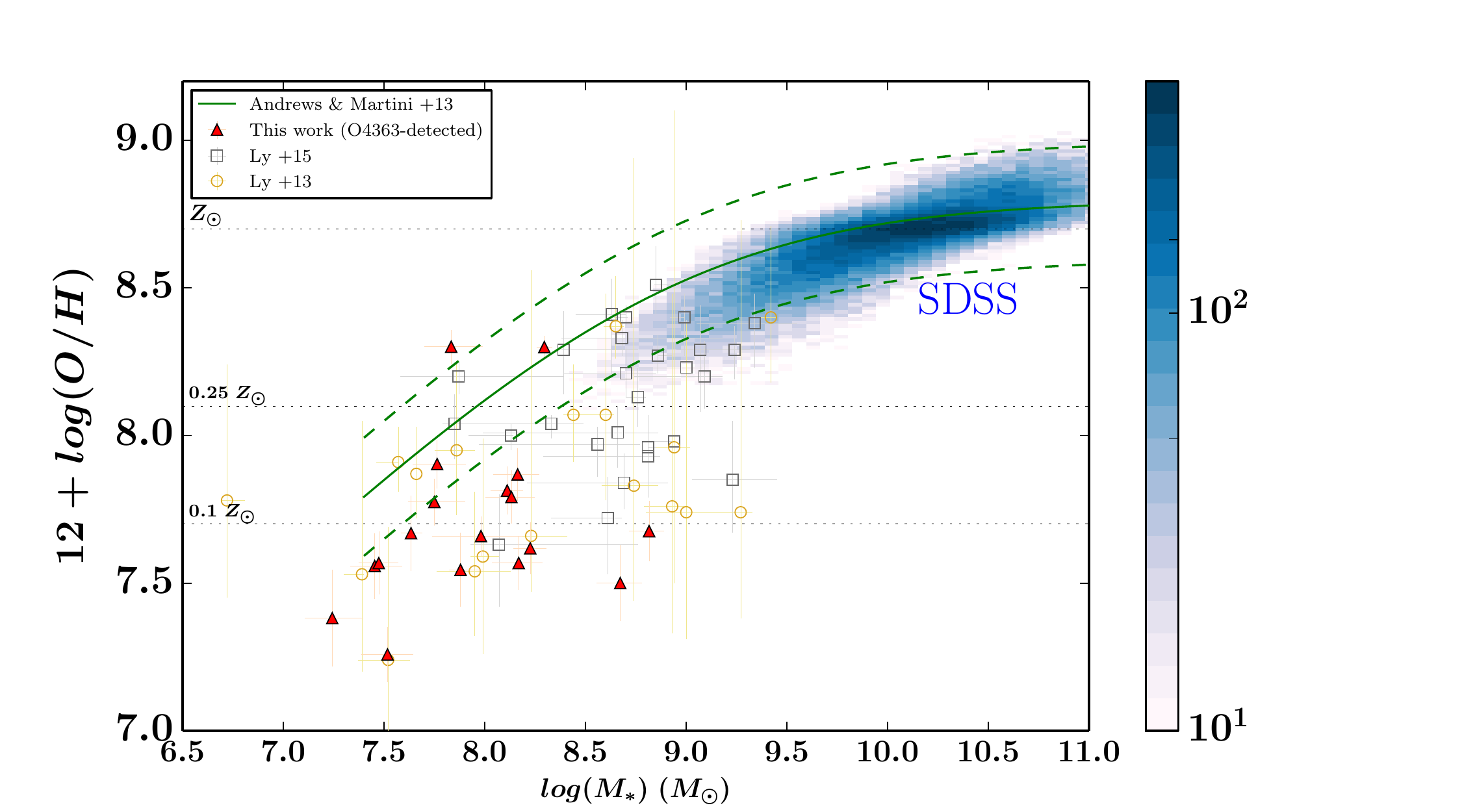}}
  \caption{\small Mass-metallicity diagram for the galaxies in our sample with [\oiii]$\lambda 4363\ $\AA\ detection (red triangles), compared to the [\oiii]$\lambda 4363\ $ galaxies with $T_e$ oxygen abundances compiled by \citet{ly14,ly15} (open circles and open squares respectively). The AM13 MZR and $\pm 1\sigma\ $ deviation are shown with green lines.
  }\label{MZRdirectmethod}
   \end{figure}

In Fig. \ref{MZRdirectmethod} we display the MZR of 22 galaxies in VUDS sample for which the auroral line [\oiii]$\lambda 4363\ $ has been detected. This subset is consistent in the low mass range ($10^7 < M_\ast < 10^{8.5}$) with the sample of [\oiii]$\lambda 4363$\ selected galaxies compiled by \citet{ly14,ly15}. From this plot we see that VUDS SFDGs with auroral line detection have lower metallicities at fixed mass compared to the median sample. Indeed, \citet{telford16} show that, at constant $M_\ast$, metal poor galaxies have higher S/N of the oxygen lines in their optical spectra, in particular the auroral line, improving its detection. Therefore, our selection based on [\oiii]$4363$ tend to include those objects with the lowest metallicities in VUDS. They should have higher [\oiii]$\lambda$4363/[\oiii]$\lambda$5007 ratios, and the auroral line may be detected easier. 

Finally, we studied the average sSFRs of the galaxies with [\oiii]$\lambda$4363 detection and, among them, the mean value of XMPs. We find that they are higher by $0.5$ and $0.4$ dex (respectively) than the whole population of SFDGs ($sSFR_{mean}=10^{-8.5}$).

\subsection{Comparison with the MZR of similar SFDGs samples}

\begin{figure*}[t!]
  \centering
  \includegraphics[angle=0,width=18.5cm,trim={0.5cm 0.5cm 0.5cm 1.5cm},clip]{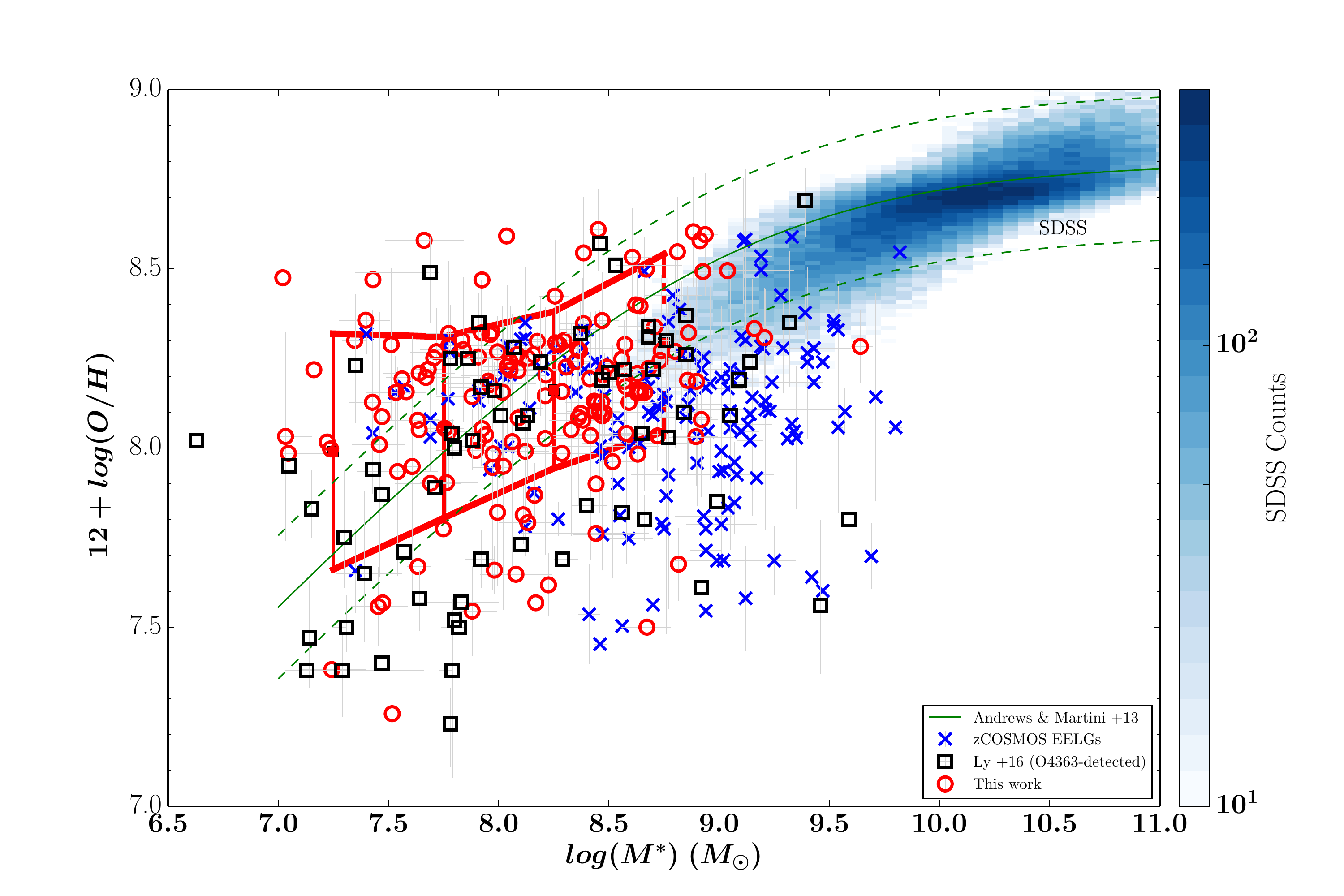} 
  \caption{\small Comparison between the MZR derived for VUDS SFDGs (red circles) with the mass-metallicity diagram for the galaxy samples compiled by \citet{ly16} (black squares) and \citet{amorin15} (blue crosses). The MZR by \citet{andrewsmartini13} with $\pm 1$-$\sigma$ scatter are shown with green lines and the SDSS galaxies are presented in a 2-D histogram with number density colorbar.
  }\label{MZR_comparison}
   \end{figure*}

Recently, \citet{ly16} (LY16) studied the mass-metallicity relation at low-mass and intermediate redshift with a sample of 164 emission-line galaxies (uniformly distributed in $0.1<z<1$) and abundances derived using the direct method for a subset of them (66) with [\oiii]$\lambda 4363$ \AA\ $3$-$\sigma$ detection. Likewise, zCOSMOS EELGs compiled by A15 represent a very complementary sample to compare with, since their spectra is taken with the same instrument, they span the same $z$ range and have similar properties to our galaxies, in particular to the EELG fraction. Here we recalculate the metallicity of these two datasets with our code HCm using the emission line fluxes retrieved from their catalogue. Overall, we find a consistency between HCm metallicities and the A15 values. 

In Fig. \ref{MZR_comparison} we present the MZR including our SFDGs and the other two comparison samples. We see that the distribution of A15 galaxies is mainly consistent with VUDS SFDGs, though they are more massive, as we have shown before. At lower masses ($< 10^{8.4} M_\odot$), they are in agreement with the dispersion obtained for the VUDS galaxies, while at higher masses a significant fraction of them is below our relation, because the sample is biased towards more extreme objects. We notice in the same figure that the galaxies studied by LY16 are more consistent with the slope of the MZR found by AM13 than ours. The differences with respect to our MZR may be due to the bias towards $O4363$-detected galaxies in LY16 sample, and this suggestion is supported also by Fig. \ref{MZRdirectmethod}, as we have discussed before.

\section{Discussion}\label{discussion}

\subsection{Implications for galaxy assembly and evolution}

The information on metallicity and gas fraction $f_{gas}$ can be used to constrain chemical evolution models when applied to low-mass galaxies. In this paper we use simple models of chemical evolution in order to investigate from a broad point of view the mechanisms that influence the ISM abundances of our SFDGs. We consider the closed-box model, in which galaxies do not interact with their environment, and two open-box models in which gas flows are allowed to flow in and out from a galaxy. In these models, the gas is well mixed at any time of the galaxy evolution (instantaneous mixing approximation) and newly born stars with $M_\ast > 1 M_\odot$ die and produce a stellar wind almost immediately (instantaneous recycling approximation). Gas outflows and inflows are assumed to occur a a constant fraction of the SFR, respectively $\eta$ and $\Lambda$ (so that $\dot{M}_{loss} = \eta \times$SFR and $\dot{M}_{acc} = \Lambda \times$SFR). The outflowing gas in the models has the same metallicity as the remaining gas reservoir, and we always consider pristine (metal-free) gas accretion.

We adopt the general solution for these simple models derived by \citet{kudritzki15} (hereafter K15) and relating the oxygen mass fraction $O_m$ (defined as the mass fraction of oxygen in the ISM: $O_m = M_O/M_{gas}$) and the gas fraction $f_{gas}$ :
\begin{equation}\label{general_eq}
O_m(t)=\frac{y_O}{\Lambda} \left(1-\left[\frac{(1+\alpha)}{f_{gas}}-\alpha \right]^\omega \right)
\end{equation}
where: 
$\omega$=$\Lambda$/((1-R)(1+$\alpha$)), $\alpha$ is defined as $\alpha$=($\eta$-$\Lambda$)/(1-R), t is the time and R is the fraction of stellar mass returned to the ISM through stellar winds, which is assumed here to have a constant value R$=0.18$ \citep{ascasibar15,sanchezalmeida15}. The factor $y_O$ is the oxygen yield, i.e. the mass of oxygen ejected by a generation of stars divided by the mass of the same generation that remains in stellar remnants and long-lived stars.

Closed-box models correspond to $\Lambda=0$ and $\eta=0$. In this case, the equation relating $O_m$ and $f_{gas}$ is simpler (K15) :
\begin{equation}
O_m(t)=\frac{y_O}{(1-R)}ln\left(\frac{1}{f_{gas}}\right)
\end{equation}
In open box models where only outflows of gas are allowed, $\Lambda=0$ and $\eta \neq 0$ and Eq. \ref{general_eq} is not defined, so we apply the following expression (K15):
\begin{equation}
O_m(t)=\frac{y_O}{(1-R)} \left(\frac{1}{1+\eta/(1-R)} ln\left[\frac{1}{f_{gas}}\left(1+\frac{\eta}{1-R}\right)-\frac{\eta}{1-R}\right]\right)
\end{equation}
In order to obtain equations that are consistent with the metallicity used in this paper (i.e. the oxygen abundance defined in terms of number densities, $12+$log(N(O)/N(H)), we use N(O)/N(H)$=$$O_m \times M_{gas}/(16M_{H})$ (from K15), assuming that $M_{gas} \sim 1.35 \times M_H$, wich takes into account the contribution of helium \citep{ascasibar15}. 

\begin{figure}[t!]
  \resizebox{\hsize}{!}{\includegraphics[angle=0,width=9.4cm,trim={0.5cm 0.cm 1.5cm 1.cm},clip]{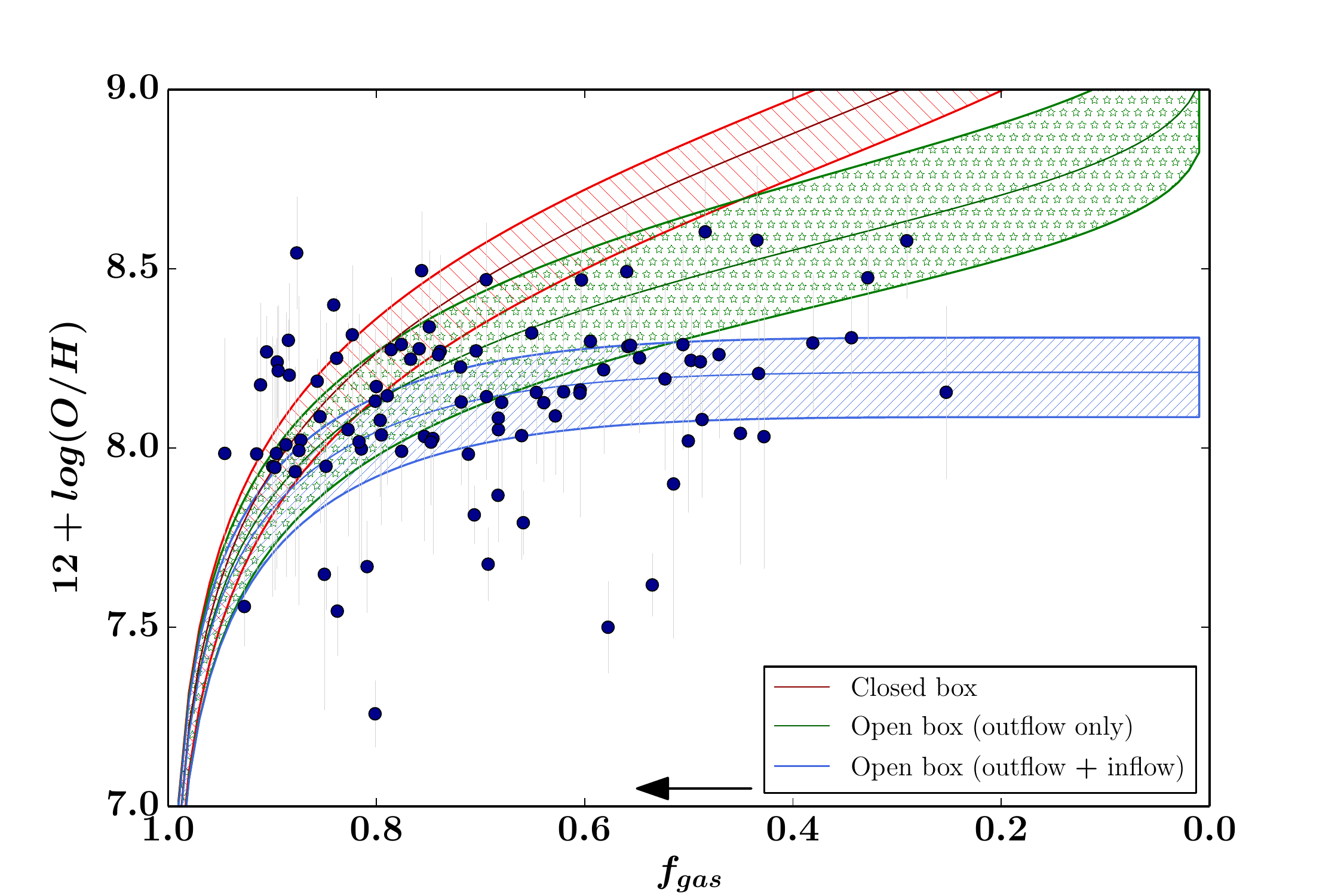}}
  \caption{\small Diagram relating the metallicity 12+log(O/H) and the gas fraction $f_{gas}$ for our subsample of SFDGs. The curves represent: \textit{(red)} closed-box model with oxygen yield $y_O=0.008$ ($0.006$-$0.010$); \textit{(green)} open box model with outflows for $y_O=0.008$ ($0.006$-$0.010$) and outflow rate $\eta=3.2$ \citep{sanchezalmeida15} ($2<\eta<4$); \textit{(blue)} open box model with outflows and inflows, choosing $y_O=0.008$ ($0.006$-$0.010$), $\eta=3.2$ and inflow rate $\Lambda=\eta+0.95$. As discussed in the text, we adnowledge an average error for $f_{gas}$ of $\sim 0.2\,$ dex.
  }\label{chemicalevolution}
\end{figure}

In Fig. \ref{chemicalevolution} we show three different regions representing a closed-box model with typical oxygen yields for SF galaxies ranging $0.006$$<$$y_O$$<$$0.010$ \citep{zahid12} (red area), and a model considering outflows only, with varying $y_O$ and for different outflow rates in the range $2$$<$$\eta$$<$$4$ \citep{wuyts12,sanchezalmeida15}. In the last case, we also include inflows of pristine gas, choosing a fixed inflow rate of $\Lambda=\eta+0.95$ \citep{wuyts12} and changing $y_O$ in the same intervals of other models. At high $f_{gas}$ the different regions overlap and all the three models are equally possible, while at lower $f_{gas}$ they separate allowing to discriminate among the three regimes. 

The distribution of our galaxies shows that they populate a large region in 12+log(O/H) -$f_{gas}$ plane, both at high and low gas fractions. As we have mentioned before, SFDGs at higher $f_{gas}$ can be reproduced equally well by any of the three models. They may be still in an early stage of their evolution where few amounts of gas have been converted into stars, or simply the star formation proceeds slowly in these systems consuming the gas reservoir in long timescales. On the opposite side at low $f_{gas}$ ($\la 0.8 $), our galaxies show metallicities which are too low to be reproduced by a wide range of closed-box models with physically motivated oxygen yields taken from literature. All galaxies at lower $f_{gas}$ are consistent with open box models allowing the galaxies to exchange gas with their environment, with a slightly preference for 'outflow+inflow' models. Unless we choose very high outflow rates $\eta$, which are allowed in principle according to recent models ($2<\eta<100$ is reported in \citet{ascasibar15}), or we decrease appreciably the value of the oxygen yield $y_O$, the SFDGs with the lowest metallicity can be explained only by assuming the presence of nearly-pristine inflows of gas occurring at a constant rate.    

We also notice that some galaxies lie on the left of the closed-box model, in a phase-space that is not allowed in principle, because interactions with the environment through metal-rich outflows or metal-poor inflows always decrease the metallicity of the ISM compared to an isolated galaxy (which gives the highest possible value). The latter result was also found by other similar works on $z \sim 1$ and $M_\ast \sim 10^9 M_\odot$ SF galaxies \citep[e.g.]{wuyts12} and, assuming that the metallicity is correct within the uncertainties, it can be due in our case to the large errors on the gas fractions. Considering a typical uncertainty of $\sim 0.2$ dex or even higher (difficult to quantify and resulting from a combination of the assumptions made in Eq. \ref{sigma_SFR_eq}, the KS law and the measurement of $r_e$), the points may shift to higher $f_{gas}$ and so they become compatible with the models. On the other side, some SFDGs show lower abundances at fixed $f_{gas}$ compared to the bulk of our SFDGs. We checked that those galaxies with $12$+log(O/H)$<7.9$ and below the 'outflow+inflow' model are all EELGs. 

Because of the crucial role of the gas fraction, direct measurements of the gas mass $M_{gas}$ for our galaxies (as already discussed in Section \ref{gasfractions}) are certainly needed to improve constraining analytical models from the observations, and in general to reduce the uncertainty on both the slope and the normalization on the KS law for SFDGs below $z \sim 1$. We also remind the reader that we are considering simple scenarios of chemical evolution and more sophisticated models, e.g. considering the additional effect of metal-enriched gas inflows \citep[e.g.][]{spitoni10} or relaxing the instantaneous mixing and recycling approximation, will certainly help to understand the role of our SFDGs in cosmic evolution.

This analysis suggests that open-box models are better in agreement with our observations compared to closed-box models. However, we remark the large uncertainties of $f_{gas}$ and the need of direct measurements of HI content for VUDS SFDGs.

\subsection{Implications for reionization studies}

\begin{figure}[t!]
  \resizebox{\hsize}{!}{\includegraphics[angle=0,width=9.8cm,trim={1.2cm 0.cm 2cm 1.cm},clip]{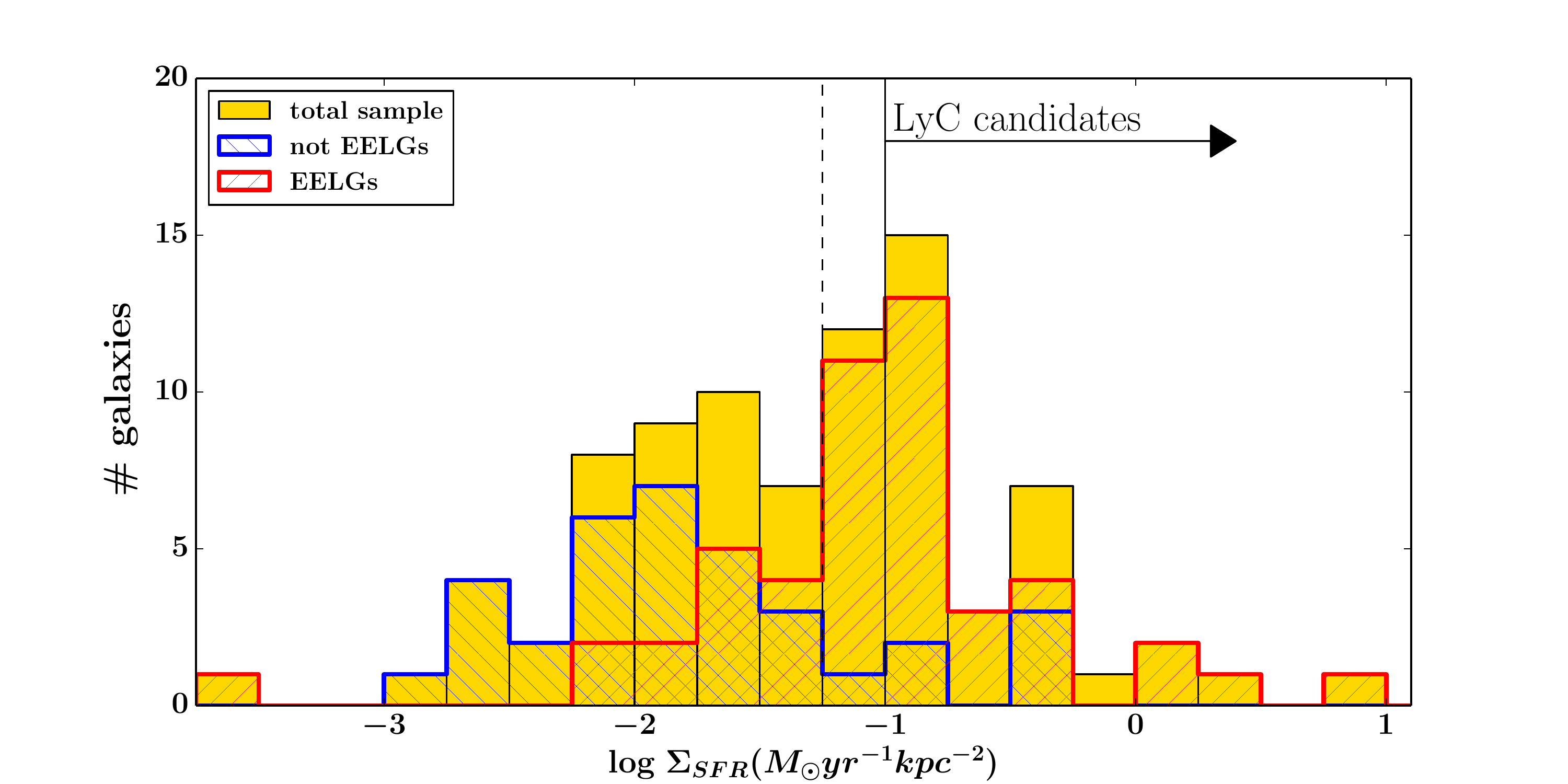}}
  \caption{\small Histogram of log($\Sigma_{SFR}$) for VUDS SFDGs, showing the limit for outflows \citep{sharma16} with a blue dotted vertical line. The majority of the galaxies are included in the range $0.001<\Sigma_{SFR}<1$, with a few outliers on both sides. In particular, we find two galaxies with very high $\Sigma_{SFR}$, $1.4$ and $2.9$ $M_\odot yr^{-1} kpc^{-2}$ respectively.}\label{models}
\end{figure}\label{lyc}

Large efforts are currently being invested to identify the first luminous sources which reionized the neutral intergalactic medium at redshifts higher than $6$ \citep[e.g.][]{pentericci14,robertson15,giallongo15,grazian16}. One of the main contributors to cosmic reionization are faint SFDGs with a significant fraction of escaping Lyman continuum photons (i.e. emission at wavelengths $\lambda < 912$ \AA). Although this is still an open question, SFDGs could be the best representatives of the LyC leaking galaxies, since the observed faint-end UV LF appears to be steeper at high-z \citep[e.g.][]{bouwens15}, and they can contribute more than $20 \%$ to the total ionizing flux \citep{dressler15,vanzella16}. However, direct observations of the LyC photons are extremely challenging and only a few direct detections have been found at low \citep[e.g.][]{izotov16} and intermediate ($z \sim 3$) redshift \citep{debarros16,vanzella16}.

In recent years, local candidates to LyC leakers have been selected among EELGs \citep[e.g. Green Peas;][]{jaskotoey14}, which later turned out to yield successful detections using UV spectroscopy \citep{izotov16}. They present similar properties to those expected at very high redshift, namely strong emission lines (with high EW) in the optical spectra, low masses, small sizes, low metallicities and high ionizations, as well as strongly ionized outflows, which appear ubiquitous amongst galaxies with high SFR per unit area \citep[$\Sigma_{SFR}$ $\geq 0.1 M_\odot yr^{-1} kpc^{-2}$;][]{heckman11,sharma16}. Many of these properties have been found also in a fraction of our VUDS SFDGs. 

In our sample, $30$ SFDGs show $\Sigma_{SFR}$ (calculated as equation \ref{KSeq1} and assuming \citet{chabrier03} IMF) higher than the lower limit to produce strong outflows mentioned before, as shown in Fig. \ref{lyc}. The majority (25) of this subset comprises EELGs, which could be the best candidates for LyC leakers in our VUDS sample. Furthermore, 7 EELGs with enhanced $\Sigma_{SFR}$ also present simultaneously high EW(\hb) ($> 100$ \AA) and high [\oiii]$5007$/[\oii]$3727$ ratio ($>4$), which suggest a very young starbursts episode.

Additional constraints on the escape fraction $f_{esc}$ of ionizing radiation can come from the stacking of GALEX data available for all our COSMOS galaxies. A spectroscopic follow-up in the UV rest-frame could also be a good option for future works, focused on the Lyman continuum and Ly-$\alpha$ emission line properties of low-z analogs of those primeval faint SFDG that are believed to be responsible of reionization during the first Myr after the Big Bang.  

\section{Summary and conclusions}\label{summaryconclusions}

Our work investigates a sample of star-forming dwarf galaxies (SFDGs) down to $M_\ast \sim 10^7 M_\odot$ at intermediate redshift, selected from the VUDS spectroscopic survey by the presence of optical emission lines. From their spectra and ancillary photometric data, we have derived important physical properties for this sample (e.g. mass, star formation rates, metallicity, ionization and sizes), adding new constrains to the low-mass end of the mass-metallicity relation. We have studied simple scenarios of chemical evolution in a crucial phase of the cosmic stellar mass assembly. Our main findings are summarized as follows: 
\begin{enumerate}
\item We have selected in the three fields of VUDS survey (COSMOS, ECDFS and VVDS-02h) a sample of $164$ star-forming dward galaxies at intermediate redshift ($0.13 < z < 0.88$) with detection of the following emission lines: [\oiii]$\lambda 3727\,$\AA, [\oiii]$\lambda 5007\,$\AA, \hb\ and \ha. 19 galaxies in this sample have additional detection of the auroral line [\oiii]$\lambda 4363\,$\AA, and we find that $56\%$ are EELGs, according to the definition EW([\oiii]$\lambda$5007 $>100$ \AA. 
\item Using deep multi-wavelength data available in the three VUDS fields we have derived stellar masses for the selected galaxies, confirming their low values ranging $10^7 \leq M_\ast/M_\odot \leq 10^9$. Star formation rates from \ha\ luminosity after corrections for dust attenuation and extinction lie in the range $10^{-3} \lesssim SFR \lesssim 10^1 M_\odot/yr$ (Chabrier IMF). Combining the two measurements, we find that our sample is representative of higher SFRs on average ($\sim +0.5$ dex) compared to the mean star-forming population (i.e. the SF 'Main Sequence') at similar redshift.
\item We have used a novel methodology (HCm) to derive the metallicity and ionization parameter which is based on the comparison of optical emission line ratios and detailed photoionization models. Its robustness has been successfully tested against the direct method and other consistent strong-line methods. Applying HCm to our SFDGs, we find that they have low metallicities ranging $7.26 <$ 12+log(O/H) $< 8.7$ ($0.04$-$ 1\ Z_\odot$) and ionization parameters in the range $-3.2 <$ log(U) $< -1.5 $. We find $12$ extremely metal-poor ($Z < 0.1\ Z_\odot$) galaxies in our sample.
\item We have performed a morphological classification of VUDS SFDGs in COSMOS and ECDFS (101), for which high resolution ($\sim 0.09\arcsec$) HST-ACS images are available in these two fields. EELGs have on average more disturbed morphologies (cometary,clumpy shapes and interacting-merging systems) compared to the remaining population.
\item Using GALFIT for fitting the HST images available for a subset of VUDS SFDGs, we find that they are overall very compact, with a median effective radius $r_e$=$1.23\ kpc$. They are also gas-rich systems, with median gas fraction (defined as $f_{gas}= M_{gas}/(M_\ast+M_{gas})$) of $0.74$, derived inverting an assumedly non variable Kennicutt-Schmidt law with exponent $n$=$1.4$. Despite the large uncertainties of the gas fraction measurements resulting both from the KS law and the size errobars, we highlighted the importance of direct estimations of the HI content in galaxies for more precise estimations of $f_{gas}$, which can be addressed in a future work.
\item We have added new constrains to the low-mass end ($M_\ast < 10^9 M_\odot$) of the mass-metallicity relation (MZR). The MZR of VUDS SFDGs is generally consistent with the local relation of \citet{andrewsmartini13}, but it shows a flatter slope which is more similar to the result of \citet{lee06} for nearby SFDGs. The average dispersion of our sample is $\sim 0.26$ dex, and it shows an increasing trend toward lower masses, also found in similar studies \citep{zahid12b}. We find that below $10^8 M_\odot$ more metal-poor galaxies show higher sSFRs, while at higher masses this differentiation vanishes. This dependence at lower masses indicates that sSFR is partly responsible for the moderate slope and for the increasing scatter of our mass-metallicity relation toward $M_\ast \sim 10^7 M_\odot$, in agreement with recent theoretical works \citep[e.g.][]{lilly13}.
\item We have compared our results with the predictions of simple chemical evolution models. Our data suggest that a closed-box model is not sufficient to reproduce the distribution of our galaxies in the metallicity-$f_{gas}$ plane, while they are more consistent with open-box models (including both outflows and inflows). Quantitative measurements of the total gas content, as well as more sophisticated models, would help to better contrain and fully investigate the chemical evolution of our SFDGs. 
\item In our sample we have found $30$ promising candidates to LyC leakers having star formation surface densities $\Sigma_{SFR} > 0.1 M_\odot yr^{-1} kpc^{-2}$. From these, 7 candidates are compact EELGs with EW(\hb) $>100$ \AA\ and EW([\oiii]$5007$) $>300$ \AA, characterized by high SSFR, low metallicity, and high ionization, as probed by their unusually high ($>4$) [OIII]$5007$/[OII]$3727$ ratios.
\end{enumerate}

\begin{acknowledgements}
We thank the anonymous referee for very detailed and constructive comments that have improved this manuscript. Precious and continuous support for the VUDS survey was provided by the ESO staff, in particular by the Paranal staff conducting the observations and Marina Rejkuba and the ESO user support group in Garching. We also thank J. S\'anchez Almeida who has provided insightful comments to the paper, and E. Daddi for helpful suggestions. This work is supported by funding from the European Research Council Advanced Grant ERC-2010-AdG-268107-EARLY and by INAF Grants PRIN 2010, PRIN 2012 and PICS 2013. RA and AF acknowledge the FP7 SPACE project “ASTRODEEP” (Ref.No: 312725), supported by the European Commission. RA acknowledges the support from the ERC Advanced Grant 695671 “QUENCH”. AC, OC, MT and VS acknowledge the grant MIUR PRIN 2010--2011.  EPM acknowledges support from the Spanish MINECO through grant AYA2013-47742-C4-1-P.DM gratefully acknowledges LAM hospitality during the initial phases of the project. This work is based on data products made available at the CESAM data center, Laboratoire d'Astrophysique de Marseille. This work partly uses observations obtained with MegaPrime/MegaCam, a joint project of CFHT and CEA/DAPNIA, at the Canada-France-Hawaii Telescope (CFHT) which is operated by the National Research Council (NRC) of Canada, the Institut National des Sciences de l'Univers of the Centre National de la Recherche Scientifique (CNRS) of France, and the University of Hawaii. This work is based in part on data products produced at TERAPIX and the Canadian Astronomy Data Centre as part of the Canada-France-Hawaii Telescope Legacy Survey, a collaborative project of NRC and CNRS.
\end{acknowledgements} 


\begin{thebibliography}{}


\bibitem[Abazajian et al.(2009)]{abazajian09} Abazajian, K.~N., Adelman-McCarthy, J.~K., Ag{\"u}eros, M.~A., et al.\ 2009, \apjs, 182, 543-558 

\bibitem[Amor{\'{\i}}n et al.(2007)]{amorin07} Amor{\'{\i}}n, R.~O., Aguerri, J.~A.~L., Cair{\'o}s, L.~M., Caon, N., \& Mu{\~n}oz-Tu{\~n}{\'o}n, C.\ 2007, Galaxy Evolution across the Hubble Time, 235, 300 

\bibitem[Amor{\'{\i}}n et al.(2009)]{amorin09} Amor{\'{\i}}n, R., Aguerri, J.~A.~L., Mu{\~n}oz-Tu{\~n}{\'o}n, C., \& Cair{\'o}s, L.~M.\ 2009, \aap, 501, 75 

\bibitem[Amor{\'{\i}}n et al.(2010)]{amorin10} Amor{\'{\i}}n, R.~O., P{\'e}rez-Montero, E., \& V{\'{\i}}lchez, J.~M.\ 2010, \apjl, 715, L128 

\bibitem[Amor{\'{\i}}n et al.(2012)]{amorin12} Amor{\'{\i}}n, R., P{\'e}rez-Montero, E., V{\'{\i}}lchez, J.~M., \& Papaderos, P.\ 2012, \apj, 749, 185 

\bibitem[Amor{\'{\i}}n et al.(2014)]{amorin14} Amor{\'{\i}}n, R., Sommariva, V., Castellano, M., et al.\ 2014, \aap, 568, L8 

\bibitem[Amor{\'{\i}}n et al.(2016)]{amorin16} Amor{\'{\i}}n, R., Mu{\~n}oz-Tu{\~n}{\'o}n, C., Aguerri, J.~A.~L., \& Planesas, P.\ 2016, \aap, 588, A23 


\bibitem[Amor{\'{\i}}n et al.(2015)]{amorin15} Amor{\'{\i}}n, R., P{\'e}rez-Montero, E., Contini, T., et al.\ 2015, \aap, 578, A105 


\bibitem[Andrews \& Martini(2013)]{andrewsmartini13} Andrews, B.~H., \& Martini, P.\ 2013, \apj, 765, 140 


\bibitem[Ascasibar et al.(2015)]{ascasibar15} Ascasibar, Y., Gavil{\'a}n, M., Pinto, N., et al.\ 2015, \mnras, 448, 2126 

\bibitem[Asplund et al.(2009)]{asplund09} Asplund, M., Grevesse, N., Sauval, A.~J., \& Scott, P.\ 2009, \araa, 47, 481

\bibitem[Atek et al.(2011)]{atek11} Atek, H., Siana, B., Scarlata, C., et al.\ 2011, \apj, 743, 121 


\bibitem[Baldwin et al.(1981)]{baldwin81} Baldwin, J.~A., Phillips, M.~M., \& Terlevich, R.\ 1981, \pasp, 93, 5 

\bibitem[Bielby et al.(2012)]{bielby12} Bielby, R., Hudelot, P., McCracken, H.~J., et al.\ 2012, \aap, 545, A23 

\bibitem[Bolzonella et al.(2010)]{bolzonella10} Bolzonella, M., Kova{\v c}, K., Pozzetti, L., et al.\ 2010, \aap, 524, A76 

\bibitem[Bothwell et al.(2013)]{bothwell13} Bothwell, M.~S., Maiolino, R., Kennicutt, R., et al.\ 2013, \mnras, 433, 1425 


\bibitem[Boulade et al.(2003)]{boulade03} Boulade, O., Charlot, X., Abbon, P., et al.\ 2003, \procspie, 4841, 72 

\bibitem[Bouwens et al.(2015)]{bouwens15} Bouwens, R.~J., Illingworth, G.~D., Oesch, P.~A., et al.\ 2015, \apj, 811, 140 


\bibitem[Brinchmann et al.(2004)]{brinchmann04} Brinchmann, J., Charlot, S., White, S.~D.~M., et al.\ 2004, \mnras, 351, 1151 


\bibitem[Bruzual \& Charlot(2003)]{bruzual03} Bruzual, G., \& Charlot, S.\ 2003, \mnras, 344, 1000 


\bibitem[Cair{\'o}s et al.(2007)]{cairos07} Cair{\'o}s, L.~M., Caon, N., Garc{\'{\i}}a-Lorenzo, B., et al.\ 2007, \apj, 669, 251 


\bibitem[Calzetti et al.(2000)]{calzetti00} Calzetti, D., Armus, L., Bohlin, R.~C., et al.\ 2000, \apj, 533, 682 


\bibitem[Cappelluti et al.(2009)]{cappelluti09} Cappelluti, N., Brusa, M., Hasinger, G., et al.\ 2009, \aap, 497, 635 
 
\bibitem[Cappelluti et al.(2016)]{cappelluti16} Cappelluti, N., Comastri, A., Fontana, A., et al.\ 2016, \apj, 823, 95 

\bibitem[Cardamone et al.(2009)]{cardamone09} Cardamone, C., Schawinski, K., Sarzi, M., et al.\ 2009, \mnras, 399, 1191 

\bibitem[Cardamone et al.(2010)]{cardamone10} Cardamone, C.~N., van Dokkum, P.~G., Urry, C.~M., et al.\ 2010, \apjs, 189, 270 

\bibitem[Cardelli(1988)]{cardelli88} Cardelli, J.~A.\ 1988, \apj, 335, 177 


\bibitem[Cardelli et al.(1988)]{cardelli88b} Cardelli, J.~A., Clayton, G.~C., \& Mathis, J.~S.\ 1988, \apjl, 329, L33 

\bibitem[Ceverino et al.(2015)]{ceverino15} Ceverino, D., Dekel, A., Tweed, D., \& Primack, J.\ 2015, \mnras, 447, 3291 


\bibitem[Chabrier(2003)]{chabrier03} Chabrier, G.\ 2003, \pasp, 115, 763 


\bibitem[Charlot \& Longhetti(2001)]{charlotlonghetti01} Charlot, S., \& Longhetti, M.\ 2001, \mnras, 323, 887 


\bibitem[Chiappetti et al.(2013)]{chiappetti13} Chiappetti, L., Clerc, N., Pacaud, F., et al.\ 2013, \mnras, 429, 1652  

\bibitem[Christensen et al.(2012)]{christensen12} Christensen, L., Richard, J., Hjorth, J., et al.\ 2012, \mnras, 427, 1953 

\bibitem[Civano et al.(2012)]{civano12} Civano, F., Elvis, M., Brusa, M., et al.\ 2012, \apjs, 201, 30 

\bibitem[Cuillandre et al.(2012)]{cuillandre12} Cuillandre, J.-C.~J., Withington, K., Hudelot, P., et al.\ 2012, \procspie, 8448, 84480M 

\bibitem[Cullen et al.(2014)]{cullen14} Cullen, F., Cirasuolo, M., McLure, R.~J., Dunlop, J.~S., \& Bowler, R.~A.~A.\ 2014, \mnras, 440, 2300 

\bibitem[Curti et al.(2016)]{curti16} Curti, M., Cresci, G., Mannucci, F., et al.\ 2016, arXiv:1610.06939 

\bibitem[Daddi et al.(2007)]{daddi07} Daddi, E., Dickinson, M., Morrison, G., et al.\ 2007, \apj, 670, 156 


\bibitem[Damen et al.(2011)]{damen11} Damen, M., Labb{\'e}, I., van Dokkum, P.~G., et al.\ 2011, \apj, 727, 1 


\bibitem[Dalcanton et al.(2004)]{dalcanton04} Dalcanton, J.~J., Yoachim, P., \& Bernstein, R.~A.\ 2004, \apj, 608, 189 

\bibitem[De Barros et al.(2016)]{debarros16} De Barros, S., Vanzella, E., Amor{\'{\i}}n, R., et al.\ 2016, \aap, 585, A51 

\bibitem[Denicol{\'o} et al.(2002)]{denicolo02} Denicol{\'o}, G., Terlevich, R., \& Terlevich, E.\ 2002, \mnras, 330, 69 


\bibitem[Dressler et al.(2015)]{dressler15} Dressler, A., Henry, A., Martin, C.~L., et al.\ 2015, \apj, 806, 19 


\bibitem[Elbaz et al.(2007)]{elbaz07} Elbaz, D., Daddi, E., Le Borgne, D., et al.\ 2007, \aap, 468, 33 


\bibitem[Ellison et al.(2008)]{ellison08} Ellison, S.~L., Patton, D.~R., Simard, L., \& McConnachie, A.~W.\ 2008, \apjl, 672, L107 


\bibitem[Elvis et al.(2009)]{elvis09} Elvis, M., Civano, F., Vignali, C., et al.\ 2009, \apjs, 184, 158 


\bibitem[Erb et al.(2006)]{erb06} Erb, D.~K., Shapley, A.~E., Pettini, M., et al.\ 2006, \apj, 644, 813 


\bibitem[Ferland et al.(2013)]{ferland13} Ferland, G.~J., Porter, R.~L., van Hoof, P.~A.~M., et al.\ 2013, \rmxaa, 49, 137 


\bibitem[Filho et al.(2016)]{filho16} Filho, M.~E., S{\'a}nchez Almeida, J., Amor{\'{\i}}n, R., et al.\ 2016, \apj, 820, 109 


\bibitem[Finlator \& Dav{\'e}(2008)]{finlatordave08} Finlator, K., \& Dav{\'e}, R.\ 2008, \mnras, 385, 2181 


\bibitem[Fontana et al.(2006)]{fontana06} Fontana, A., Salimbeni, S., Grazian, A., et al.\ 2006, \aap, 459, 745 


\bibitem[Garilli et al.(2010)]{garilli10} Garilli, B., Fumana, M., Franzetti, P., et al.\ 2010, \pasp, 122, 827 

\bibitem[Gawiser et al.(2006)]{gawiser06} Gawiser, E., van Dokkum, P.~G., Herrera, D., et al.\ 2006, \apjs, 162, 1 

\bibitem[Giallongo et al.(2015)]{giallongo15} Giallongo, E., Grazian, A., Fiore, F., et al.\ 2015, \aap, 578, A83 

\bibitem[Gil de Paz \& Madore(2005)]{gildepazmadore05} Gil de Paz, A., \& Madore, B.~F.\ 2005, \apjs, 156, 345 

\bibitem[Grazian et al.(2015)]{grazian15} Grazian, A., Fontana, A., Santini, P., et al.\ 2015, \aap, 575, A96 

\bibitem[Grazian et al.(2016)]{grazian16} Grazian, A., Giallongo, E., Gerbasi, R., et al.\ 2016, \aap, 585, A48 

\bibitem[Guo et al.(2015)]{guo15} Guo, K., Zheng, X.~Z., Wang, T., \& Fu, H.\ 2015, \apjl, 808, L49 


\bibitem[Guo et al.(2016)]{guo16} Guo, Y., Koo, D.~C., Lu, Y., et al.\ 2016, \apj, 822, 103 


\bibitem[Guzm{\'a}n et al.(1997)]{guzman97} Guzm{\'a}n, R., Gallego, J., Koo, D.~C., et al.\ 1997, \apj, 489, 559 

\bibitem[H{\"a}gele et al.(2008)]{hagele08} H{\"a}gele, G.~F., D{\'{\i}}az, {\'A}.~I., Terlevich, E., et al.\ 2008, \mnras, 383, 209 

\bibitem[Heckman et al.(2011)]{heckman11} Heckman, T.~M., Borthakur, S., Overzier, R., et al.\ 2011, \apj, 730, 5 

\bibitem[Henry et al.(2013)]{henry13} Henry, A., Scarlata, C., Dom{\'{\i}}nguez, A., et al.\ 2013, \apjl, 776, L27 

\bibitem[Hopkins et al.(2014)]{hopkins14} Hopkins, P.~F., Kere{\v s}, D., O{\~n}orbe, J., et al.\ 2014, \mnras, 445, 581 


\bibitem[Hughes et al.(2013)]{hughes13} Hughes, T.~M., Cortese, L., Boselli, A., Gavazzi, G., \& Davies, J.~I.\ 2013, \aap, 550, A115 

\bibitem[Ilbert et al.(2005)]{ilbert05} Ilbert, O., Tresse, L., Zucca, E., et al.\ 2005, \aap, 439, 863

\bibitem[Ilbert et al.(2006)]{ilbert06} Ilbert, O., Arnouts, S., McCracken, H.~J., et al.\ 2006, \aap, 457, 841 


\bibitem[Ilbert et al.(2009)]{ilbert09} Ilbert, O., Capak, P., Salvato, M., et al.\ 2009, \apj, 690, 1236 

 
\bibitem[Ilbert et al.(2013)]{ilbert13} Ilbert, O., McCracken, H.~J., Le F{\`e}vre, O., et al.\ 2013, \aap, 556, A55 


\bibitem[Izotov et al.(2016)]{izotov16b} Izotov, Y.~I., Guseva, N.~G., Fricke, K.~J., \& Henkel, C.\ 2016, \mnras, 462, 4427 


\bibitem[Izotov et al.(2015)]{izotov15} Izotov, Y.~I., Guseva, N.~G., Fricke, K.~J., \& Henkel, C.\ 2015, \mnras, 451, 2251 


\bibitem[Izotov et al.(2016)]{izotov16} Izotov, Y.~I., Schaerer, D., Thuan, T.~X., et al.\ 2016, \mnras, 461, 3683 


\bibitem[Izotov \& Thuan(1999)]{izotovthuan99} Izotov, Y.~I., \& Thuan, T.~X.\ 1999, \apj, 511, 639 


\bibitem[Jaskot \& Oey(2014)]{jaskotoey14} Jaskot, A.~E., \& Oey, M.~S.\ 2014, Massive Young Star Clusters Near and Far: From the Milky Way to Reionization, 171 


\bibitem[Jaskot \& Oey(2013)]{jaskotoey13} Jaskot, A.~E., \& Oey, M.~S.\ 2013, \apj, 766, 91 


\bibitem[Juneau et al.(2014)]{juneau14} Juneau, S., Bournaud, F., Charlot, S., et al.\ 2014, \apj, 788, 88 


\bibitem[Juneau et al.(2011)]{juneau11} Juneau, S., Dickinson, M., Alexander, D.~M., \& Salim, S.\ 2011, \apj, 736, 104 


\bibitem[Kakazu et al.(2007)]{kakazu07} Kakazu, Y., Cowie, L.~L., \& Hu, E.~M.\ 2007, \apj, 668, 853 


\bibitem[Karachentsev et al.(2004)]{karachentsev04} Karachentsev, I.~D., Karachentseva, V.~E., Huchtmeier, W.~K., \& Makarov, D.~I.\ 2004, \aj, 127, 2031 


\bibitem[Kauffmann et al.(2003)]{kauffmann03} Kauffmann, G., Heckman, T.~M., White, S.~D.~M., et al.\ 2003, \mnras, 341, 33 


\bibitem[Kennicutt \& Evans(2012)]{kennicuttevans12} Kennicutt, R.~C., \& Evans, N.~J.\ 2012, \araa, 50, 531 


\bibitem[Kennicutt(1998)]{kennicutt98} Kennicutt, R.~C., Jr.\ 1998, \apj, 498, 541 


\bibitem[Kewley \& Dopita(2002)]{kewleydopita02} Kewley, L.~J., \& Dopita, M.~A.\ 2002, \apjs, 142, 35 


\bibitem[Kewley \& Ellison(2008)]{kewleyellison08} Kewley, L.~J., \& Ellison, S.~L.\ 2008, \apj, 681, 1183-1204 


\bibitem[Kniazev et al.(2004)]{kniazev04} Kniazev, A.~Y., Pustilnik, S.~A., Grebel, E.~K., Lee, H., \& Pramskij, A.~G.\ 2004, \apjs, 153, 429 


\bibitem[Kobayashi et al.(2007)]{kobayashi07} Kobayashi, C., Springel, V., \& White, S.~D.~M.\ 2007, \mnras, 376, 1465 


\bibitem[Kobulnicky et al.(1999)]{kobulnicky99} Kobulnicky, H.~A., Kennicutt, R.~C., Jr., \& Pizagno, J.~L.\ 1999, \apj, 514, 544 


\bibitem[Koekemoer et al.(2007)]{koekemoer07} Koekemoer, A.~M., Aussel, H., Calzetti, D., et al.\ 2007, \apjs, 172, 196 


\bibitem[Kudritzki et al.(2015)]{kudritzki15} Kudritzki, R.-P., Ho, I.-T., Schruba, A., et al.\ 2015, \mnras, 450, 342 


\bibitem[Kunth \& {\"O}stlin(2000)]{kunthostlin00} Kunth, D., \& {\"O}stlin, G.\ 2000, \aapr, 10, 1 


\bibitem[Lamareille et al.(2004)]{lamareille04} Lamareille, F., Mouhcine, M., Contini, T., Lewis, I., \& Maddox, S.\ 2004, \mnras, 350, 396 


\bibitem[Lara-L{\'o}pez et al.(2010)]{laralopez10} Lara-L{\'o}pez, M.~A., Cepa, J., Bongiovanni, A., et al.\ 2010, \aap, 521, L53 


\bibitem[Lara-L{\'o}pez et al.(2013)]{laralopez13} Lara-L{\'o}pez, M.~A., Hopkins, A.~M., L{\'o}pez-S{\'a}nchez, A.~R., et al.\ 2013, \mnras, 433, L35 


\bibitem[Le F{\`e}vre et al.(2015)]{lefevre15} Le F{\`e}vre, O., Tasca, L.~A.~M., Cassata, P., et al.\ 2015, \aap, 576, A79 


\bibitem[Le F{\`e}vre et al.(2003)]{lefevre03} Le Fevre, O., Vettolani, G., Maccagni, D., et al.\ 2003, \procspie, 4834, 173 


\bibitem[Lee et al.(2006)]{lee06} Lee, H., Skillman, E.~D., Cannon, J.~M., et al.\ 2006, \apj, 647, 970 


\bibitem[Lehmer et al.(2005)]{lehmer05} Lehmer, B.~D., Brandt, W.~N., Alexander, D.~M., et al.\ 2005, \apjs, 161, 21 

\bibitem[Leloudas et al.(2015)]{leloudas15} Leloudas, G., Schulze, S., Kr{\"u}hler, T., et al.\ 2015, \mnras, 449, 917 

\bibitem[Lequeux et al.(1979)]{lequeux79} Lequeux, J., Peimbert, M., Rayo, J.~F., Serrano, A., \& Torres-Peimbert, S.\ 1979, \aap, 80, 155 


\bibitem[Leroy et al.(2005)]{leroy05} Leroy, A., Bolatto, A.~D., Simon, J.~D., \& Blitz, L.\ 2005, \apj, 625, 763 


\bibitem[Lilly et al.(2013)]{lilly13} Lilly, S.~J., Carollo, C.~M., Pipino, A., Renzini, A., \& Peng, Y.\ 2013, \apj, 772, 119 

\bibitem[L{\'o}pez-S{\'a}nchez et al.(2012)]{lopezsanchez12} L{\'o}pez-S{\'a}nchez, {\'A}.~R., Dopita, M.~A., Kewley, L.~J., et al.\ 2012, \mnras, 426, 2630 

\bibitem[Lunnan et al.(2013)]{lunnan13} Lunnan, R., Chornock, R., Berger, E., et al.\ 2013, \apj, 771, 97 

\bibitem[Luo et al.(2008)]{luo08} Luo, B., Bauer, F.~E., Brandt, W.~N., et al.\ 2008, \apjs, 179, 19-36 

\bibitem[Ly et al.(2014)]{ly14} Ly, C., Malkan, M.~A., Nagao, T., et al.\ 2014, \apj, 780, 122 

\bibitem[Ly et al.(2015)]{ly15} Ly, C., Rigby, J.~R., Cooper, M., \& Yan, R.\ 2015, \apj, 805, 45 

\bibitem[Ly et al.(2016)]{ly16} Ly, C., Malhotra, S., Malkan, M.~A., et al.\ 2016, \apjs, 226, 5 

\bibitem[Ly et al.(2016)]{ly16} Ly, C., Malkan, M.~A., Rigby, J.~R., \& Nagao, T.\ 2016, \apj, 828, 67 

\bibitem[Madau \& Dickinson(2014)]{madaudickinson14} Madau, P., \& Dickinson, M.\ 2014, \araa, 52, 415 


\bibitem[Maiolino et al.(2008)]{maiolino08} Maiolino, R., Nagao, T., Grazian, A., et al.\ 2008, \aap, 488, 463 

\bibitem[Mannucci et al.(2009)]{mannucci09} Mannucci, F., Cresci, G., Maiolino, R., et al.\ 2009, \mnras, 398, 1915 

\bibitem[Mannucci et al.(2010)]{mannucci10} Mannucci, F., Cresci, G., Maiolino, R., Marconi, A., \& Gnerucci, A.\ 2010, \mnras, 408, 2115 

\bibitem[Mannucci et al.(2011)]{mannucci11} Mannucci, F., Salvaterra, R., \& Campisi, M.~A.\ 2011, \mnras, 414, 1263 

\bibitem[Marino et al.(2013)]{marino13} Marino, R.~A., Rosales-Ortega, F.~F., S{\'a}nchez, S.~F., et al.\ 2013, \aap, 559, A114 


\bibitem[Maseda et al.(2014)]{maseda14} Maseda, M.~V., van der Wel, A., Rix, H.-W., et al.\ 2014, \apj, 791, 17 

\bibitem[Mauduit et al.(2012)]{mauduit12} Mauduit, J.-C., Lacy, M., Farrah, D., et al.\ 2012, \pasp, 124, 714 

\bibitem[McGaugh(1991)]{mcgaugh91} McGaugh, S.~S.\ 1991, \apj, 380, 140 


\bibitem[Moll{\'a} et al.(2009)]{molla09} Moll{\'a}, M., Garc{\'{\i}}a-Vargas, M.~L., \& Bressan, A.\ 2009, \mnras, 398, 451 

\bibitem[Moustakas et al.(2010)]{moustakas10} Moustakas, J., Kennicutt, R.~C., Jr., Tremonti, C.~A., et al.\ 2010, \apjs, 190, 233-266 

\bibitem[Nagao et al.(2006)]{nagao06} Nagao, T., Maiolino, R., \& Marconi, A.\ 2006, \aap, 459, 85 


\bibitem[Nakajima \& Ouchi(2014)]{nakajima14} Nakajima, K., \& Ouchi, M.\ 2014, \mnras, 442, 900 


\bibitem[Noeske et al.(2007)]{noeske07} Noeske, K.~G., Weiner, B.~J., Faber, S.~M., et al.\ 2007, \apjl, 660, L43 

\bibitem[Olmo-Garcia et al.(2016)]{olmogarcia16} Olmo-Garcia, A., Sanchez Almeida, J., Munoz-Tunon, C., et al.\ 2016, arXiv:1611.07426 

\bibitem[Onodera et al.(2016)]{onodera16} Onodera, M., Carollo, C.~M., Lilly, S., et al.\ 2016, \apj, 822, 42 


\bibitem[Osterbrock(1989)]{osterbrock89} Osterbrock, D.~E.\ 1989, Research supported by the University of California, John Simon Guggenheim Memorial Foundation, University of Minnesota, et al.~Mill Valley, CA, University Science Books, 1989, 422 p.  

\bibitem[P{\'e}rez-Montero(2014)]{perezmontero14} P{\'e}rez-Montero, E.\ 2014, \mnras, 441, 2663 

\bibitem[P{\'e}rez-Montero \& Contini(2009)]{perezmonterocontini09} P{\'e}rez-Montero, E., \& Contini, T.\ 2009, \mnras, 398, 949 

\bibitem[P{\'e}rez-Montero \& D{\'{\i}}az(2005)]{perezmonterodiaz05} P{\'e}rez-Montero, E., \& D{\'{\i}}az, A.~I.\ 2005, \mnras, 361, 1063 

\bibitem[Papaderos \& {\"O}stlin(2012)]{papaderosostlin12} Papaderos, P., \& {\"O}stlin, G.\ 2012, \aap, 537, A126 

\bibitem[Papaderos et al.(1996)]{papaderos96} Papaderos, P., Loose, H.-H., Thuan, T.~X., \& Fricke, K.~J.\ 1996, \aaps, 120, 207 


\bibitem[Peeples et al.(2008)]{peeples08} Peeples, M.~S., Pogge, R.~W., \& Stanek, K.~Z.\ 2008, \apj, 685, 904-914


\bibitem[Pelupessy et al.(2004)]{pelupessy04} Pelupessy, F.~I., van der Werf, P.~P., \& Icke, V.\ 2004, \aap, 422, 55 

\bibitem[Peng et al.(2002)]{peng02} Peng, C.~Y., Ho, L.~C., Impey, C.~D., \& Rix, H.-W.\ 2002, \aj, 124, 266

\bibitem[Peng et al.(2010)]{peng10} Peng, C.~Y., Ho, L.~C., Impey, C.~D., \& Rix, H.-W.\ 2010, \aj, 139, 2097 

\bibitem[Pentericci et al.(2014)]{pentericci14} Pentericci, L., Vanzella, E., Fontana, A., et al.\ 2014, \apj, 793, 113 

\bibitem[Pettini \& Pagel(2004)]{pettinipagel04} Pettini, M., \& Pagel, B.~E.~J.\ 2004, \mnras, 348, L59 


\bibitem[Pierre et al.(2004)]{pierre04} Pierre, M., Valtchanov, I., Altieri, B., et al.\ 2004, \jcap, 9, 011 


\bibitem[Pilyugin et al.(2010)]{pilyugin10} Pilyugin, L.~S., V{\'{\i}}lchez, J.~M., Cedr{\'e}s, B., \& Thuan, T.~X.\ 2010, \mnras, 403, 896 

\bibitem[Press \& Schechter(1974)]{pressschechter74} Press, W.~H., \& Schechter, P.\ 1974, \apj, 187, 425 


\bibitem[Reddy et al.(2010)]{reddy10} Reddy, N.~A., Erb, D.~K., Pettini, M., Steidel, C.~C., \& Shapley, A.~E.\ 2010, \apj, 712, 1070 


\bibitem[Ribeiro et al.(2016)]{ribeiro16} Ribeiro, B., Le F{\`e}vre, O., Tasca, L.~A.~M., et al.\ 2016, \aap, 593, A22 


\bibitem[Robertson et al.(2015)]{robertson15} Robertson, B.~E., Ellis, R.~S., Furlanetto, S.~R., \& Dunlop, J.~S.\ 2015, \apjl, 802, L19 


\bibitem[Rola et al.(1997)]{rola97} Rola, C.~S., Terlevich, E., \& Terlevich, R.~J.\ 1997, \mnras, 289, 419 

\bibitem[Salim et al.(2014)]{salim14} Salim, S., Lee, J.~C., Ly, C., et al.\ 2014, \apj, 797, 126 

\bibitem[S{\'a}nchez Almeida et al.(2014)]{sanchezalmeida14b} S{\'a}nchez Almeida, J., Elmegreen, B.~G., Mu{\~n}oz-Tu{\~n}{\'o}n, C., \& Elmegreen, D.~M.\ 2014, \aapr, 22, 71 

\bibitem[S{\'a}nchez Almeida et al.(2014)]{sanchezalmeida14} S{\'a}nchez Almeida, J., Morales-Luis, A.~B., Mu{\~n}oz-Tu{\~n}{\'o}n, C., et al.\ 2014, \apj, 783, 45 

\bibitem[S{\'a}nchez Almeida et al.(2015)]{sanchezalmeida15} S{\'a}nchez Almeida, J., Elmegreen, B.~G., Mu{\~n}oz-Tu{\~n}{\'o}n, C., et al.\ 2015, \apjl, 810, L15 

\bibitem[S{\'a}nchez Almeida et al.(2016)]{sanchezalmeida16} S{\'a}nchez Almeida, J., P{\'e}rez-Montero, E., Morales-Luis, A.~B., et al.\ 2016, \apj, 819, 110 


\bibitem[S{\'a}nchez et al.(2012)]{sanchez12} S{\'a}nchez, S.~F., Kennicutt, R.~C., Gil de Paz, A., et al.\ 2012, \aap, 538, A8 


\bibitem[S{\'a}nchez et al.(2013)]{sanchez13} S{\'a}nchez, S.~F., Rosales-Ortega, F.~F., Jungwiert, B., et al.\ 2013, \aap, 554, A58 


\bibitem[S{\'a}nchez-Janssen et al.(2013)]{sanchezjanssen13} S{\'a}nchez-Janssen, R., Amor{\'{\i}}n, R., Garc{\'{\i}}a-Vargas, M., et al.\ 2013, \aap, 554, A20 

\bibitem[Santini et al.(2009)]{santini09} Santini, P., Fontana, A., Grazian, A., et al.\ 2009, \aap, 504, 751 


\bibitem[Santini et al.(2012)]{santini12} Santini, P., Fontana, A., Grazian, A., et al.\ 2012, \aap, 538, A33 


\bibitem[Savaglio(2010)]{savaglio10} Savaglio, S.\ 2010, Progenitors and Environments of Stellar Explosions, 39 

\bibitem[Schreiber et al.(2015)]{schreiber15} Schreiber, C., Pannella, M., Elbaz, D., et al.\ 2015, \aap, 575, A74 

\bibitem[Scoville et al.(2007)]{scoville07} Scoville, N., Aussel, H., Brusa, M., et al.\ 2007, \apjs, 172, 1 

\bibitem[S{\'e}rsic(1968)]{sersic68} S{\'e}rsic, J.~L.\ 1968, Boletin de la Asociacion Argentina de Astronomia La Plata Argentina, 13, 20 

\bibitem[Shapley et al.(2005)]{shapley05} Shapley, A.~E., Coil, A.~L., Ma, C.-P., \& Bundy, K.\ 2005, \apj, 635, 1006 


\bibitem[Sharma et al.(2016)]{sharma16} Sharma, M., Theuns, T., Frenk, C., et al.\ 2016, arXiv:1606.08688 


\bibitem[Skillman et al.(1989)]{skillman89} Skillman, E.~D., Kennicutt, R.~C., \& Hodge, P.~W.\ 1989, \apj, 347, 875 

\bibitem[Smit et al.(2014)]{smit14} Smit, R., Bouwens, R.~J., Labb{\'e}, I., et al.\ 2014, \apj, 784, 58 

\bibitem[Sparre et al.(2015)]{sparre15} Sparre, M., Hayward, C.~C., Feldmann, R., et al.\ 2015, arXiv:1510.03869 


\bibitem[Spitoni et al.(2010)]{spitoni10} Spitoni, E., Calura, F., Matteucci, F., \& Recchi, S.\ 2010, \aap, 514, A73 

\bibitem[Stark et al.(2017)]{stark17} Stark, D.~P., Ellis, R.~S., Charlot, S., et al.\ 2017, \mnras, 464, 469 

\bibitem[Stasi{\'n}ska(2010)]{stasinska10} Stasi{\'n}ska, G.\ 2010, Stellar Populations - Planning for the Next Decade, 262, 93 

\bibitem[Stasi{\'n}ska et al.(2015)]{stasinska15} Stasi{\'n}ska, G., Izotov, Y., Morisset, C., \& Guseva, N.\ 2015, \aap, 576, A83 


\bibitem[Stoughton et al.(2002)]{stoughton02} Stoughton, C., Lupton, R.~H., Bernardi, M., et al.\ 2002, \aj, 123, 485 

\bibitem[Taniguchi et al.(2007)]{taniguchi07} Taniguchi, Y., Scoville, N., Murayama, T., et al.\ 2007, \apjs, 172, 9 

\bibitem[Tasca et al.(2015)]{tasca15} Tasca, L.~A.~M., Le F{\`e}vre, O., Hathi, N.~P., et al.\ 2015, \aap, 581, A54 

\bibitem[Tasca et al.(2016)]{tasca16} Tasca, L.~A.~M., Le Fevre, O., Ribeiro, B., et al.\ 2016, arXiv:1602.01842 


\bibitem[Tassis et al.(2008)]{tassis08} Tassis, K., Kravtsov, A.~V., \& Gnedin, N.~Y.\ 2008, Dark Galaxies and Lost Baryons, 244, 256 

\bibitem[Taylor et al.(2009)]{taylor09} Taylor, E.~N., Franx, M., van Dokkum, P.~G., et al.\ 2009, \apjs, 183, 295 

\bibitem[Telford et al.(2016)]{telford16} Telford, O.~G., Dalcanton, J.~J., Skillman, E.~D., \& Conroy, C.\ 2016, \apj, 827, 35  


\bibitem[Terlevich et al.(1991)]{terlevich91} Terlevich, R., Melnick, J., Masegosa, J., Moles, M., \& Copetti, M.~V.~F.\ 1991, \aaps, 91, 285 


\bibitem[Th{\"o}ne et al.(2015)]{thone15} Th{\"o}ne, C.~C., de Ugarte Postigo, A., Garc{\'{\i}}a-Benito, R., et al.\ 2015, \mnras, 451, L65 

\bibitem[Tremonti et al.(2004)]{tremonti04} Tremonti, C.~A., Heckman, T.~M., Kauffmann, G., et al.\ 2004, \apj, 613, 898 


\bibitem[Troncoso et al.(2014)]{troncoso14} Troncoso, P., Maiolino, R., Sommariva, V., et al.\ 2014, \aap, 563, A58 

\bibitem[van der Wel et al.(2011)]{vanderwel11} van der Wel, A., Straughn, A.~N., Rix, H.-W., et al.\ 2011, \apj, 742, 111 

\bibitem[Vanzella et al.(2016)]{vanzella16} Vanzella, E., de Barros, S., Vasei, K., et al.\ 2016, \apj, 825, 41 

\bibitem[Whitaker et al.(2014)]{whitaker14} Whitaker, K.~E., Franx, M., Leja, J., et al.\ 2014, \apj, 795, 104 


\bibitem[Wuyts et al.(2012)]{wuyts12} Wuyts, E., Rigby, J.~R., Sharon, K., \& Gladders, M.~D.\ 2012, \apj, 755, 73 


\bibitem[Wuyts et al.(2013)]{wuyts13} Wuyts, S., F{\"o}rster Schreiber, N.~M., Nelson, E.~J., et al.\ 2013, \apj, 779, 135 


\bibitem[Xue et al.(2012)]{xue12} Xue, Y.~Q., Wang, S.~X., Brandt, W.~N., et al.\ 2012, \apj, 758, 129 


\bibitem[Zahid et al.(2012)]{zahid12} Zahid, H.~J., Dima, G.~I., Kewley, L.~J., Erb, D.~K., \& Dav{\'e}, R.\ 2012, \apj, 757, 54 

\bibitem[Zahid et al.(2012)]{zahid12b} Zahid, H.~J., Bresolin, F., Kewley, L.~J., Coil, A.~L., \& Dav{\'e}, R.\ 2012, \apj, 750, 120

\bibitem[Zamojski et al.(2007)]{zamojski07} Zamojski, M.~A., Schiminovich, D., Rich, R.~M., et al.\ 2007, \apjs, 172, 468 

\bibitem[Zhao et al.(2010)]{zhao10} Zhao, Y., Gao, Y., \& Gu, Q.\ 2010, \apj, 710, 663 

\end{thebibliography}




\clearpage
\newpage

\begin{appendix}
\section{Derivation of metallicity with other methods} \label{appendixa}

We explore in this section the main differences and similarities between the metallicities based on the code HCm and those derived with the widely-used strong-line methods. Strong-line methods are based on the physical properties of HII regions, in which a relation has been found between emission-line intensities and oxygen abundances \citep[see][]{perezmonterodiaz05}. The calibrations can be derived empirically or through photoionization models. Empirical calibrations are provided by many authors, e.g. \citet{pettinipagel04,marino13,maiolino08,perezmonterocontini09}. These methods are generally consistent with direct observations but, on the contrary, can be sistematically different when good measurements of the O$4363\,$ line are available \citep{stasinska10}. On the other side, photoionization models typically require numerous constraints, e.g. not only the emission line ratios, but also the stellar content and the nebular gas distribution. One important problem is that many of the most widely used calibrations based on photoionization models \citep[e.g.][]{mcgaugh91,kewleydopita02} also give systematic overabundances with respect to the direct method. These differences can be up to $0.7$\ dex depending on the models and the $Z$ range \citep{kewleyellison08,moustakas10,stasinska10,lopezsanchez12}. 

In this work we compare our results with the calibrations of \citet{maiolino08}, \citet{mcgaugh91}, \citet{pettinipagel04} and \citet{marino13}. After providing a brief description of each method and showing the results, we also provide polynomial fittings which can be used for conversion between those calibrations and HCm-consistent metallicities. Then we will check the consistency of the MZR derived in section \ref{MZR} under the use of different methods for deriving the oxygen abundance. 

\subsection{Comparison with strong-line calibrations}

\begin{figure*}[t]
\centering
   \includegraphics[angle=0,width=7.0cm]{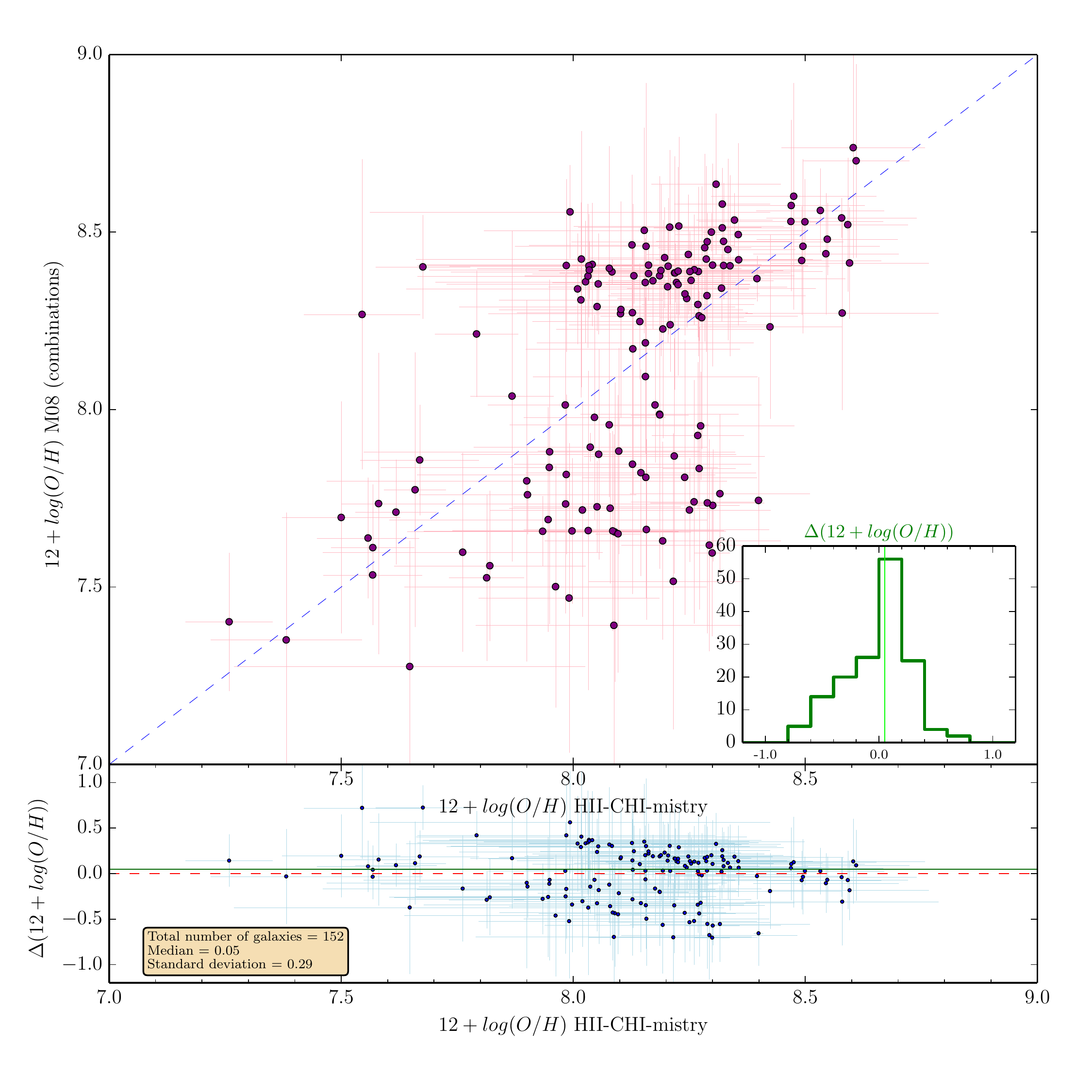} 
   \includegraphics[angle=0,width=7.0cm]{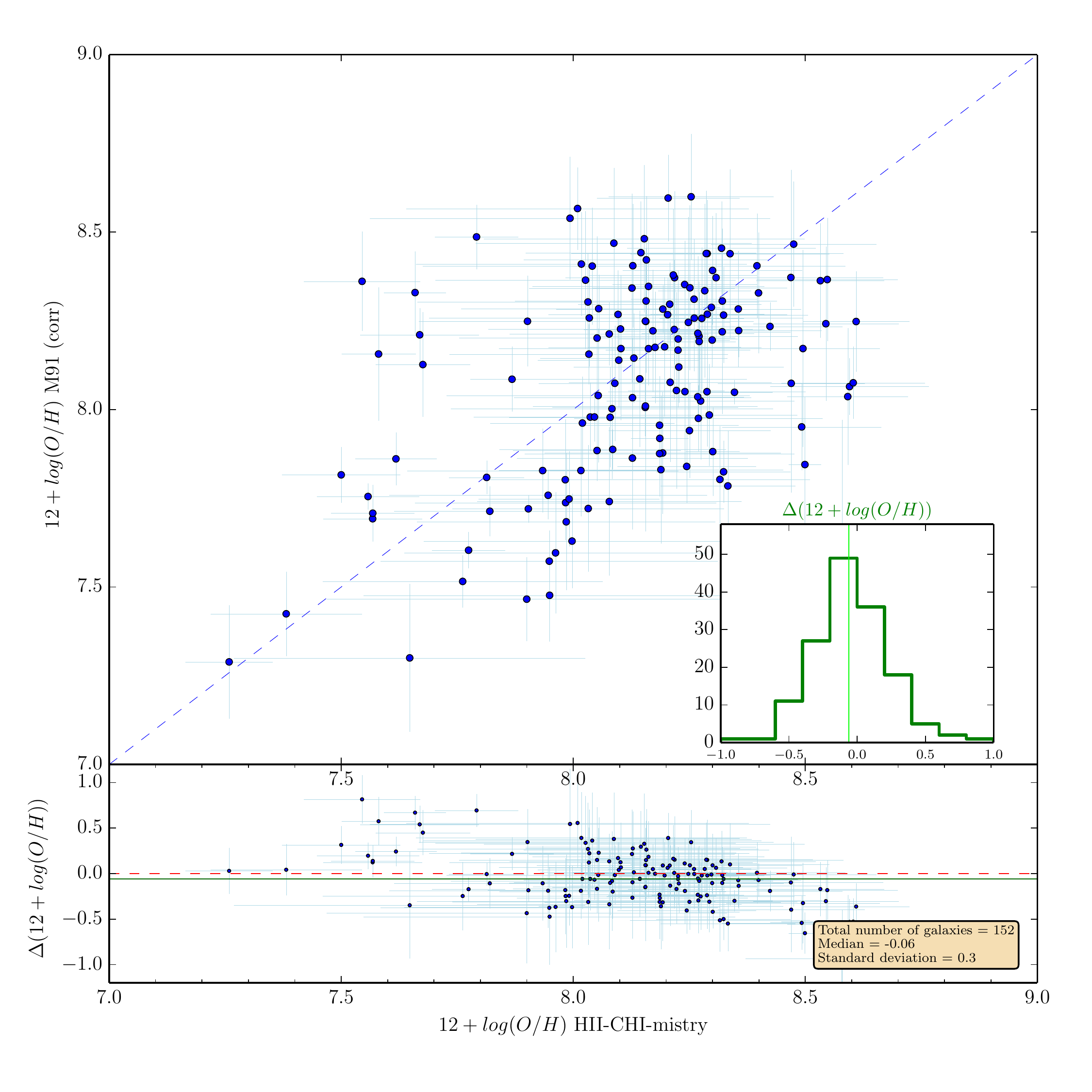} 
   \includegraphics[angle=0,width=7.0cm]{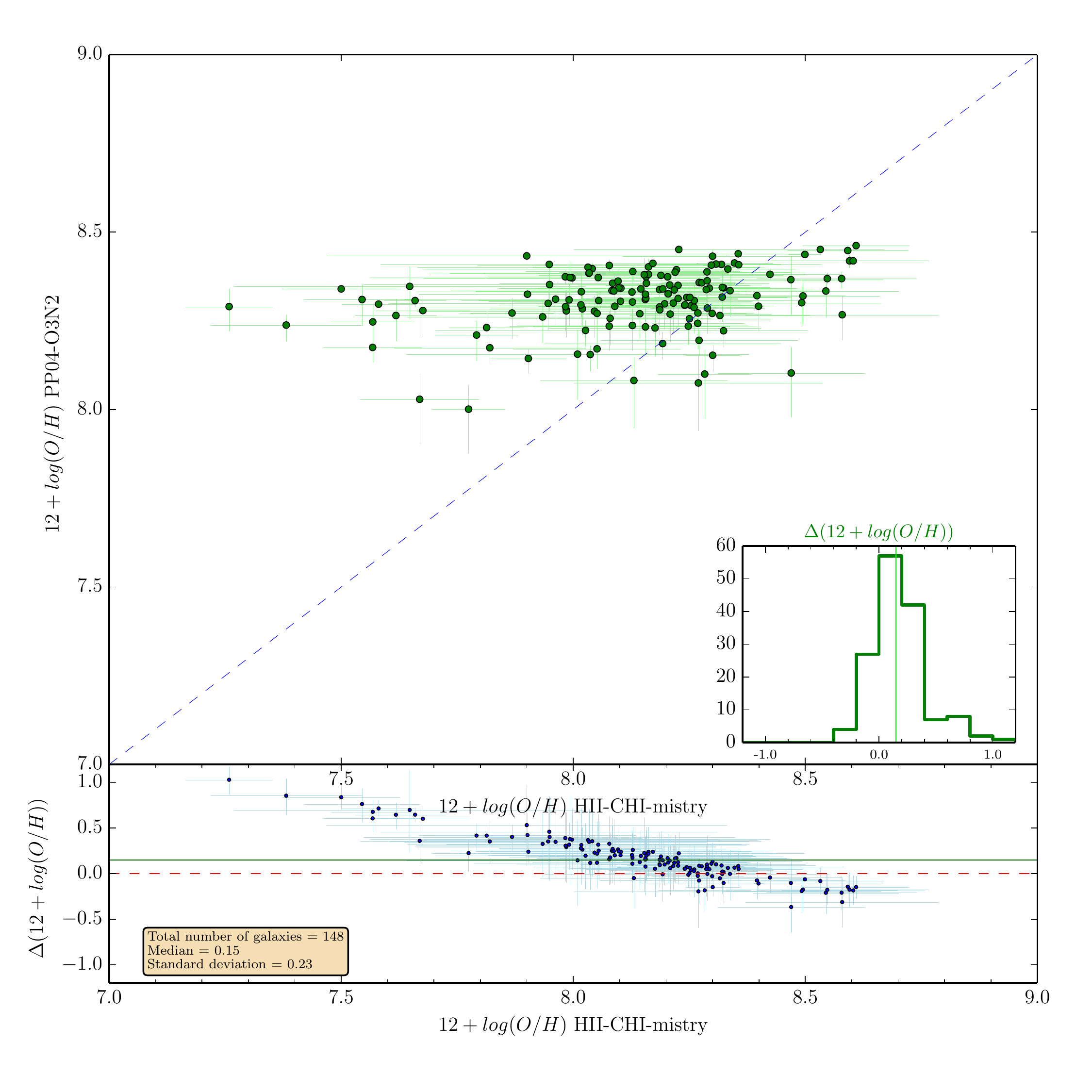} 
   \includegraphics[angle=0,width=7.0cm]{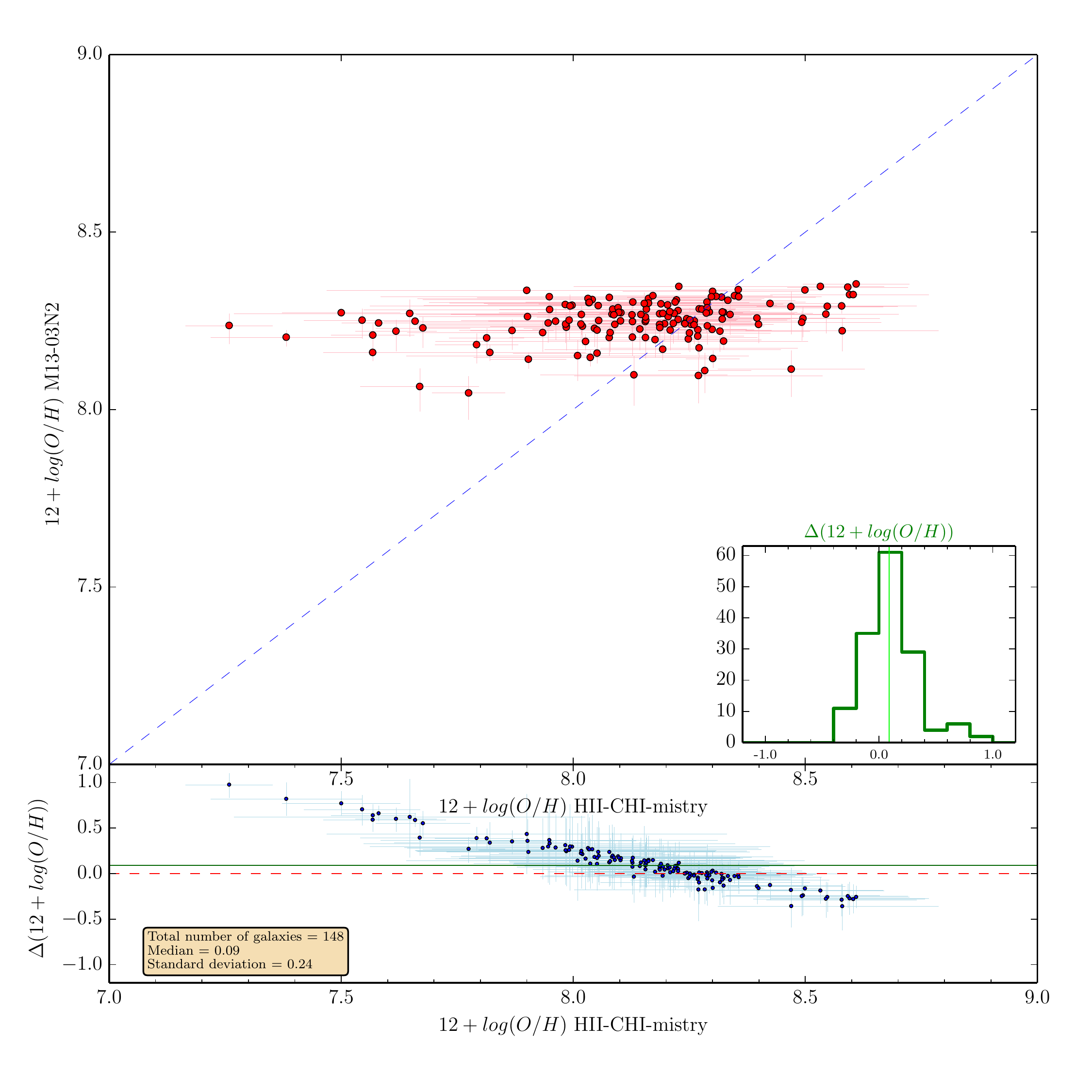}
   \caption{\small Comparison between the metallicity derived with the code HCm and different strong-line calibrations. \textit{Top}: a combination of calibrations by \citet{maiolino08} and corrected \citet{mcgaugh91} calibration. \textit{Bottom}: calibrations by \citet{pettinipagel04} and \citet{marino13}. A histogram is shown in each panel displaying the difference between the matallicities estimated with each of the 4 strong-line methods and HCm. In the bottom part the difference is plotted as a function of HCm values.}
   \label{strong-line}
\end{figure*} 

The method of \citet{maiolino08} (M08) is based on two different calibrations for the lower and higher metallicity branches. At high metallicity $\Delta (12+log(O/H)) > 8.35$ it is based on photoionization models from \citet{kewleydopita02}, while at lower metallicity it adopts an empirical calibration derived from a sample of metal-poor galaxies from \citet{nagao06} with auroral line detection and $T_e$-based metallicities. The comparison in Fig. \ref{strong-line} between M08 metallicities and our values shows some systematics around $12+log(O/H) \simeq 8$, likely due to the $R_{23}$ degeneracy: at metallicities $12+log(O/H) \leq 8$ M08 calibrations tend to give lower values ($\sim 0.3-0.4$ dex, with peaks of $-0.6$ dex) than ours, while at higher metallicities ($8 \leq 12+log(O/H) \leq 8.4$) produces higher values by the same amount. These different trends smear out when we consider the whole metallicity range, and the two medians are very close ($\Delta (12+log(O/H)) = 0.05$). The overall dispersion is about $\sim 0.3$ dex in the full range of Z. 

The \citet{mcgaugh91} calibration (M91) is based on detailed H II region models. The M91 method used for comparison is corrected to include variations of the ionization parameter inside the nebula, and adopts the [\nii]/[\oii] line ratio to break the $R_{23}$ degeneracy, applying then analytic expressions for the M91 lower and upper branches given in \citet{kobulnicky99}. In the upper-right part of Fig. \ref{strong-line} we show the comparison between the results obtained with HCm and the corrected version of M91 calibration (hereafter M$91_{corr}$). Even though both methods are based on similar photoionization models, the dispersion of the points around the $1$:$1$ relation is still significant ($\sim 0.25$ dex). We can find galaxies with $12+log(O/H)_{HCm}$ around $8$ (where $R_{23}$ peaks) having discrepancies with respect to M$91_{corr}$ up to $0.6-0.7$ dex. These discrepancies can be justified both with the way the ionization parameter enters the model estimation and with modifications in the theoretical modeling of stellar atmospheres at higher redshifts, where we find typically more massive and more metal-poor ionizing stars compared to Local Universe. 

Finally, PP04 and M13 O3N2 methods were derived indipendently by \citet{pettinipagel04} and \citet{marino13} to fit the observed relationships between O3N2 parameters, defined as O3N2 = [\oiii]/\hb)/([\nii]/\ha), and the $T_e$-based metallicities for samples of extragalactic HII regions. The first adopts the catalogue of 137 objects compiled by \citet{denicolo02}, while M13 include $603$ HII regions extracted from the literature \citep[e.g.][]{pilyugin10} and from the CALIFA survey \citep[see][]{sanchez12}. Therefore we can refer to the PP04 and M13 methods as empirical $T_e$-based calibrations. The comparison of Fig. \ref{strong-line} (bottom) shows that metallicities derived using PP04 and M13 calibrations span a narrow range $8.0 \leq (12+log(O/H) \leq 8.4$. Even though the dispersion is similar to previous calibrations ($\simeq 0.24\ $ dex and $\simeq 0.15\ $) we see that they give systematic overabundances in the low metallicity range, with differences as high as $\simeq + 0.7\ $ dex at $12+log(O/H) < 7.6$. 

\subsection{MZR derived with strong-line methods}

\begin{figure}[t!]
  \resizebox{\hsize}{!}{\includegraphics[angle=0,width=9.4cm]{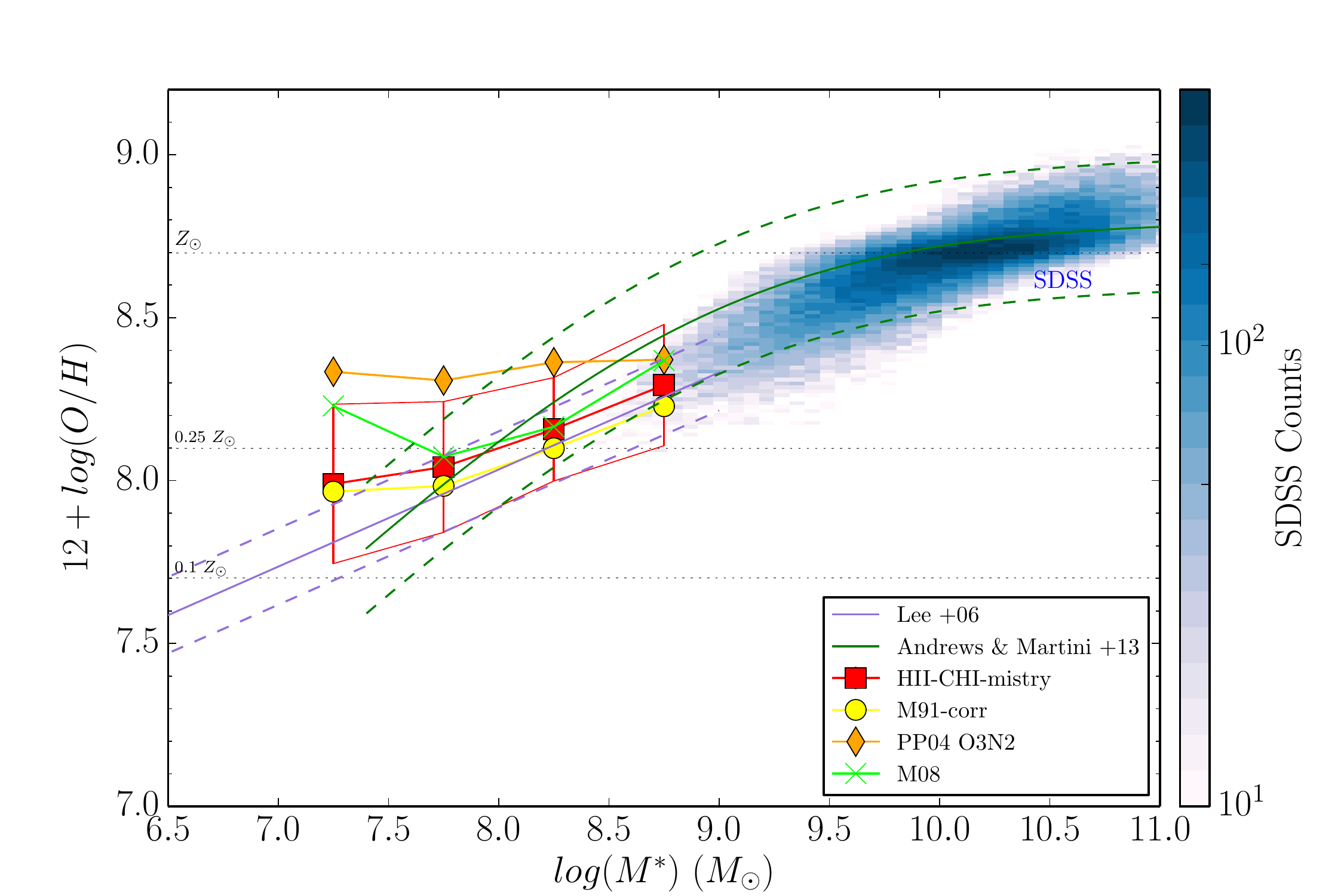}}
  \caption{\small Comparison between the MZR derived with different strong line calibrations: PP04 O3N2 and M13 (orange diamonds), M08 (green crosses) and M91$_{corr}$ (yellow circles). We see that HCm and M91 relations show a similar trend in the whole metallicity range. Green lines are the \citet{andrewsmartini13} MZR and $\pm 1\sigma$ uncertainties.
  }\label{MZR-strongline}
   \end{figure}
   
We study the effects of using different metallicity calibrations on the mass-metallicity relation. We consider the strong-line methods presented in previous section (M08, M91, PP04 and M13) and plot for each galaxy the stellar mass vs. the new oxygen abundances (see Fig. \ref{MZR-strongline}. As we have seen before, PP04 and M13 give systematic overabundances at lower masses when compared to HCm, so that their medians are above our $1\sigma$ limits. The M08 calibration is mainly consistent with our results, but gives higher abundances ($\sim 0.25\ $ dex) in the lowest and highest mass bins. Finally, the $MZR_{M91-corr}$ shows a trend very similar to our MZR. 
Overall, it turns out from Fig. \ref{MZR-strongline} that choosing a different metallicity calibration affects both the slope and the normalization of the mass-metallicity relation. Even though our MZR overlaps on a large area the MZR of AM13 (showing consistent metallicity ranges within the errors at fixed stellar mass), we can not recover its slope using any of the three other metallicity calibrators. In particular, PP04 and M08 have the opposite effect, with higher abundances toward lower masses.

\clearpage

\begin{table*}
\section{Tables}\label{appendixb}
\caption{{\normalsize Basic information about VUDS selected galaxies}}
\label{Table1}
\centering
\tiny
\begin{tabular}{|cccccc|}
\noalign{\smallskip}
\hline\hline
\noalign{\smallskip}
VUDS ID & $\alpha (J2000$) & $\delta (J2000$) & z &  $i_{AB}$ & $I_{AB,0}$\\
(1) & (2) & (3) & (4) & (5) & (6) \\
\hline
\noalign{\smallskip}
\vspace{0.15mm}
  510086862 & 149.661556 & 1.872843 & 0.1268  & 24.801 & -13.522   \\
  510376669 & 149.87759	 & 1.786568 & 0.531   &	24.573 & -16.459   \\
  510146174 & 149.790243 & 1.988252 & 0.4781  & 24.995 & -16.689   \\
  510165275 & 150.125865 & 2.0236   & 0.4787  & 24.368 & -17.282   \\
  510175664 & 150.113065 & 2.042952 & 0.5018  & 24.976 & -17.181   \\
  510229076 & 150.049327 & 2.137676 & 0.2197  & 24.867 & -14.698   \\
  510329403 & 150.13033  & 1.772111 & 0.1845  & 24.639 & -15.012   \\
  510330378 & 150.08737  & 1.763878 & 0.3374  & 24.374 & -16.478   \\
  510352169 & 149.99831  & 1.786842 & 0.6282  & 23.792 & -17.877   \\
  510353245 & 149.98308  & 1.782065 & 0.257   & 24.527 & -15.539   \\
\noalign{\smallskip}                                                         
\hline                                                                          
\hline
\end{tabular}                                             
\normalsize                      
\begin{list}{}{}
\item Columns: (1) VUDS identification number; (2) and (3) equatorial coordinates (J2000), right ascension and declination, respectively; (4) VUDS spectroscopic redshift; (5) $i$-band magnitude, used for the selection of galaxies in VUDS (6) Rest-frame absolute magnitude in the i band coming from SED fitting, as described in section \ref{masses}.
(The entire version of this table for the full sample of SFDGs is available {\it \emph{online}}). 
\end{list}
\end{table*}


\begin{sidewaystable}{ht}


\rule{-1.15cm}{20.101cm}    
\tiny
\begin{tabular}{|ccccccccccccccccc|}
\noalign{\smallskip}
\hline\hline
\noalign{\smallskip}
VUDS ID & [\oii]  & W([\oii]) & [\neiii]  & \hg  & [\oiii] & \hb  & W([\hb]) & [\oiii]  & [\oiii]  & W([\oiii]) &  \ha  & W(\ha) &  [\nii]  &  [\sii]   & [\sii]  & c(\hb) \\
 & $\lambda 3727$ &  &  $\lambda 3868$ & & $\lambda 4363$ &  &  &  $\lambda 4959$  & $\lambda 5007$ &  & &  & $\lambda 6583$ & $\lambda 6716$ & $\lambda 6731$ & \\
(1) & (2) & (3) & (4) & (5) & (6) & (7) & (8) & (9) & (10) & (11) & (12) & (13) & (14) & (15) & (16) & (17) \\
\hline
\noalign{\smallskip}
\vspace{0.05mm}
510086862  & 13.7 $\pm$ 1.5  & 80.3  $\pm$ 4 & 3.4  $\pm$ 0.9    & -  & - & 5.5   $\pm$ 0.8  & 31.5  $\pm$ 3 	 &  3.3   $\pm$ 1.0  & 10.1  $\pm$ 1.0  & 52.0    $\pm$ 7.0  & 12.5 $\pm$ 2 & 80	$\pm$ 15  & 1.5  $\pm$ 0.4    & -  & -   & 0.06 $^c$ \\
510146174  & 9.8  $\pm$ 1      & 158	$\pm$ 60 & 2.1	$\pm$ 0.4   & 3	 $\pm$ 0.5   & -  & 7.4   $\pm$ 1.5	& 106	$\pm$ 37 &  11.9  $\pm$ 1.6   & 39.6  $\pm$ 0.8  & 560.0   $\pm$ 150.0      & -  & -  & -   & -  & -                     & 0.47 $^c$ \\
510165275  & 20.8 $\pm$ 1.5    & 75	$\pm$ 13 & 9.4	$\pm$ 0.6      & 12	 $\pm$ 1.5   & 3.2   $\pm$ 0.5      & 23.5  $\pm$ 0.7	& 150	$\pm$ 15  &  35.5  $\pm$ 1.5  & 102.4 $\pm$ 1.3	  & 661.0   $\pm$ 170.0  & -  & - & -    & - & - & 0.23 $^b$ \\
510175664  & 12.1 $\pm$ 1.4    & 92	$\pm$ 30  & 4.7	$\pm$ 1       & -  & -   & 14.7  $\pm$ 1.2	& 135	$\pm$ 70 &  17.1  $\pm$ 0.9 & 48.9  $\pm$ 1.2	  & 648.0   $\pm$ 250.0      & -  & -  & -  & -    & -                           & 0.13 $^c$ \\
510229076  & 6.4  $\pm$ 0.8    & 49	$\pm$ 17 & 2	$\pm$ 0.2  & 3.3 $\pm$ 0.6   & - & 5.8 $\pm$ 1.6 & 44.5 $\pm$ 15 & 5.4 $\pm$ 0.5 & 17.4  $\pm$ 1.4 & 109.0 $\pm$ 36.0 & 25.1 $\pm$ 1.6 & 116 $\pm$ 36 & - & 4.9 $\pm$ 1 & 0.6 $\pm$ 0.4  & 0.59 $^a$ \\
510329403  & 6.8  $\pm$ 1.3    & 59	$\pm$ 12   & 2.5  $\pm$ 0.9    & -   & -  & 3.6   $\pm$ 0.8	& 12.7  $\pm$ 2.7 &  4.1   $\pm$ 0.9 & 11.4  $\pm$ 1.0	  & 37.0    $\pm$ 7.0  & 12   $\pm$ 1.5	& 50	$\pm$ 11    & -  & -  & -        & 0.23 $^a$  \\
510330378  & 15   $\pm$ 2      & 130	$\pm$ 60 & -    & 3	 $\pm$ 0.6   & -    & 13.5  $\pm$ 1.1	& 66	$\pm$ 13 &  23.3  $\pm$ 1.5  & 67.5  $\pm$ 1.5 & 274.0   $\pm$ 44.0    & 42   $\pm$ 4  & 260 $\pm$ 120 & 3.7  $\pm$ 1.1 & -  & - & 0.14 $^a$  \\
510352169  & 26.4 $\pm$ 2.6    & 50	$\pm$ 11  & 29.7  $\pm$ 3.4    & -   & -    & 50.8  $\pm$ 4.0	& 195	$\pm$ 34  &  94.2  $\pm$ 5.0  & 247.0 $\pm$ 3.0	  & 670.0   $\pm$ 190.0  & -  & - & -     & -     & -                            & 0.36 $^c$  \\
510353245  & 12   $\pm$ 2      & 24.3	$\pm$ 5 & -   & -    & -    & 9.0   $\pm$ 1.6	& 25	$\pm$ 6  &  13.0  $\pm$ 2.5 & 39.5  $\pm$ 2.5  & 122.0   $\pm$ 24.0    & 17.1 $\pm$ 2	& 116	$\pm$ 46       & -   & -  & -                    & 0.31 $^c$ \\
510376669  & 9.9  $\pm$ 0.7    & 44	$\pm$ 11	& 6.6	$\pm$ 0.7      & 7.2	 $\pm$ 1     & 3.1   $\pm$ 0.4 & 24.0 $\pm$ 1.6 & 184 $\pm$ 60 &  55.6  $\pm$ 1.5  & 116.0 $\pm$ 2.0  & 950.0   $\pm$ 300.0  & -  & -   & -  & -  & -    & 0.19 $^c$ \\
\noalign{\smallskip}                                                         
\hline                                                                          
\hline
\end{tabular}  
\normalsize 

\caption{{Emission line fluxes and EWs of selected galaxies in VUDS}}
\label{Table2}
\normalsize

\begin{list}{}{}
\item Columns: 
(1) VUDS identification number; (2) to (16) Fluxes and EWs of the emission lines measured with the task \textit{splot} of IRAF. The fluxes are given in $10^{-18} erg\ cm^{-2} s^{-1}$, while the equivalent widths (W) are in \AA. No extinction correction has been applied to these fluxes. (17) {Reddening constant derived from $(a)$ \ha/\hb\ or $(b)$ \hg/\hb\ ratios, whenever possible, or $(c)$ the SED best-fitting for those galaxies where $(a)$ and $(b)$ cannot be measured or where they produce a negative extinction (i.e., \ha/\hb$<$\,2.82 or \hg/\hb$<$\,0.47, assuming Case B recombination with  $T_e=$\,2$\times$10$^4$K, $n_e=$100\,cm$^{-3}$)}.
(The entire version of this table for the full sample of ELGs is available {\it \emph{online}}).
\end{list}

\end{sidewaystable}

\begin{table*}[h!]
\caption{{\normalsize Derived physical properties of intermediate-redshift ELGs in VUDS}}
\label{Table3}
\centering
\begin{tabular}{lccccccc}
\noalign{\smallskip}
\hline\hline
\noalign{\smallskip}
VUDS ID & SFR & $M_\ast$ & $12+log(O/H)$ & $r_e$ &   $q$   & $\mu_{gas}$ & $f_{gas}$  \\
(1) & (2) & (3) & (4) & (5) & (6) & (7) & (8)  \\
\hline
\hline
\noalign{\smallskip}
\vspace{0.15mm}

510086862  & -2.45       $\pm$ 0.06    & 7.02 $\pm$ 0.01  &  8.47   $\pm$ 0.20	& 0.34  $\pm$ 0.02  & 0.44    $\pm$   0.03            &  12.06    & 0.33      \\
510146174  & -0.22 	 $\pm$ 0.08    & 7.84 $\pm$ 0.2  &  8.27   $\pm$ 0.14	& 0.50  $\pm$ 0.03  & 0.69    $\pm$   0.05            &  276.24   & 0.79      \\
510165275  & -0.11	 $\pm$ 0.01    & 7.45 $\pm$ 0.14  &  7.56   $\pm$ 0.11	& 0.68  $\pm$ 0.01  & 0.27    $\pm$   0.02            &  212.59   & 0.93      \\
510175664  & -0.34	 $\pm$ 0.03    & 8.44 $\pm$ 0.35 &  7.76   $\pm$ 0.22	& 0.18  $\pm$ 0.01  $^A$   & 0.61    $\pm$   0.10     &  ...	  & ...       \\ 
510229076  & -1.20	 $\pm$ 0.03    & 7.03 $\pm$ 0.12  &  8.03   $\pm$ 0.30	& 0.24  $\pm$ 0.01	   & 0.68    $\pm$   0.05     &  155.08   & 0.75     \\
510329403  & -1.97	 $\pm$ 0.05    & 7.16 $\pm$ 0.17  &  8.22   $\pm$ 0.27	& 0.90  $\pm$ 0.06	   & 0.53    $\pm$   0.03     &  6.71	  & 0.58     \\
510330378  & -0.84	 $\pm$ 0.04    & 7.64 $\pm$ 0.21  &  8.05   $\pm$ 0.17	& 0.52  $\pm$ 0.07	   & 0.69    $\pm$   0.05     &  93.44    & 0.68     \\
510352169  & 0.77	 $\pm$ 0.03    & 8.00 $\pm$ 0.09  &  7.82   $\pm$ 0.24	& 0.42  $\pm$ 0.01  $^A$   & 0.52    $\pm$   0.02     &  ...	  & ...      \\ 
510353245  & -1.35	 $\pm$ 0.05    & 7.56 $\pm$ 0.13  &  8.19   $\pm$ 0.28	& 0.50  $\pm$ 0.02	   & 0.65    $\pm$   0.04     &  43.02    & 0.52     \\
510376669  & 0.02	 $\pm$ 0.03    & 7.47 $\pm$ 0.1  &  7.57   $\pm$ 0.11	& 0.53  $\pm$ 0.01  $^A$   & 0.15    $\pm$   0.03     &  ...	  & ...	     \\

\noalign{\smallskip}							     											       
\hline							      			
\hline
\end{tabular}	
\normalsize								
\begin{list}{}{}
\item Notes:   
(1) VUDS ID; 
(2) Star formation rates (in log($M_\odot$/yr) from \ha\ or \hb\ luminosity assuming \citet{chabrier03} IMF and a theoretical ratio $\ha/\hb=2.82$; uncertainties in SFR account for the propagation of errors in line fluxes and reddening;
(3) Stellar masses (in log$M_\odot$) and 1$\sigma$ uncertainties from SED fitting, as described in section \ref{masses} 
(4) Gas-phase metallicity derived from the code HII-CHI-mistry. The code also returns an error for the estimation; 
(5) Effective radius (in $kpc$), derived applying GALFIT to \textit{F814W}-band ACS-HST images as described in section \ref{morphology-sizes}. 
We put a flag (A,B,C,D,E) that tells how well did GALFIT converge to a particular S\'ersic profile: [A] Fit converged but with fixed S\'ersic index n; [B] Galaxy fit with fixed radius, fixed S\'ersic index and fixed axis ratio; [C] GALFIT did not converge within the expected number of iterations (unstable solution); [D] No GALFIT results due to lack of imaging data; [E] GALFIT did not converge at all, no physical solution (galaxy too faint, too irregular). If no flag are present then GALFIT converged with no errors.
(6) Ratio between the minor and major axis of the ellipse enclosing half of the total luminosity of the galaxy in the \textit{F814W} filter;
(7) and (8) Gas surface density ($\mu_{gas}$, in $M_\odot pc^{-2}$) and gas fraction $f_{gas}$, which were derived from the SFR surface densities assuming a Kennicutt-Schmidt law with exponent $1.4$, as described in section \ref{morphology-sizes}. For $\mu_{gas}$ and $f_{gas}$ we should consider an uncertainty of at least $0.2$-$0.3$ dex ($\sim$ a factor of two), which comes both from the uncertainties of the size and also from the KS law (which has an average dispersion over different types of galaxies of $\sim 0.2\ $ dex).
\end{list}
\end{table*}

%
\end{appendix}
\clearpage

\end{document}